\documentclass[a4paper,11pt]{article}
\pdfoutput=1 

\usepackage{jheppub} 

\usepackage[T1]{fontenc} 
\usepackage{fontawesome}
\usepackage[table]{xcolor}
\usepackage[caption=false]{subfig}
\usepackage{epsfig}
\usepackage{colortbl}
\usepackage{tikz-feynman}
\usepackage{hepnames}
\usepackage{adjustbox}
\usepackage{multirow}
\usepackage{slashed}
\usepackage{lipsum}
\usepackage{chngcntr}
\usepackage{bigints}
\usepackage[normalem]{ulem}
\usepackage{amsmath}
\usepackage{textcomp}
\usepackage{cleveref}
\usepackage{chemarrow}
\usepackage{booktabs}
\usepackage{float}
\usepackage{amsfonts}
\usepackage{hhline}
\newcommand{\LK}[1]{{\color{purple} #1}}

\usepackage{silence}
\tikzfeynmanset{warn luatex=false}

\DeclareUnicodeCharacter{2212}{\ensuremath{-}}

\title{\boldmath Benchmark Dark SMEFT Scenarios from the Interplay of Flavor Physics and Dark Matter Searches}
\author[a]{Lipika Kolay,}
\author[b]{Soumitra Nandi,}
\author[c]{and  Ipsita Ray.}
\affiliation[a]{Department of Physics, Indian Institute of Technology Guwahati,\\North Guwahati, Assam-781039, India,}
\affiliation[b]{Department of Physics, Indian Institute of Technology Gandhinagar,\\Ahmedabad, Gujarat-382055, India,}
\affiliation[c]{Physique des Particules, Universit\'e de Montr\'eal, 1375 Avenue Th\'er\`ese-Lavoie-Roux, Montr\'eal, QC, Canada  H2V 0B3}
\emailAdd{lipika.kolay@iitgn.ac.in, soumitra.nandi@iitg.ac.in, ipsitaray02@gmail.com
}
\abstract{We investigate the interplay between low-energy flavor physics and dark matter (DM) phenomenology within the dimension-6 dark Standard Model Effective Field Theory (dSMEFT), featuring a fermionic DM candidate. Focusing on a minimal set of effective operators, we analyze both single- and multi-operator scenarios under current flavor and DM constraints. We show that correlated multi-operator configurations are required to simultaneously satisfy all existing bounds, leading to tightly constrained benchmark regions with implications for future collider searches. We further present representative phenomenological predictions and discuss possible ultraviolet completions of the effective framework.}

\keywords{Dark Matter, Effective Field Theory, Low-energy flavor Observables.}

\begin{document}
\maketitle
\flushbottom

\section{Introduction} \label{sec:intro}
Despite its remarkable success, the Standard Model (SM) remains incomplete, most notably because it does not explain the existence of dark matter (DM). This motivates the search for physics beyond the Standard Model (BSM), which can be explored through both high-energy collider experiments and precision low-energy observables. Effective field theory (EFT) provides a systematic and model-independent framework to describe the low-energy effects of heavy new physics, with the Standard Model Effective Field Theory (SMEFT) serving as a prominent example \cite{Grzadkowski:2010es, Buchmuller:1985jz}.

Since SMEFT contains only SM degrees of freedom, it cannot accommodate a DM candidate. This limitation motivates dark-sector EFTs, where additional dark fields remain dynamical below the cutoff scale while heavier states are integrated out. Depending on the symmetry and energy regime, these frameworks include the dark Standard Model effective field theory (dSMEFT) and the dark Low Energy Effective Field Theory (dLEFT) \cite{Aebischer:2022wnl, Song:2023jqm, Brod:2017bsw, Brod:2017bsw, Liang:2023yta, DelNobile:2011uf, DeSimone:2013gj, DeSimone:2016fbz, Bhattacharya:2021edh, Harnik:2008uu, Criado:2021trs, Crivellin:2014qxa}. By incorporating fermionic, scalar, or vector DM degrees of freedom in a model-independent manner, dSMEFT provides a powerful framework to study the interplay between DM phenomenology and low-energy precision observables.

Low-energy flavor observables and DM searches provide complementary probes of BSM physics. In particular, flavor-changing neutral current (FCNC) processes, which are loop-suppressed in the SM, are highly sensitive to new physics and impose stringent constraints on effective interactions. Precision measurements of the $b\to s\mu^+\mu^-$ transition \cite{LHCb:2014cxe, LHCb:2015svh, LHCb:2020lmf, LHCb:2021zwz, LHCb:2021xxq, LHCb:2022vje, ATLAS:2018gqc, LHCb:2020gog, Belle:2016fev, Belle:2019oag, BELLE:2019xld}, the rare decay $B_s\to\mu^+\mu^-$ \cite{LHCb:2021awg}, neutral $B_{(q)}$ ($q=d,s$) meson mixing \cite{HFLAV:2022esi, Belle-II:2023bps}, the invisible decay $B^+\to K^+\nu\bar{\nu}$ \cite{Belle-II:2023esi}, and the $b\to d\,\mu^+\mu^-$ transition probed through $B^\pm\to\pi^\pm\mu^+\mu^-$ decays \cite{LHCb:2012de, LHCb:2015hsa} provide stringent bounds on flavor-violating interactions. At the same time, the dSMEFT offers a model-independent framework for studying DM candidates, including weakly interacting massive particles (WIMPs) \cite{Bai:2010hh, Cheung:2010zf, Zheng:2010js, Haisch:2012kf, Abdallah:2014hon, Demetriou:2025ewa, Fuyuto:2024oii, Jahedi:2026dzy} and feebly interacting massive particles (FIMPs) \cite{Barducci:2026teq, Borah:2025ema, Barman:2020plp}. Existing dSMEFT studies have extensively investigated relic abundance, direct and indirect detection, and collider signatures \cite{Bai:2010hh, Cheung:2010zf, Zheng:2010js, Fox:2011pm, Haisch:2012kf, Abdallah:2014hon, Beniwal:2022rde, Demetriou:2025ewa}, while flavor-violating operators have largely been constrained using flavor observables mostly in model-independent analyses \cite{Buras:2024ewl, Kumar:2024ivx, Kala:2026pvl, Kolay:2026mgv, Kolay:2026bjm}. However, the same Wilson coefficients (WCs) that contribute to FCNC processes also govern DM annihilation and scattering rates, implying that flavor constraints alone do not guarantee a viable DM scenario. Combined with the Planck measurement of the DM relic abundance \cite{Planck:2018vyg} and limits from direct and indirect detection experiments \cite{XENON:2023cxc, XENON:2025vwd, LZ:2022lsv, LZ:2024zvo, PandaX:2023xgl, PandaX:2024qfu, PandaX:2023ejt, Fermi-LAT:2015att, Fermi-LAT:2016afa, HESS:2016mib, Silverwood:2014yza}, this motivates a correlated analysis of flavor and DM observables. Such an approach is particularly relevant for heavy DM scenarios, where new states may evade direct production but leave observable virtual effects in low-energy processes. In this work, we focus on flavor-violating dSMEFT operators that contribute to the decays and mixings of the $B_d$ and $B_s$-mesons, which provide a rich set of precisely measured observables, and identify the regions of parameter space that simultaneously satisfy flavor, cosmological, and astrophysical constraints.

The remainder of this paper is organized as follows. In Sec.~\ref{sec:theory}, we introduce the dSMEFT framework and motivate the choice of the dimension-6 operator basis considered in this work. In Sec.~\ref{sec:phenomenology}, we discuss the implications of these operators for low-energy flavor observables and DM phenomenology, including relic abundance, direct detection, and indirect detection. In Sec.~\ref{sec:DM_flavor_combined}, we present the results of our combined analysis of flavor and DM constraints, discuss the resulting parameter space, and examine possible simplified and UV-complete scenarios that can generate the effective operators considered in this work. Finally, Sec.~\ref{sec:summary} summarizes our main findings and conclusions.

\vspace{1.5cm}

\section{dSMEFT Framework} \label{sec:theory}
We are interested in the dimension-6 SMEFT operators extended with a fermionic DM field \( \chi \). This dSMEFT operator basis respects the full SM gauge symmetry. In this work, we consider the DM \( \chi \) to be a singlet Dirac fermion. The stability of the DM is ensured by imposing a discrete \( \mathbb{Z}_{2} \) symmetry, under which the DM field is odd, while all the SM particles are even. The DM field is electrically neutral and transforms as a singlet under the SM gauge group. We consider this dark fermion to lie in the sub-TeV region.  

\subsection{Operator basis of dSMEFT}
In this dSMEFT basis, a total of five types of dimension-6 operators can be constructed, as shown in Table~\ref{tab:operator_lists}. The operators are all invariant under the SM gauge group \( \rm SU(3)_{c} \times SU(2)_{L} \times U(1)_{Y} \), as required for a consistent dSMEFT construction. Here, \( p, r \) denote the generation indices for quark and lepton fields, while \( a, b \) correspond to the dark fermion indices. For the dark fermion \( \chi \), we take \( a = b \), i.e., we assume a single DM generation in our analysis. One may also consider more general scenarios where \( a \neq b \), which would introduce an additional dark sector fermion. The presence of such an extra fermion can give rise to a variety of new physics scenarios. It may act as a co-annihilation partner, facilitate assisted DM mechanisms, serve as a mediator between the DM and the SM sector, or itself constitute a second DM candidate \cite{Liang:2023yta, Mescia:2024rki, Belfatto:2025ids}.  

We adopt the following notation for the SM fermion fields:  
\( q \equiv \) left-handed quark doublets and \( \ell \equiv \) left-handed lepton doublets.  
The fields \( e, \, d, \, u \) denote the right-handed charged leptons, down-type quarks, and up-type quarks, respectively. These fields have the structure  
\[
\ell = 
\begin{pmatrix}
\nu_{L} \\ e_{L} 
\end{pmatrix}, \quad 
q = 
\begin{pmatrix}
u_{L} \\ d_{L} 
\end{pmatrix}, \quad 
e = e_{R}, \quad 
u = u_{R}, \quad 
d = d_{R}.
\]  

For dSMEFT with a Dirac fermion \( \chi \), as can be seen from Table~\ref{tab:operator_lists}, the allowed interactions involve only vector or axial-vector currents of the DM. Scalar and tensor currents do not arise in this case due to gauge invariance and the chiral structure of the SM fields: left-handed fermions transform as \( \rm SU(2)_{L} \) doublets, whereas right-handed fermions are singlets. Consequently, contractions such as \( \delta_{AB} \, \bar{u}_{A} u_{B} \), with $(A, B) \in (L, R)$, vanish, and a similar argument excludes tensor structures as well.
\begin{table}[t]
	\centering
	\rowcolors{1}{green!15!yellow!10}{cyan!10}
	\renewcommand{\arraystretch}{1.9}
	\begin{tabular}{|c| c  c|}
		\hline
        \cellcolor{cyan!10} & \cellcolor{cyan!10} &\cellcolor{cyan!10} \\
		\multirow{-2}{*}{\shortstack{ Dimension}} & \multirow{-2}{*}{Name}  & \multirow{-2}{*}{Operator} \\
		\hline
		\hline
		\cellcolor{green!5}& $ \mathcal{Q}^{V,A}_{\ell \chi } $ & $ \big( \bar{\ell}_{p} \gamma_{\mu}\ell_{r} \big)\big(\bar{\chi}_{a} \gamma^{\mu}P_{A}\chi_{b}\big) $ \\
        \cellcolor{green!5}& $ \mathcal{Q}^{V,A}_{q \chi } $ & $ \left( \bar{q}_{p} \gamma_{\mu} q_{r} \right)\left(\bar{\chi}_{a} \gamma^{\mu} P_{A} \chi_{b}\right) $ \\
    	\cellcolor{green!5}& $ \mathcal{Q}^{V,A}_{e \chi } $ & $ \left( \bar{e}_{p} \gamma_{\mu}e_{r} \right)\left(\bar{\chi}_{a} \gamma^{\mu} P_{A} \chi_{b}\right) $ \\
		\cellcolor{green!5}& $ \mathcal{Q}^{V,A}_{u\chi } $ & $ \left( \bar{u}_{p} \gamma_{\mu} u_{r} \right)\left(\bar{\chi}_{a} \gamma^{\mu}P_{A} \chi_{b}\right) $ \\
        \multirow{-5}{*}{ \cellcolor{green!5} \shortstack{dSMEFT \\ dim-6}}& $ \mathcal{Q}^{V,A}_{d\chi } $ & $ \left( \bar{d}_{p} \gamma_{\mu} d_{r}
        \right)\left(\bar{\chi}_{a} \gamma^{\mu}P_{A} \chi_{b}\right) $ \\	
		\hline
	\end{tabular}
	\caption{Dimension-6 $\Delta B=\Delta L = 0$ operators in dSMEFT relevant to this work. $``A"$ denotes the chirality ($``L"$ or $``R"$) of the dark fermion.}\label{tab:operator_lists}
\end{table}
In this analysis, we take only one dark fermion $ \chi $, so we will not use any generation index for $ \chi $. In the present analysis, we have assumed that all dSMEFT WCs are real. 
\paragraph{\underline{Motivation for the Choice of Operators:} }
We focus on the flavor-violating $\bar{b}\Gamma_{\mu} q\bar{\chi}\Gamma^{\mu}\chi$ ($\Gamma^{\mu}$ stands for the respective current, $q \equiv s,d$) operators for several well-defined theoretical and phenomenological reasons. 
First, the $b \to s$ transition constitutes the most experimentally explored flavor-changing neutral current mode in the quark sector, with a large number of high-precision measurements available across multiple decay channels \cite{FlavourLatticeAveragingGroupFLAG:2024oxs, ParticleDataGroup:2024cfk, Mannel:2023egv}, allowing the corresponding WCs to be constrained in a statistically robust and over-constrained manner \cite{Hurth:2023jwr, Mahmoudi:2023upg, Biswas:2020uaq}. 
Second, many well-motivated new-physics scenarios predict that non-standard interactions dominantly involve the third generation \cite{Allwicher:2023shc, Allwicher:2024ncl, DAlise:2024qmp}, arising from mass-hierarchical couplings \cite{Weinberg:2020zba, Yang:2024znv, Fuentes-Martin:2020bnh}, flavor-symmetry breaking patterns \cite{Fuentes-Martin:2022xnb, Barbieri:2017tuq}, or loop-induced effects mediated by heavy quarks \cite{Belfatto:2025ids, Arcadi:2021glq}, rendering $b\to s$ transitions a natural portal to physics beyond the SM. 
Third, flavor-violating interactions involving light quarks are stringently constrained by kaon and other light-quark meson observables \cite{Chang:2017wpl, Polonsky:2024pcc}, typically pushing the scale of new physics far above the TeV range. In addition, light-quark transitions suffer from limited experimental coverage and larger non-perturbative QCD uncertainties at low energies \cite{Buchalla:2001ux, FlavourLatticeAveragingGroupFLAG:2024oxs}, whereas the $b\to s$ sector benefits from a much richer dataset and better theoretical control \cite{Dingfelder:2016twb}, still allowing sizeable new-physics effects and leaving phenomenologically viable parameter space.

Moreover, compared to flavor-conserving DM-quark operators, the $\bar{b}\Gamma_\mu \,s\,\bar{\chi} \Gamma^\mu \chi$ interaction enjoys a structural advantage in that it does not induce tree-level DM-nucleon scattering, thereby avoiding the strongest current direct-detection bounds and contributing only at one-loop level \cite{DEramo:2016gos}. 
Sensitivity to electric dipole moment constraints is also reduced, since flavor-violating operators generate such contributions only at higher loop levels and require additional flavor insertions, permitting CP-violating phases without immediate exclusion. 
Finally, rare and invisible $B$-meson decays provide clean experimental signatures with limited SM backgrounds, enabling a transparent interpretation of potential deviations within a minimal and controlled effective-field-theory framework.

In addition, we also consider a lepton-sector interaction involving muons, motivated by the fact that muonic observables are among the most precisely measured in the flavor sector and play a central role in constraining physics beyond the SM. Restricting the analysis to muons avoids the extremely stringent bounds associated with electron couplings, while still allowing for sizeable and phenomenologically relevant effects in rare flavor processes. Moreover, muon-specific interactions provide complementary sensitivity to new physics and enable correlated tests across quark and lepton sectors within a unified effective-field-theory framework.

\section{Phenomenological Implications of the dSMEFT Operators} \label{sec:phenomenology}
As motivated in the previous section, this work focuses on the dimension-6 dSMEFT operators listed in Table~\ref{tab:operator_lists}, with particular attention to interactions involving third-generation quarks and second-generation charged leptons. These operators induce contributions to a wide range of low-energy flavor observables, notably $B$- and $B_s$-meson decays and mixing phenomena, and also affect top-quark observables, including top-FCNC decays. In the following subsections, we systematically analyse the modifications induced by these higher-dimensional interactions and derive the resulting constraints on the associated WCs at a given NP scale.

\subsection{Low-energy flavor Observables} \label{sec:semileptonic}
In this work, we consider a range of low-energy observables in the $B$-meson sector, including neutral meson mixing, rare decays, semileptonic decays, and invisible decay modes. Our analysis is restricted to processes that receive contributions from the dSMEFT operators through either tree-level amplitudes or one-loop diagrams involving at most two operator insertions. The effects of these higher-dimensional interactions are incorporated at low energies by matching the dSMEFT operators onto the appropriate effective operator basis governing the relevant $b$-quark transitions.

\subsubsection{Neutral Meson Mixing } \label{sec:mixing}
Neutral pseudoscalar meson mixing plays an important role in constraining NP models, since the corresponding experimental observables are measured with high precision.
Since we are interested in the processes involving $B$-mesons, we will get contributions to the $B_{s}-\bar{B}_{s}$ and $B^0-\bar{B}^0$ meson mixing processes. In meson mixing, the relevant observable which encodes the possible effects of heavy new physics is the frequency of oscillations $\Delta M_q$ = 2$|M_{12}^q|$ ($q \equiv d,s$), where $M_{12}^q$  corresponds to the amplitude of the mixing process and is given as \cite{Hocker:2006xb}
\begin{equation}
	M_{12}^q = M_{12}^{q,\rm SM}(1+ \Delta_q e^{2i \delta}) ~,
\end{equation}
where $ \Delta_{q}  = |M_{12}^{q,\rm NP}|/ |M_{12}^{q,\rm SM}|$ = $\Delta M_q^{\rm NP}/\Delta M_q^{\rm SM}$ and $\delta$ is the relative phase between the SM and NP amplitudes. Expressing the observables in terms of $\Delta_q$ helps to partially reduce the uncertainties associated with the meson decay constants $f_{B_q}$, the QCD correction factor $\eta_B$, and the bag parameters $B_{B_q}$.
The current theoretical and experimental status of the mixing observables are \cite{Albrecht:2024oyn, LHCb:2023sim, Belle-II:2023bps} : 
\begin{align}\label{eq:mixing_SM_and_exp}
	{\rm \Delta M_{s}^{ Exp}} &= 17.765 (6) \, \rm ps^{-1}\,, \qquad
	{\rm \Delta M_{s}^{ SM}} = 18.23 (63) \,  \rm ps^{-1}\,,\nonumber \\
	{\rm \Delta M_{d}^{ Exp}} &= 0.5065 (19) \, \rm ps^{-1}\,, ~~~~~
	{\rm \Delta M_{d}^{ SM}} = 0.535(21)  \,  \rm ps^{-1}\,.\
\end{align}
Since the SM mixing phase in the $B_s$--$\bar{B}_s$ system is negligibly small, and our NP scenario does not introduce any additional CP-violating phase, we use $\Delta_s$ as the relevant observable, where $\Delta_{s} =  -0.02545\pm 0.03368$, obtained using the values in Eq.~\eqref{eq:mixing_SM_and_exp}. In contrast, the SM mixing phase in the $B^0$--$\bar{B}^0$ system is sizable and must be accounted for. Therefore, for the $B^0$ system, we use the mass difference $\Delta M_d$ as the corresponding observable.
It is worth noting that the updated SM prediction is consistent with the corresponding experimental measurement given in Eq.~\eqref{eq:mixing_SM_and_exp} for the $B_s^0$-$\bar{B}_s^0$ mixing observable within $1\sigma$. In contrast, the $B^0$--$\bar{B}^0$ mixing observable exhibits a $1.35\,\sigma$ tension between the SM prediction and the experimental measurement. 



\begin{figure}
	\centering
	\subfloat[]{\begin{tikzpicture}
			\begin{feynman}
				\vertex (a1){\( b\)};
				\vertex [square dot,blue,above right=1.8cm of a1](a2){};
				\vertex [above left=1.8cm of a2](a3){\(d_{i}\)};
				\vertex [square dot,blue,right=1.8cm of a2](a4){};
				\vertex [above right=1.8cm of a4](a5){\(b\)};
				\vertex [below right=1.8cm of a4](a6){\(d_{i}\)};

				\diagram* {
					(a1) -- [fermion,arrow size=1.2pt] (a2) -- [fermion,arrow size=1.2pt] (a3),
					(a2) --[fermion,arrow size=1.2pt,  half left, looseness=1.5,edge label'={\(\chi\)}](a4) --[fermion,arrow size=1.2pt,  half left, looseness=1.5,edge label'={\(\chi\)}](a2),
					(a5) --[fermion,arrow size=1.2pt ](a4) --[fermion,arrow size=1.2pt](a6),
				};
			\end{feynman}
	\end{tikzpicture}}
	\caption{Feynman diagrams of the neutral meson mixing process of the B-meson, with underlying quark transition $b \, \bar{d}_{i} \leftrightarrow \bar{b} \, d_{i}$. The blue square-dots represent the dim-6 operators.}
	\label{fig:meson_mixing_vertex}
\end{figure}
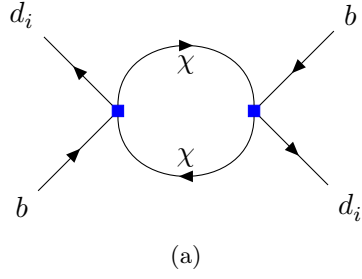

In our scenario, neutral meson mixing receives contributions at the one-loop level through the exchange of the DM particle $\chi$. The corresponding Feynman diagram is shown in Fig.~\ref{fig:meson_mixing_vertex}. After integrating out the loop degrees of freedom, the effective Lagrangian relevant for neutral meson mixing can be written (following the notation of Eq.~\eqref{eq:eff_loop_b2sll}) as
\begin{equation}
	\mathcal{L}_{\rm mix} = \sum_{A,B = \{L,R\}} \mathcal{C}_{AB} \, [\bar{d}_i \gamma_\mu P_{A} b] [\bar{d}_i \gamma^\mu P_{B} b] \,,
\end{equation}
with $i= \{1,2\}$, corresponding to the $d,s$ quarks, respectively. The loop-induced coefficients $\mathcal{C}_{AB}$ are expressed in terms of the dSMEFT WCs as
\begin{align}
	\mathcal{C}_{AB}
	&= \frac{1}{\Lambda^4}(\mathcal{C}_{q\chi,i3}^{VA} + \mathcal{C}_{d\chi,i3}^{VA}) \,
	\mathbb{C}_{AB} \,
	(\mathcal{C}_{q\chi,3i}^{VB} + \mathcal{C}_{d\chi,3i}^{VB}) \,.
\end{align}
Here, the function $\mathbb{C}_{AB}$ is given by:
\begin{align}
    \mathbb{C}_{AB} = \{\mathbb{C}_{1},\, \mathbb{C}_{2}, \,\mathbb{C}_{3}, \, \mathbb{C}_{4}\};  \quad \quad {\rm for~~~ } AB=\{LL, LR, RL, RR\} \,. 
\end{align}
The explicit expressions for the loop functions $\mathbb{C}_{i}s$ are provided in Appendix~\ref{app:loop}. The loop diagram shown in Fig.~\ref{fig:meson_mixing_vertex} is divergent. For the loop-induced contributions, ultraviolet divergences arise as a consequence of the non-renormalizable nature of the effective theory. 
The pole terms obtained in dimensional regularization can be absorbed into the renormalization of the corresponding low-energy WCs, while the accompanying logarithmic dependence on the ultraviolet scale is retained. The resulting amplitudes are then expressed in terms of the effective WCs of the relevant weak effective operators \cite{Kolay:2024wns, Kolay:2025jip, Biswas:2021pic, Kala:2025srq}. This procedure captures the dominant short-distance effects induced by the dSMEFT operators and retains the leading logarithmic dependence on the separation between the high- and low-energy scales. The finite local contributions, which are sensitive to the ultraviolet completion, are encoded in the effective WCs and treated as free parameters to be constrained by experimental data. Although we do not perform a complete renormalization-group analysis, the logarithmically enhanced terms retained in our calculation correspond to the leading scale-dependent contributions that would also emerge from the renormalization-group evolution of the effective operators between the matching and low-energy scales.

The cutoff scale $\Lambda$ is taken to be sufficiently high such that all particles in the model lie below this scale. The relevant new physics parameters entering the observables are the DM mass, the effective operator couplings, and the cutoff scale. We perform our analysis for two different values of the cutoff scale to examine the validity of the bounds on the effective couplings with respect to the choice of the hard cutoff. Throughout this work, we focus on the sub-TeV mass region of the DM.

Throughout this work, we assume that all WCs are real. The relevant matrix elements corresponding to the meson mixing are given by:
\begin{subequations}
	\begin{eqnarray}
		& \langle \bar{B}_{q}^{0} | \left(\bar{q} \gamma_{\mu} P_{L} b \right)^2| B_{q}^{0} \rangle  =   \langle \bar{B}_{q}^{0} | \left(\bar{q} \gamma_{\mu} P_{R} b\right)^2 | B_{q}^{0} \rangle = \frac{2}{3} B_{B_q} f_{B_{q}}^2 M_{B_{q}}^2 \,, \\
		& \langle \bar{B}_{q}^{0} | \left(\bar{q} \gamma_{\mu} P_{L} b \right)\left(\bar{q} \gamma_{\mu} P_{R} b \right)| B_{q}^{0} \rangle  =  \langle \bar{B}_{q}^{0} | \left(\bar{q} \gamma_{\mu} P_{R} b\right)\left(\bar{q} \gamma_{\mu} P_{L} b\right) | B_{q}^{0} \rangle = - \frac{2}{3} B_{B_q} f_{B_{q}}^2 M_{B_{q}}^2 \,.
	\end{eqnarray}
\end{subequations}
The expression for $\Delta M_q^{\rm NP}$ as a function of the effective loop contributions and coefficients of the dSMEFT operators is:
\begin{eqnarray}\label{eq:DeltaM_NP}
	\Delta M_{q}^{\rm NP} & = &   \frac{2}{3\,} B_{B_{q}} \, \eta_{B} \, f_{B_{q}}^2  m_{B_{q}}  
	\Bigg[ \frac{\mathbb{C}_{1}}{\Lambda^4}  \left( \mathcal{C}_{d\chi,i3}^{VL^2} + \mathcal{C}_{d\chi,i3}^{VR^2} + \mathcal{C}_{q\chi,i3}^{VL^2} + \mathcal{C}_{q\chi,i3}^{VR^2}
	-2 \mathcal{C}_{q\chi,i3}^{VR} \mathcal{C}_{d\chi,i3}^{VR} - 2 \mathcal{C}_{d\chi,i3}^{VL} \mathcal{C}_{q\chi,i3}^{VL}  \right) \nonumber \\
	&& + 2 \, \frac{\mathbb{C}_{2}}{\Lambda^4}  \left( \mathcal{C}_{d\chi,i3}^{VL}- \mathcal{C}_{q\chi,i3}^{VL} \right) \left(\mathcal{C}_{d\chi,i3}^{VR} - \mathcal{C}_{q\chi,i3}^{VR})\right) \Bigg] \,,
\end{eqnarray}
where $\mathcal{C}_{q\chi,i3}^{VL(R)}$ ($\mathcal{C}_{d\chi,i3}^{VL(R)}$) is the coefficient of $ \mathcal{Q}^{VA}_{q \chi } $ ($ \mathcal{Q}^{VA}_{d \chi } $) operators. $i = 1(2)$ for $d(s)$ quarks respectively. The inputs for the decay constants and bag parameters used in this analysis are provided in Table \ref{tab:input}. 

\paragraph{\underline{Parameter Space from Meson mixing} :} 

\begin{table}[t]
	\begin{center} 
		\begin{tabular}{| l l| l l |} 
			\hline  
			\hline 
			\rule[-2mm]{0mm}{7mm}
			$\mu_b$                           &  4.8 GeV 
			& 
			$\alpha_{\rm em}$                 & $ 1/137$\\ 
			$\alpha_s (\mu_b)$                & 0.214  
			&
			$G_F$                             & 1.166 $\times$ $10^{-5}$ ~GeV$^{-2}$ \\
			$\eta_B$                          & $ 0.551 \pm 
			0.007$                & $\Gamma_t$        
			& $1.42^{+0.19}_{-0.15}$ GeV \\
			$f_{B_s}$                         & $230.3 \pm 1.3$ MeV \cite{FlavourLatticeAveragingGroupFLAG:2024oxs} 
			& 
			$f_{B_d}$                         & $190.0 \pm 1.3$ MeV \cite{FlavourLatticeAveragingGroupFLAG:2024oxs} \\
			$B_{B_s}$                         & $1.232 \pm 0.053$ \cite{Dowdall:2019bea}            & $B_{B_d}$ 
			& $1.222 \pm 0.061$ \cite{Dowdall:2019bea} \\
			\hline
		\end{tabular} 
		\caption{Inputs used in the analysis.}
		\label{tab:input}
	\end{center} 
\end{table}

\begin{figure}[t!]
	\centering
	\subfloat[]{\includegraphics[width=0.5\linewidth]{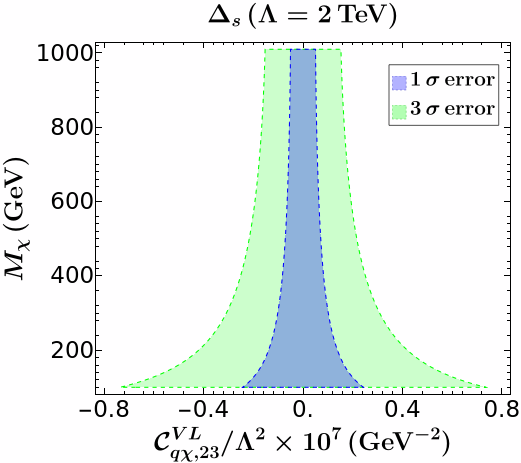}}~
	\subfloat[]{\includegraphics[width=0.5\linewidth]{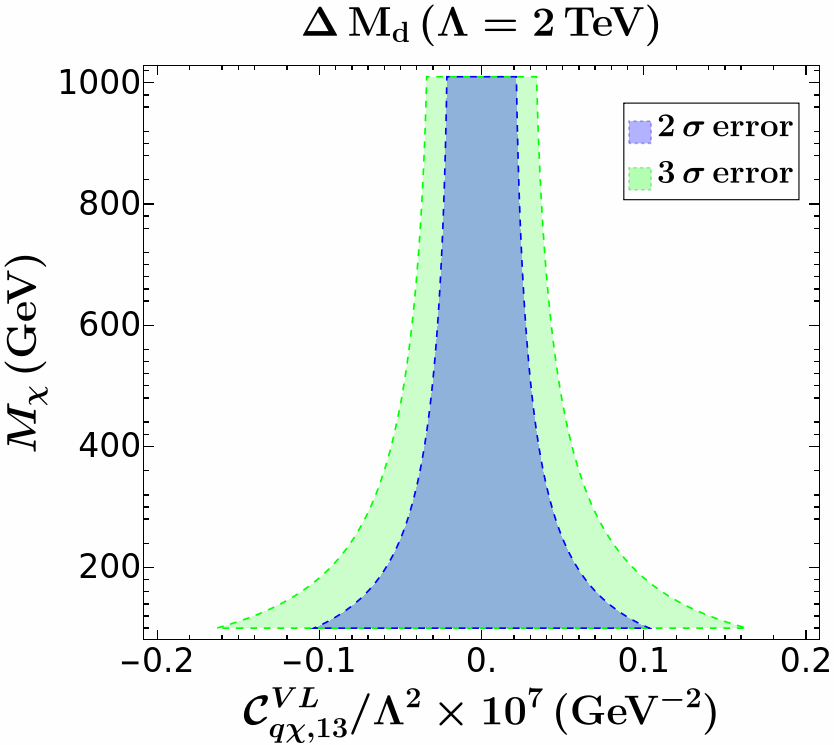}}
	\caption{Allowed parameter space in the mass-coupling plane from the neutral meson mixing processes $B^{0}_{(s)}-\bar{B}^{0}_{(s)}$. The constraints are shown for the $1\sigma$ and $3\sigma$ uncertainties of the observable $\Delta_s$, and the $2\sigma$ and $3\sigma$ uncertainties of the observable $\Delta M_d$. }
	\label{fig:mixing_oneOp}
\end{figure}
Figure~\ref{fig:mixing_oneOp} shows the parameter space allowed by the observables $\Delta_s$ (left panel) and $\Delta M_d$ (right panel) in the one-operator (normalized with scale) scenario, where the variation is shown with $M_{\chi}$. From Eq.~\eqref{eq:DeltaM_NP}, we see that  both the observables depend on the dSMEFT WCs in a similar manner. Unlike the SM contribution, there is no additional suppression factor for $\Delta M_d$. Since the allowed NP contribution to $\Delta M_{d}^{\rm NP}$ is much smaller than that to $\Delta M_{s}^{\rm NP}$, as can be seen from Eq.~\eqref{eq:mixing_SM_and_exp}, the allowed values of the WCs $\mathcal{C}_{q(d)\chi,13}^{VL(R)}$ are expected to be smaller than those of $\mathcal{C}_{q(d)\chi,23}^{VL(R)}$. This behavior is reflected in Fig.~\ref{fig:mixing_oneOp}. As discussed earlier, since $\Delta M_d$ does not agree with the SM prediction at $1\sigma$, we have varied it within its $2\sigma$ and $3\sigma$ uncertainties, whereas $\Delta M_s$ is varied within its $1\sigma$ and $3\sigma$ uncertainties. We find that the allowed values of the WCs are of the order of $\sim \mathcal{O}(10^{-8})\,\mathrm{GeV}^{-2}$ in both cases, although the allowed ranges differ. For $B_s^0-\bar{B}_s^0$ mixing, using the $3\sigma$ uncertainty of the observable, we obtain $|\mathcal{C}_{q\chi,23}^{VL}/\Lambda^2| \leq 0.8 \times 10^{-7}\,\mathrm{GeV}^{-2}$, whereas for $B^0-\bar{B}^0$ mixing, the corresponding bound is $|\mathcal{C}_{q\chi,13}^{VL}/\Lambda^2| \leq 0.15 \times 10^{-7}\,\mathrm{GeV}^{-2}$. Here, the mentioned bound is for $M_{\chi} \sim 100$ GeV. For higher $M_{\chi}$, the allowed region is even more constrained. Similar bounds are obtained for the other WCs, as can be seen from Eq.~\eqref{eq:DeltaM_NP}. The plots are shown for the cutoff scale $\Lambda = 2$~TeV. Since the dependence on the cutoff scale is logarithmic, the bounds remain nearly unchanged for $\Lambda = 1$~TeV, with only slight modifications.

\begin{figure}
	\centering
	\subfloat[]{\includegraphics[width=0.5\linewidth]{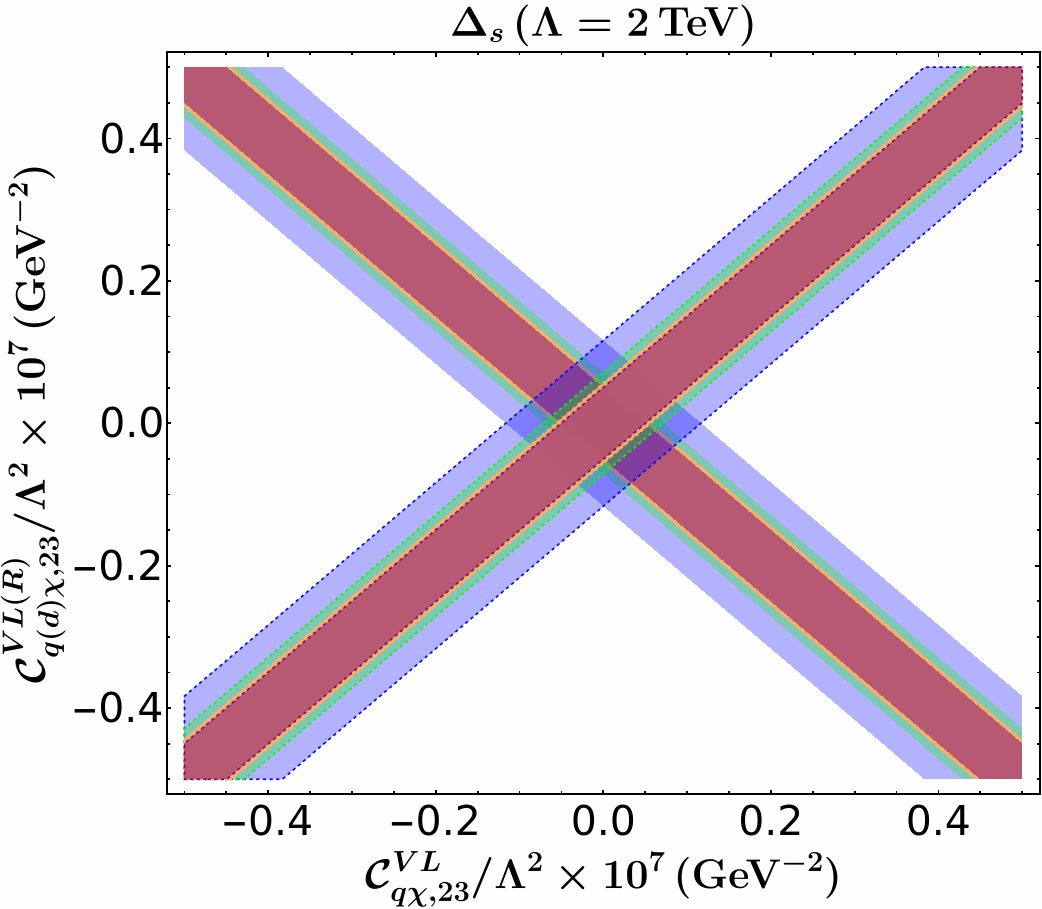}}~~
	\subfloat[]{\includegraphics[width=0.5\linewidth]{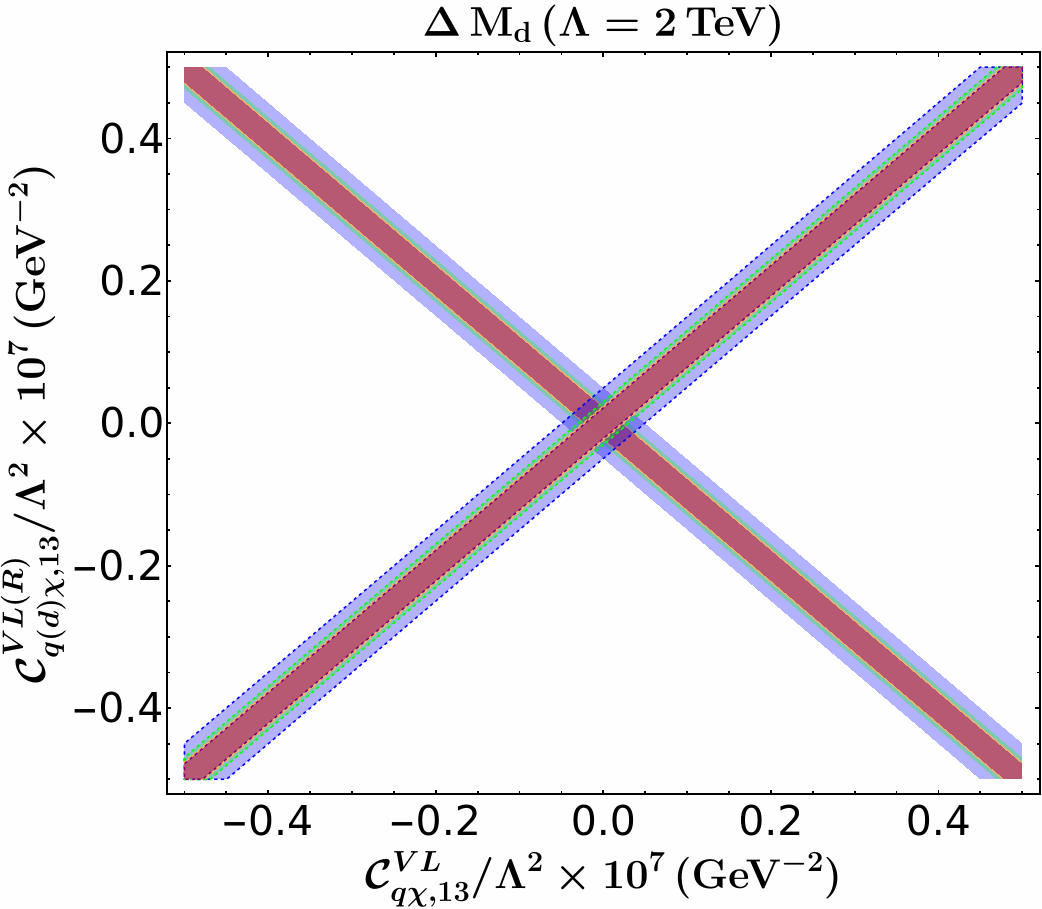}} \\
	\includegraphics[width=0.95\linewidth]{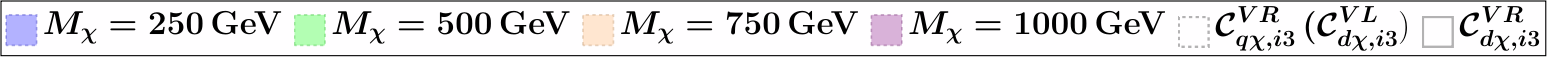}
	\caption{Allowed parameter space in the coupling planes from the meson mixing observables in the two-operator scenarios. The different colored bands correspond to different DM mass values. The regions with solid boundaries (showing a negative correlation) correspond to the case with $\mathcal{C}_{d\chi,i3}^{VR}$ on the $y$-axis, whereas the regions with dashed boundaries (showing a positive correlation) correspond to the two cases with either $\mathcal{C}_{q\chi,i3}^{VR}$ or $\mathcal{C}_{d\chi,i3}^{VL}$ on the $y$-axis. The observable $\Delta_s$ is varied within its $1\sigma$ uncertainty, whereas $\Delta M_d$ is varied within its $2\sigma$ uncertainty. }
	\label{fig:mixing_twoOp}
\end{figure}

Figure~\ref{fig:mixing_twoOp} shows the allowed parameter space from the neutral meson mixing observables in the two-operator scenario. Unlike the previous case, here both operators are varied simultaneously. On the $x$-axis, we vary the WC $\mathcal{C}_{q\chi,i3}^{VL}/\Lambda^2$, whereas on the $y$-axis  $\mathcal{C}_{q(d)\chi,i3}^{VL(R)}/\Lambda^2$ are varied, as indicated in the plot legends. In a two-operator scenario $(X,\,Y)$, the NP contribution to the mass difference varies as $\Delta M_q^{\rm NP}\propto (X\pm Y)^2$. Therefore, unlike the previous case, where each operator was independently constrained by the observable, here a correlation between the two operators emerges, allowing them to take relatively large values provided they vary simultaneously. The dependence $\propto (X-Y)^2$ exhibits a positive correlation between the couplings. In our case, the combinations $(\mathcal{C}_{q\chi,i3}^{VL}-\mathcal{C}_{q\chi,i3}^{VR})$, $(\mathcal{C}_{q\chi,i3}^{VL}-\mathcal{C}_{d\chi,i3}^{VL})$ [shown in fig \ref{fig:mixing_twoOp}], and $(\mathcal{C}_{d\chi,i3}^{VL}-\mathcal{C}_{d\chi,i3}^{VR})$ exhibit positive correlations. On the other hand, the dependence $\propto (X+Y)^2$ exhibits a negative correlation between the couplings. In our case, the combinations $(\mathcal{C}_{q\chi,i3}^{VL}-\mathcal{C}_{d\chi,i3}^{VR})$ [shown in Fig.~\ref{fig:mixing_twoOp}] and $(\mathcal{C}_{q\chi,i3}^{VR}-\mathcal{C}_{d\chi,i3}^{VL})$ exhibit negative correlations. For both observables, the couplings are varied within the range $|\mathcal{C}_{q(d)\chi,i3}^{VL(R)}|/\Lambda^2 \leq 0.4 \times 10^{-7}\,\mathrm{GeV}^{-2}$. From the figure, however, we observe that for a given value of the $x$-axis variable, a smaller region along the $y$-axis is allowed in the case of $B^0-\bar{B}^0$ mixing. This again demonstrates that $\Delta M_d$ constrains the parameter space more stringently. 



\subsubsection{One-loop Contributions to the $b \to d_{i} \ell \ell$ processes}
As motivated in the previous section, our study focuses on processes involving third-generation quarks. The transitions $b \to d_{i} \, \mu^+ \, \mu^-$, with $d_{i} = (d,s)$, are particularly useful due to the large number of precise experimental measurements available in this sector, allowing for a detailed study of these processes. In our setup, these transitions arise at the one-loop level with two insertions of the dSMEFT operators, since no tree-level contributions to such flavor-changing neutral current processes are generated within this framework. In this section, we therefore focus on the loop-level contributions responsible for these transitions.

\begin{figure}[t!]
		\centering 
		\subfloat[]{\begin{tikzpicture}
			\begin{feynman}
			\vertex (a1){\( b\)};
			\vertex [square dot,blue,above right=1.8cm of a1](a2){};
			\vertex [above left=1.8cm of a2](a3){\(d_{i}\)};
			\vertex [square dot,blue,right=1.5cm of a2](a4){};
			\vertex [above right=1.8cm of a4](a5){\(\ell\)};
			\vertex [below right=1.8cm of a4](a6){\(\ell\)};
			
			\diagram* {
				(a1) -- [fermion,arrow size=1.2pt] (a2) -- [fermion,arrow size=1.2pt] (a3),
				(a2) --[fermion,arrow size=1.2pt,  half left, looseness=1.5,edge label'={\(\chi\)}](a4) --[fermion,arrow size=1.2pt,  half left, looseness=1.5,edge label={\(\chi\)}](a2),
				(a6) --[fermion,arrow size=1.2pt](a4) --[fermion,arrow size=1.2pt](a5),
				
			};
			\end{feynman}
			\end{tikzpicture}} \\
		\caption{One-loop Feynman diagrams with two insertions of the dim-6 dSMEFT operators modifying the semileptonic $b \to d_{i} \ell \ell$ processes. The blue square dots represent the higher-dimensional operators. $\chi$ is the fermionic DM.}
		\label{fig:Feyn_b2sll_loop}
	\end{figure}
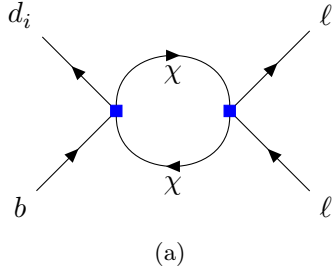
For the dSMEFT listed above, the contribution to $b \to d_{i}\, \ell \ell$  will be via the Feynman diagram shown in Fig.~\ref{fig:Feyn_b2sll_loop}. The loop effects generate local operators with the same quantum numbers and Lorentz structure as those already present in the low-energy effective theory. The resulting $b \to d_{i} \, \ell \ell $ effective four-fermion operators, along with the respective coefficients, can be written as
\begin{equation}\label{eq:eff_loop_b2sll}
    \mathcal{L}_{\rm eff}^{b \to d_{i} \ell \ell} = \sum_{A,B}\mathcal{C}_{AB} (\bar{d}_{i} \gamma_{\mu} P_{A} b) (\bar{\ell} \gamma^{\mu} P_{B} \ell)\,, 
\end{equation}
where, $A, B = \{L,R\}$ represent the chirality. The effective coefficients of the above Eq.~\eqref{eq:eff_loop_b2sll} are given by:
\begin{align}
    \mathcal{C}_{AB}&=\frac{1}{\Lambda^4}(\mathcal{C}_{q\chi,3i}^{VA} + \mathcal{C}_{d\chi,3i}^{VA}) \mathbb{C}_{AB} (\mathcal{C}_{\ell\chi,22}^{VB} + \mathcal{C}_{e\chi,22}^{VB}) \,,
\end{align}
The index $i=\{1,2\}$ for $d$ and $s$ quarks in the final state, respectively. 
Here, the function $\mathbb{C}_{AB}$ is given by:
\begin{align}
    \mathbb{C}_{AB} = \{\mathbb{C}_{1},\, \mathbb{C}_{2}, \,\mathbb{C}_{3}, \, \mathbb{C}_{4}\};  \quad \quad {\rm for~~~ } AB=\{LL, LR, RL, RR\} \,. 
\end{align}
The explicit expressions for the loop functions $\mathbb{C}_{AB}$ are provided in Appendix~\ref{app:loop}.

For our choice of operators, the $b \to d_{i} \ell \ell$ transition is generated by a single one-loop diagram with the DM field $\chi$ running in the loop. In addition to this, one could also consider contributions from wavefunction renormalisation diagrams. However, for vector-type interactions, these vanish since the corresponding closed fermion loop evaluates to zero.  
One may also consider dLEFT operators, where scalar interactions between the SM fermions and the DM field are present. Such interactions would generate effective scalar and tensor currents for the $b \to d_{i} \ell \ell$ process, leading to a different set of contributions compared to the present setup. 
In the next subsection, we will show how these one-loop diagrams impact various FCNC processes.

\subsubsection{Rare Decays of $B$ meson} 
The one-loop level contributions in the FCNC processes discussed above will contribute to various low energy rare and invisible $B$ and $B_s$ meson semileptonic decays. 
The processes $b \to s(d)\,\ell^{+}\ell^{-}$ play a central role in probing new physics, since they are loop-suppressed in the SM and a large amount of precise experimental data is available from multiple experiments.
The low energy effective Hamiltonian describing the $b\to d_{i} \ell^+\ell^-$ ($d_{i} \equiv d,s$) transitions at the scale $\mu \sim m_b$ after integrating out the heavy degrees of freedom 
can be written as \cite{Bobeth:1999mk,Becirevic:2012fy, Altmannshofer:2008dz} : 
\begin{equation} \label{eq:Heff}
	{\cal H}_{eff}^{b \to d_i \ell \ell} = - \frac{4\,G_F}{\sqrt{2}}\left(
	\lambda_t {\cal H}_{eff}^{(t)} + \lambda_u {\cal
		H}_{eff}^{(u)}\right) \,, 
\end{equation}
where $G_F$ is the Fermi constant, $\lambda_j=V_{jb}V_{jd_i}^\ast (j=u,t)$ are the CKM elements and
\begin{eqnarray}
	{\cal H}^{(t)}_{eff} & = & C_1 \mathcal O_1^c + C_2 \mathcal O_2^c  + \sum_{i=3}^{6} C_i 
	\mathcal O_i + \sum_{i=7,8,9,10,P,S} \biggl(C_i \mathcal   O_i + C'_i  
	\mathcal  O'_i\biggr)\ \,, \\
	{\cal H}_{eff}^{(u)} 
	& = & 
	C_1 (\mathcal O_1^c-\mathcal O_1^u)  + C_2(\mathcal O_2^c-\mathcal
	O_2^u)\,.
\end{eqnarray}
 The operator basis is known as weak effective theory (WET). 
The operators relevant to our analysis are given as:
\begin{subequations}
	\begin{align}\label{eq:basisOps}
		\mathcal O_{9} &= \frac{e^2}{16 \pi^2} 
		(\bar{d_i} \gamma_{\mu} P_L b)(\bar{\ell} \gamma^\mu \ell) ,&
		\mathcal O_{9}^\prime &= \frac{e^2}{16 \pi^2} 
		(\bar{d_i} \gamma_{\mu} P_R b)(\bar{\ell} \gamma^\mu \ell) , \\
		\mathcal O_{10} &=\frac{e^2}{16 \pi^2}
		(\bar{d_i}  \gamma_{\mu} P_L b)(  \bar{\ell} \gamma^\mu \gamma_5 \ell) ,&
		\mathcal O_{10}^\prime &=\frac{e^2}{16 \pi^2}
		(\bar{d_i}  \gamma_{\mu} P_R b)(  \bar{\ell} \gamma^\mu \gamma_5 \ell) ,
	\end{align}
\end{subequations}
where $\ell = \mu$. Here, $ C_i $ and $C'_i$s are the WCs corresponding to the operators $ \mathcal{O}_i $ and $ \mathcal{O}'_i$s, respectively. The details of the operator basis can be found from \cite{Buras:1995iy, Becirevic:2012fy, Bobeth:1999mk, Altmannshofer:2008dz} which includes the current-current ($i=1,2$), QCD penguin ($i=3,..,6$), electromagnetic dipole ($\mathcal O_7$), chromomagnetic dipole ($\mathcal O_8$), and the semileptonic operators $\mathcal O_9^{(\prime)}$ and $\mathcal O_{10}^{(\prime)}$, respectively. The operators $\mathcal{O}_{1-10}$ are present in the SM, whereas the operators $\mathcal O_i'$ and $\mathcal O_{S,P}$ are present mainly in the new physics scenarios. The values of the WCs in the SM at $\mu_b$ scale can be found in Ref. \cite{Mahmoudi:2024zna}\,. 

The $b \to d_i \ell \ell$ transition receives contributions from one-loop diagrams with two insertions of dimension-6 operators, as shown in Fig.~\ref{fig:Feyn_b2sll_loop}. Within our chosen operator basis, the loop calculation generates contributions exclusively to the vector and axial-vector semileptonic operators. In terms of the low-energy effective Hamiltonian, this implies that new physics effects appear only in the WCs $\mathcal{C}_{9}^{(\prime)}$ and $\mathcal{C}_{10}^{(\prime)}$, while no contributions arise for the dipole operators $\mathcal{C}_{7}$ and $\mathcal{C}_{8}$, since the operator structure does not induce electromagnetic or chromomagnetic interactions at this level.
After performing the loop integrations and simplifying the resulting expressions, the WCs can be written in terms of the loop contributions $\mathbb{C}_{1}$ and $\mathbb{C}_{2}$ given in Eq.~\eqref{eq:loop_contribution}. The new physics contributions to the WCs will be added to the SM values: $C_{i} = C_{i}^{\rm SM} + C_{i}^{\rm NP}$. In terms of those loop functions, the low-energy WCs can be written as:

\small
\begin{align}
C_{9}^{\rm NP}  = &  \frac{K_{X}}{\Lambda^4} \left[ (\mathcal{C}^{VL}_{\ell \chi,22} + \mathcal{C}^{VL}_{e \chi,22}) (\mathbb{C}_{1} \mathcal{C}_{q\chi,i3}^{VL} + \mathbb{C}_{2} \mathcal{C}_{q\chi,i3}^{VR}) + (\mathcal{C}^{VR}_{\ell \chi,22} + \mathcal{C}^{VR}_{e \chi,22}) (\mathbb{C}_{1} \mathcal{C}_{q\chi,i3}^{VR} + \mathbb{C}_{2} \mathcal{C}_{q\chi,i3}^{VL})\right], \nonumber \\ 
C_{9}^{\prime ~\rm NP} = &  \frac{K_{X}}{\Lambda^4} \left[ (\mathcal{C}^{VL}_{\ell \chi,22} + \mathcal{C}^{VL}_{e \chi,22}) (\mathbb{C}_{1} \mathcal{C}_{d\chi,i3}^{VL} + \mathbb{C}_{2} \mathcal{C}_{d\chi,i3}^{VR}) + (\mathcal{C}^{VR}_{\ell \chi,22} + \mathcal{C}^{VR}_{e \chi,22}) (\mathbb{C}_{1} \mathcal{C}_{d\chi,i3}^{VR} + \mathbb{C}_{2} \mathcal{C}_{d\chi,i3}^{VL})\right],\nonumber\nonumber \\ 
C_{10}^{\rm NP}  = &  \frac{K_{X}}{\Lambda^4} \left[ (\mathcal{C}^{VL}_{e \chi,22} -  \mathcal{C}^{VL}_{\ell \chi,22}) (\mathbb{C}_{1} \mathcal{C}_{q\chi,i3}^{VL} + \mathbb{C}_{2} \mathcal{C}_{q\chi,i3}^{VR}) + (\mathcal{C}^{VR}_{e \chi,22} - \mathcal{C}^{VR}_{\ell \chi,22}) (\mathbb{C}_{1} \mathcal{C}_{q\chi,i3}^{VR} + \mathbb{C}_{2} \mathcal{C}_{q\chi,i3}^{VL})\right], \nonumber \\
C_{10}^{\prime ~\rm NP}  = &  \frac{K_{X}}{\Lambda^4} \left[ (\mathcal{C}^{VL}_{e \chi,22} -  \mathcal{C}^{VL}_{\ell \chi,22}) (\mathbb{C}_{1} \mathcal{C}_{d\chi,i3}^{VL} + \mathbb{C}_{2} \mathcal{C}_{d\chi,i3}^{VR}) + (\mathcal{C}^{VR}_{e \chi,22} - \mathcal{C}^{VR}_{\ell \chi,22}) (\mathbb{C}_{1} \mathcal{C}_{d\chi,i3}^{VR} + \mathbb{C}_{2} \mathcal{C}_{d\chi,i3}^{VL})\right] \,, \label{eq:WCb2sll}
\end{align}

with 
\begin{equation}
K_{X} = -\frac{1}{4 \sqrt{2} G_{F}} \frac{1}{V_{tb} V_{td_i}^{*} } \frac{16 \pi^2}{e^2} \,.
\end{equation}
The index `$i$' in the above equation stands for the generation index with $ i = \{1,2\}$ for the $b \to d \ell \ell$ and $b \to s \ell \ell $ transitions, respectively. For the leptons, in our analysis, we have mainly focused on the decays to muons due to the availability of a plethora of experimental data, as mentioned previously. 

The most relevant modes include the semileptonic processes $B \to K^{(*)}\,\ell^{+}\ell^{-}$ and $B_s \to \phi\,\ell^{+}\ell^{-}$, for which measurements of differential branching fractions, CP asymmetries, and a variety of angular observables have been reported by experiments such as LHCb, Belle, ATLAS, and CMS \cite{LHCb:2014cxe,LHCb:2015svh,LHCb:2020lmf,LHCb:2021zwz,LHCb:2021xxq,Belle:2016fev,Belle:2019oag,BELLE:2019xld,ATLAS:2018gqc,LHCb:2020gog}. In addition, ratios of branching fractions such as $R_{K^{(*)}}$, which test lepton flavor universality, provide particularly clean probes of possible new physics effects. The currently available measurements are largely consistent with the SM within uncertainties \cite{LHCb:2022vje}, thereby imposing strong constraints on new physics scenarios. The observables considered here are discussed in detail in \cite{Biswas:2020uaq}.

The $b \to d\,\ell^{+}\ell^{-}$ transitions probe the same underlying dynamics but are further suppressed by the CKM matrix elements, leading to significantly smaller branching fractions. The available experimental data are more limited, although several measurements have been reported and continue to improve. The most relevant channels include the semileptonic decays $B^{\pm} \to \pi^{\pm}\,\mu^{+}\mu^{-}$, for which the measurements of the differential branching fractions play an important role in constraining the parameter space \cite{LHCb:2012de,LHCb:2015hsa}.  These processes remain sensitive to contributions from dSMEFT operators, leading to potential modifications in the corresponding observables. While the current measurements are statistically limited, they still provide meaningful constraints on new physics scenarios and serve as an important cross-check of flavor structures beyond the SM. Further details on these transitions can be found in Refs. \cite{Biswas:2022lhu, Bause:2022rrs}.   

Besides the semileptonic modes, the rare dileptonic decays of neutral mesons, such as $P \to \mu^{+}\mu^{-}$, provide exceptionally sensitive probes of new physics. In the SM, these decays are highly suppressed by the absence of tree-level FCNCs, loop suppression, CKM hierarchies, and, for pseudoscalar mesons, helicity suppression. In our analysis, with our choice of operators, we get contributions to the rare decays $B_{s}^{0} \to \mu^{+} \mu^{-}$ and $B^{0} \to \mu^{+} \mu^{-}$. The current experimental measurements and SM predictions of these decays are given in Table \ref{tab:rare_decays_val}. 
\begin{table}[t]
    \centering
    \renewcommand{\arraystretch}{1.1}
 	\setlength{\tabcolsep}{4pt}
 	\resizebox{0.7\textwidth}{!}{%
    \begin{tabular}{|c|c|c|}
    \hline
    Process & Experimental Measurement & SM predictions  \\
    \hline \hline 
    $B_{s}^{0} \to \mu^{+} \mu^{-}$ & $(3.83 \pm 0.43) \times 10^{-9}$ \cite{CMS:2022mgd}  & $(3.66 \pm 0.14) \times 10^{-9} $\cite{Beneke:2019slt} \\ \hline
    & $(1.20 \pm 0.78) \times 10^{-10}  $ \cite{LHCb:2021vsc} & \\
    \multirow{-2}{*}{$B^{0} \to \mu^{+} \mu^{-}$}& $<1.5 \times 10^{-10}$ \cite{CMS:2022mgd} & \multirow{-2}{*}{$(1.03 \pm 0.05) \times 10^{-10} $\cite{Beneke:2019slt}} \\
    \hline
    \end{tabular}}
    \caption{Experimental measurements and SM predictions of the rare dileptonic decays of the $B$ mesons.}
    \label{tab:rare_decays_val}
\end{table}

The expression for the branching ratio of the $B_{q} \to \mu^+ \mu^-$ transition in terms of the WCs of the WET basis can be written as:
\begin{equation}
\begin{split} \label{eq:rare_BR_formula}
\mathcal{B}(B_q \rightarrow \mu^+ \mu^-) = & \tau_{B_q} f_{B_q}^2  m_{B_q} \frac{G_F^2 \alpha^2}{64 \pi^3} |V^*_{tq}V_{tb}|^2 \beta_{\mu}(m_{B_q}^2) \left[   \frac{m_{B_q}^2}{m_b^2} |C_s - C'_s|^2 \left(1-\frac{4m_{\mu}^2}{m_{B_q}^2}\right) \right.\\& \left.  + \bigg|\frac{m_{B_q}}{m_b}(C_p - C'_p) + 2\frac{m_{\mu}}{m_{B_q}} (C_{10} - C'_{10})\bigg|^2 \right]\,, 
\end{split}
\end{equation}
where $f_{B_q}$ and $\tau_{B_q}$ are the decay constant and lifetime of the meson $B_{q}$, respectively.

\subsubsection{Invisible decays of mesons}
Invisible decay modes provide important probes of new physics due to their clean theoretical structure and their loop-suppressed SM contributions. In SM, the invisible decays are the semileptonic decays of the mesons with a neutrino pair in the final state, i.e., decays like $ P \to P' \nu \bar{\nu}$. Recent data from the Belle-II collaboration on the branching fraction of $B^{+} \to K^{+} \nu \bar{\nu}$ decay shows a $2.7 \, \sigma$ deviation from the SM prediction \cite{Belle-II:2023esi}. These channels are theoretically very clean to study any possible new physics effects in these processes. 

The SM prediction and experimental results available for the invisible decays are given in Table~\ref{tab:invisible_data}. These predictions are based on the $B \to K$ and $B \to K^{*}$ form factors estimated using the lattice QCD (HPQCD \cite{Parrott:2022rgu}, FNAL/MILC \cite{Bailey:2015dka}) and LCSR \cite{Bharucha:2015bzk}, respectively. 
\begin{table}[t]
	\begin{center}
	\renewcommand{\arraystretch}{1.6}
    \resizebox{0.9\textwidth}{!} {
	\begin{tabular}{|ccc|}
	    \hline
		Process & SM Prediction & Experimental Value \\
		\hline
		\hline
		$ B^+ \to K^+ \nu\bar{\nu} $ & $ \left(5.06 \pm 0.14 \pm 0.28\right) \times 10^{-6} $ \cite{Becirevic:2023aov} & $ \left(2.3 \pm 0.5^{+0.5}_{-0.4}\right) \times 10^{-5} $ \cite{Belle-II:2023esi} \\
				
		$ B^0 \to K_{S}^0 \nu\bar{\nu} $ & $ \left( 2.05 \pm 0.07 \pm 0.12 \right) \times 10^{-6} $ \cite{Becirevic:2023aov} & $ < 1.3 \times 10^{-5} $ \cite{Belle:2017oht}\\
				
		$ B^+ \to K^{*+} \nu \bar{\nu} $ & $ \left (10.86 \pm 1.30 \pm 0.59\right) \times 10^{-6} $ \cite{Becirevic:2023aov} & $< 6.1 \times 10^{-5} $ \cite{Belle:2017oht} \\
						
		& & $ < 1.8 \times 10^{-5} $ \cite{Belle:2017oht} \\
        \multirow{-2}{*}{ $ B^0 \to K^{*0} \nu \bar{\nu} $}  & \multirow{-2}{*}{$ \left( 9.05 \pm  1.25 \pm  0.55 \right) \times 10^{-6} $ \cite{Becirevic:2023aov}} & $ \left( 3.0 \pm 2.75\right) \times 10^{-5} $  \cite{BaBar:2013npw} \\
				
		$ B^+ \to \pi^+ \nu\bar{\nu} $ & $ \left(9.70 \pm 2.10 \right) \times 10^{-6} $ \cite{Kamenik:2009kc} & $ < 1.4  \times 10^{-5} $ \cite{Belle:2017oht} \\
				
		$ B^0 \to \pi^{0} \nu\bar{\nu} $ & $ (0.73\pm 0.21) \times 10^{-7} $ \cite{Straub:2018kue} & $ < 0.9 \times 10^{-5}  $ \cite{Belle:2017oht} \\
				
		$ B^+ \to \rho^{+} \nu\bar{\nu} $ & $ (3.92\pm 0.79) \times 10^{-7} $ \cite{Straub:2018kue} & $ < 3 \times 10^{-5} $ \cite{Belle:2017oht} \\
				
		$ B^0 \to \rho^0 \nu\bar{\nu} $ & $ \left(1.82 \pm 0.31 \right) \times 10^{-7} $ \cite{Straub:2018kue} & $ < 4 \times 10^{-5} $ \cite{Belle:2017oht}\\		
		\hline
		\end{tabular} }
		\end{center}
		\caption{Updated SM predictions and the measured values or upper limit on the branching ratios of the invisible decays of $B$ mesons.}\label{tab:invisible_data}
	\end{table}
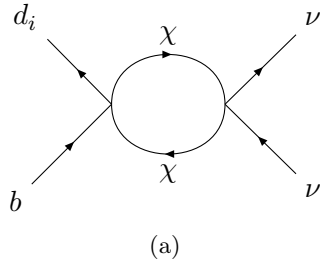
\begin{figure}[h!!!!]
	\centering
	\subfloat[]{\begin{tikzpicture}
		\begin{feynman}
			\vertex (a1){\( b\)};
			\vertex [above right=1.8cm of a1](a2);
			\vertex [above left=1.2cm of a2](a3){\(d_{i}\)};
			\vertex [right=1.5cm of a2](a4);
			\vertex [above right=1.3cm of a4](a5){\(\nu\)};
			\vertex [below right=1.3cm of a4](a6){\(\nu\)};
			
		\diagram* {
			(a1) -- [fermion,arrow size=0.8pt] (a2) -- [fermion,arrow size=0.8pt] (a3),
			(a2) --[fermion,arrow size=0.8pt,  half left, looseness=1.5,edge label={\(\chi\)}](a4) --[fermion,arrow size=0.8pt,  half left, looseness=1.5,edge label={\(\chi\)}](a2),
			(a6) --[fermion,arrow size=0.8pt](a4) --[fermion,arrow size=0.8pt](a5),
				
		};
		\end{feynman}
    \end{tikzpicture}}
    \caption{Feynman diagram contributing to the invisible decay $b \to d_{i}\, \nu \bar{\nu} $.}\label{fig:Feynan_invisible}
\end{figure}

In our case, for the operator set given in Table~\ref{tab:operator_lists}, modifications arise in the quark-level transition $b \to d_{i}\,\nu \bar{\nu}$. At the meson level, this leads to contributions to the invisible decay modes $B \to (K^{(*)}, \pi, \rho) \nu \bar{\nu}$. In our $\chi^{2}$ analysis, we include only those processes for which experimental measurements are available, and exclude modes for which only upper bounds exist. Since no experimental measurements are currently available for the $b \to d\,\nu \bar{\nu}$ transitions, these processes are not included in our numerical analysis. 

Similar to the effective Hamiltonian for the $ b \to s \ell \ell  $ processes, we can choose an effective basis for the $  b \to s \nu \bar{\nu} $ decay. Here, $\nu$ denotes the SM neutrino, which is left-handed only.
The effective Hamiltonian for this decay mode can be written as \cite{Altmannshofer:2009ma,Colangelo:1996ay}
\begin{equation}\label{eq:eff_hamilton_b2snunubar}
\mathcal{H}_{\rm eff} ^{b \to s \bar{\nu} \nu}= - \frac{4 G_{F}}{\sqrt{2}} V_{tb} V_{ts}^{*} (C_L^{\nu} \mathcal{O}_L^{\nu} + C_R^{\nu} \mathcal{O}_R^{\nu})\, + \rm h.c.,
\end{equation}
with the operator structure\footnote{For the invisible particles to be SM left-handed neutrino only these two operators are possible, since no scalar and tensor operator will survive due to the fixed chirality of neutrinos.}: 
\begin{eqnarray}
\mathcal{O}_{L}^{\nu} = \frac{e^2}{16 \pi^2}  \left( \bar{s} \gamma_{\mu} P_{L}b \right) \left(\bar{\nu} \gamma^{\mu}(1-\gamma_5)\nu\right) , & \ \ \ \ &  \mathcal{O}_{R}^{\nu}=  \frac{e^2}{16 \pi^2}  \left( \bar{s} \gamma_{\mu} P_{R}b \right) \left(\bar{\nu} \gamma^{\mu}(1-\gamma_5)\nu\right). 
\end{eqnarray}
The operator $\mathcal{O}_R^{\nu}$ arises only in the presence of NP. In the dSMEFT framework, the contributions to these invisible decay channels will be via the one-loop diagrams shown in Fig.~\ref{fig:Feynan_invisible}. 
From the matching condition, we get contributions to both $C_{L}^{\nu}$ and $C_{R}^{\nu}$. 
The total contributions to the WCs can be written as: 
\begin{eqnarray}
    C_{L}^{\nu} = C_{L}^{\nu, \rm SM} + C_{L}^{\nu, \rm NP} , \quad  \quad C_{R}^{\nu} = C_{R}^{\nu, \rm NP} \,.
\end{eqnarray}
The value of the WC in the SM is given by \cite{Chen:2024jlj,Buras:2014fpa}: 
\begin{equation}
C_{L}^{\nu, \rm SM} = - 6.32 \pm 0.07, \ \ \ \ C_{R}^{\nu, \rm SM} =0.
\end{equation} 
When we write the decay width for the invisible channel $ B \to K^{(*)} \bar{\nu} \nu $, the invisible state $\nu\bar{\nu}$ contains all three flavors of neutrino. The differential decay width of $B \to K \nu \bar{\nu}$ decay in terms of the above WCs can be written as \cite{Altmannshofer:2009ma} :
\begin{equation}\label{eq:br_B2Knunubar}
	\frac{d \Gamma (B^{+} \to K^{+} \bar{\nu}  \nu  )}{d q^2} =  \frac{G_{F}^2 \alpha^2 }{ 256  \pi^5 m_{B}^3} |V_{tb} V_{ts}^{*}|^2 \lambda^{3/2} ( m_{B}^2 , m_{K}^2, q^2) \left[f_{+}^{K}(q^2)\right]^2 |C_{L}^{\nu} + C_{R}^{\nu}|^2 \,,
	\end{equation}
where $f_{+}^K(q^2)$ is the vector form factor for the $B \to K$ transition. The differential decay width for a longitudinally and transversely polarized $K^*$ meson can be written as \cite{Altmannshofer:2009ma}:
\begin{eqnarray}
	\frac{d\Gamma_L}{dq^2}=&& P_{C} \frac{ |C_{L}^{\nu} - C_{R}^{\nu}|^2 \left( 
	A_1 (q^2) (m_B + m_{K^{*}})^2 \left(m_B^2 - m_{K^{*}}^2 - q^2 \right) 
	- A_2 (q^2) \lambda(m_B^2, m_{K^{*}}^2, q^2) \right)^2 }{m_{K^{*}}^2 (m_B + m_{K^{*}})^2}\,, \nonumber \\
	\frac{d\Gamma_T}{dq^2}= && P_{C} \frac{8 m_B^2 q^2 \left( [A_1(q^2)]^2 |C_{L}^{\nu} - C_{R}^{\nu}|^2 (m_B + m_{K^{*}})^4 
	+ [V(q^2)]^2 |C_{L}^{\nu} + C_{R}^{\nu}|^2 \lambda(m_B^2, m_{K^{*}}^2, q^2) \right) }{m_B^2 (m_B + m_{K^{*}})^2}\,, \nonumber \\
	\end{eqnarray}
where $A_1 (q^2)$, $A_2(q^2)$ and $V(q^2)$ are the $B \to K^*$ form factors and $P_{C}$ is given as: 
\begin{eqnarray}\label{eq:common_factor_invisible}
P_{C} = && \frac{ G_F^2  |V_{tb}  V_{ts}^{*}|^2 \alpha^2 
\sqrt{\lambda(m_B^2, m_{K^{*}}^2, q^2)}} {2^{10} \pi^5  m_B^3 } \,.
\end{eqnarray}
Therefore, the total differential decay width of the process is given as the sum of the decay widths of all the polarisations.
\begin{eqnarray}
    \frac{d\Gamma (B \to K^{*} \bar{\nu} \nu )}{dq^2} = \frac{d\Gamma_T}{dq^2}+ \frac{d\Gamma_L}{dq^2}\,.
\end{eqnarray}
In our case, we get NP contributions to both the operators $\mathcal{O}_{L,R}^{\nu}$. The corresponding WCs in terms of the loop contributions $\mathbb{C}_{1}$ and $\mathbb{C}_{2}$ given in Eq.~\eqref{eq:loop_contribution} and dSMEFT coefficients can be written as:
\begin{subequations}
\begin{align}
	C_{L}^{\nu}& =   \frac{K_{c}^{\nu \bar{\nu}}}{\Lambda^4} \left [ \mathbb{C}_{1} \left(\mathcal{C}_{q\chi, i3 }^{VL} \mathcal{C}_{\ell \chi}^{VL} + \mathcal{C}_{q\chi, i3}^{VR} \mathcal{C}_{\ell \chi}^{VR} \right) + \mathbb{C}_{2} \left( \mathcal{C}_{q\chi,i3}^{VL} \mathcal{C}_{\ell \chi}^{VR} + \mathcal{C}_{q\chi,i3}^{VR} \mathcal{C}_{\ell \chi}^{VL} \right) \right ]\,, \\
	C_{R}^{\nu}& =   \frac{K_{c}^{\nu \bar{\nu}}}{\Lambda^4}  \left[ \mathbb{C}_{1} \left(\mathcal{C}_{d\chi, i3}^{VL} \mathcal{C}_{\ell \chi}^{VL} + \mathcal{C}_{d\chi,i3}^{VR} \mathcal{C}_{\ell \chi}^{VR} \right) + \mathbb{C}_{2} \left( \mathcal{C}_{d\chi,i3}^{VL} \mathcal{C}_{\ell \chi}^{VR} + \mathcal{C}_{d\chi,i3}^{VR} \mathcal{C}_{\ell \chi}^{VL} \right) \right]\,.  \label{eq:WCb2snunu}
\end{align}   \end{subequations} 
The matching constant is given by: $K_{c}^{\nu \bar{\nu}} = - \frac{1}{4 \sqrt{2} G_{F}} \frac{1}{V_{tb}V_{td_{i}}^*} \frac{16 \pi^2}{e^2}$, where $i =\{ 1,2\}$, for $d_{i} = (d, s)$, respectively.  

Besides these, there is the long-distance contributions to the charged $B \to K^{(*)} \nu \bar{\nu}$ mode arising from the weak annihilation mediated by the on-shell $\tau$-lepton. It provides a significant enhancement to the branching ratios, which was first noted in Ref. \cite{Kamenik:2009kc}. The branching ratios for these tree contributions are given as \cite{Kamenik:2009kc, Becirevic:2023aov}
\begin{align}
\mathcal{B}(B \to K \nu \bar{\nu})_{\rm tree} &= \frac{\tau_B G_{F}^4 |V_{ud} V_{us}|^2 f_K^2 f_B^2 m_{\tau}}{128 \pi^2 m_B^3 \Gamma_{\tau}} (m_B^2 -m_{\tau}^2)^2 (m_{\tau}^2-m_K^2)^2\,,  \\
\mathcal{B}(B \to K^* \nu \bar{\nu})_{\rm tree} &= \frac{\tau_B G_{F}^4 |V_{ud} V_{us}|^2 f_{K^*}^2 f_B^2 m_{\tau}}{128 \pi^2 m_B^3 \Gamma_{\tau}} (m_B^2 -m_{\tau}^2)^2 (m_{\tau}^2-m_{K^*}^2)^2 \bigg(1+ \frac{2m_{K^*}^2}{m_{\tau}^2} \bigg) ~,
\end{align}
where $\Gamma_{\tau}$ is the total  width
of the intermediate $\tau$ lepton.

Note that in invisible decays, the final state is inferred only through missing energy, and hence the flavor of the neutrino cannot be identified experimentally. Therefore, the experimental bounds correspond to the sum over all three neutrino flavors, each contributing to the missing energy signature. Since different neutrino flavors can couple to different dSMEFT WCs, we compute the branching ratio for a single flavor (taken to be the muon neutrino) and rescale it by a factor of 3 for comparison with experimental measurements. A similar argument applies to scenarios involving other invisible particles in the final state. In particular, for our operator set, the transitions $b \to s(d)\,\chi \bar{\chi}$ can also contribute to invisible decay modes. However, such contributions are kinematically allowed only if the dark fermion $\chi$ is sufficiently light, satisfying $M_{\chi} \leq (m_{P} - m_{M})/2$, where $P$ and $M$ denote the initial and final state mesons, respectively.     

\subsection{Bounds from FCNC observables }\label{sec:final_paramSpace}
To constrain the WCs of the dSMEFT operators, we make use of the available low-energy FCNC data from the $b \to s \ell^+\ell^-$, $b \to s \nu\bar{\nu}$, and $b \to d \ell^+\ell^-$ transitions. As shown in Eqs.~\eqref{eq:WCb2sll} and \eqref{eq:WCb2snunu}, the corresponding observables depend on products of quark- and lepton-sector couplings. From the possible operator combinations probed by these processes, we consider several representative scenarios and assume the fermionic dark matter particle $\chi$ to be heavier than the top quark. Consequently, the dSMEFT operators contribute to $b \to s \nu\bar{\nu}$ only through virtual effects, whereas the decay $b \to s \chi\bar{\chi}$ is kinematically forbidden. In this setup, $\chi$ may interact through either left-handed or right-handed currents, enabling a separate investigation of the two chiral structures.

For the invisible decay mode, the experimentally observed final state corresponds to an undetected neutrino pair, 
\begin{equation}
\nu\bar{\nu}=\sum_{\ell=e,\mu,\tau}\nu_\ell\bar{\nu}_\ell .
\end{equation}
Although, in principle, the decay rate depends on the three flavor-specific dSMEFT couplings
 $$\mathcal{C}^{VL(R)}_{\ell\chi,11}, \qquad \mathcal{C}^{VL(R)}_{\ell\chi,22}, \qquad \mathcal{C}^{VL(R)}_{\ell\chi,33},$$ 
 the flavor of the final-state neutrinos cannot be experimentally identified. Consequently, the observable branching ratio is sensitive only to the flavor-summed rate, making it impossible to disentangle the individual flavor contributions. We therefore adopt the simplifying and well-motivated assumption of flavor-universal couplings, 
 
 $$\mathcal{C}^{VL(R)}_{\ell\chi,11} = \mathcal{C}^{VL(R)}_{\ell\chi,22} = \mathcal{C}^{VL(R)}_{\ell\chi,33}, $$
 such that 
 $$\mathcal{C}^{VL(R)}_{\ell\chi,11} + \mathcal{C}^{VL(R)}_{\ell\chi,22} + \mathcal{C}^{VL(R)}_{\ell\chi,33} = 3\,\mathcal{C}^{VL(R)}_{\ell\chi,22}.$$
Accordingly, the total branching ratio is obtained by multiplying the single-flavor result, computed for \(\nu_\mu\bar{\nu}_\mu\), by a factor of three.

\begin{table}[t!]
	\centering
	\rowcolors{1}{blue!7}{green!7}
	\renewcommand{\arraystretch}{1.4}
	\setlength{\tabcolsep}{13pt}
	\begin{tabular}{|c|c|c|c|}
		\hline
		\rowcolor{violet!10} 
		\multicolumn{4}{|c|}{$b \to s \ell \ell $ \& $b \to s \nu \bar{\nu}$}\\
		\hline
		\hline
		\rowcolor{blue!7}
		Hard Cutoff & DM mass & Couplings & Fit Result \\
		$\Lambda$ [TeV] & $M_{\chi}$ [GeV] & $\mathcal{C}_{q(d) \chi} \, \mathcal{C}_{\ell \chi} / \Lambda^4 $ &  $[\rm GeV^{-4}]$ \\
		\hline 
		\cellcolor{lime!10}&  & $\mathcal{C}_{q\chi, 23}^{VL(R)} \, \mathcal{C}_{\ell \chi ,  22}^{VL(R)}$ & $(-14.46 \pm 5.07) \times 10^{-14}$ \\
		\cellcolor{lime!10}&\multirow{-2}{*}{\cellcolor{green!10}250} & $\mathcal{C}_{d\chi, 23}^{VL(R)} \, \mathcal{C}_{\ell \chi ,  22}^{VL(R)}$ & $(-7.33 \pm 5.03) \times 10^{-14}$ \\
		\cline{2-4}
		\cline{2-4}
		\cellcolor{lime!10}& \cellcolor{blue!7} & $\mathcal{C}_{q\chi, 23}^{VL(R)} \, \mathcal{C}_{\ell \chi ,  22}^{VL(R)}$ & $(-7.23 \pm 2.52) \times 10^{-14}$ \\
		\cellcolor{lime!10}&\multirow{-2}{*}{500} & $\mathcal{C}_{d\chi, 23}^{VL(R)} \, \mathcal{C}_{\ell \chi ,  22}^{VL(R)}$ & $(-3.67 \pm 2.35)\times 10^{-14}$\\
		\cline{2-4}
		\cline{2-4}
		\cellcolor{lime!10}&  & 
		$\mathcal{C}_{q\chi, 23}^{VL(R)} \, \mathcal{C}_{\ell \chi ,  22}^{VL(R)}$ & $(-7.74 \pm 2.70) \times 10^{-14}$ \\
		\multirow{-6}{*}{\cellcolor{lime!10}1} &\multirow{-2}{*}{\cellcolor{green!10}750} &  
		$\mathcal{C}_{d\chi, 23}^{VL(R)} \, \mathcal{C}_{\ell \chi ,  22}^{VL(R)}$ & $(-3.93 \pm 2.57) \times 10^{-14}$ \\
		\hline \hline
		\cellcolor{lime!10}& \cellcolor{blue!7} & 
		$\mathcal{C}_{q\chi, 23}^{VL(R)} \, \mathcal{C}_{\ell \chi ,  22}^{VL(R)}$ & $(-9.64 \pm 3.37) \times 10^{-14}$ \\
		\cellcolor{lime!10}& \multirow{-2}{*}{250} 
		& $\mathcal{C}_{d\chi, 23}^{VL(R)} \, \mathcal{C}_{\ell \chi ,  22}^{VL(R)}$ & $(-4.89 \pm 3.34) \times 10^{-14}$ \\
		\cline{2-4}
		\cline{2-4}
		\cellcolor{lime!10} &  & 
		$\mathcal{C}_{q\chi, 23}^{VL(R)} \, \mathcal{C}_{\ell \chi ,  22}^{VL(R)}$ & $(-3.62 \pm 1.19) \times 10^{-14}$ \\
		\cellcolor{lime!10} &\multirow{-2}{*}{\cellcolor{green!10}500} &  
		$\mathcal{C}_{d\chi, 23}^{VL(R)} \, \mathcal{C}_{\ell \chi ,  22}^{VL(R)}$ & $(-1.83 \pm 1.14)\times 10^{-14}$\\
		\cline{2-4}
		\cline{2-4}
		\cellcolor{lime!10}    & \cellcolor{blue!7} & 
		$\mathcal{C}_{q\chi, 23}^{VL(R)} \, \mathcal{C}_{\ell \chi ,  22}^{VL(R)}$ & $(-2.27 \pm 0.81) \times 10^{-14}$ \\
		\cellcolor{lime!10} & \multirow{-2}{*}{750} &  
		$\mathcal{C}_{d\chi, 23}^{VL(R)} \, \mathcal{C}_{\ell \chi ,  22}^{VL(R)}$ & $(-1.15 \pm 0.82) \times 10^{-14}$ \\
		
		\cline{2-4}
		\cline{2-4}
		\cellcolor{lime!10}  & &   $\mathcal{C}_{q\chi, 23}^{VL(R)} \, \mathcal{C}_{\ell \chi ,  22}^{VL(R)}$ & $(-1.81 \pm 0.64) \times 10^{-14}$ \\
		\multirow{-8}{*}{\cellcolor{lime!10} 2} & \multirow{-2}{*}{\cellcolor{green!10}1000} &  
		$\mathcal{C}_{d\chi, 23}^{VL(R)} \, \mathcal{C}_{\ell \chi ,  22}^{VL(R)}$ & $(-0.92 \pm 0.64) \times 10^{-14}$\\
		\hline
	\end{tabular} 
	\caption{Fit results from  global fit of $ b \to s \ell \ell $ and $b \to s \nu \nu $ observables. The operator $\mathcal{O}_{e \chi}$ will not contribute to the invisible decay. The results are presented for the product couplings scaled by $\Lambda^4$ corresponding to DM mass $M_{\chi}$ = (250, 500, 750, 1000) GeV and the hard cutoff scales $\Lambda = (1,\,2)$ TeV.}
	\label{tab:fitresults}
\end{table}

Considering the $b\to s \ell \ell (\nu \nu)$ sector, we incorporate the data on the branching ratios, isospin asymmetry, and the angular observables in $B\to K^{(*)}\mu^+\mu^-$ and $B_s \to \phi\mu^+\mu^-$ decays, discussed in Sec.~\ref{sec:intro}. In addition, we also consider the inputs on the LFUV observables $R_K$ and $R_{K^*}$ and the measurement of the branching ratios on $B_s \to \mu^+ \mu^-$ and $B \to K \nu \bar{\nu}$ modes. As we have mentioned earlier, the relevant inputs on the differential rates and the angular observables are taken from the refs. \cite{LHCb:2014cxe, LHCb:2015svh, LHCb:2020lmf, LHCb:2021zwz, LHCb:2021xxq, LHCb:2022vje, Belle:2016fev, Belle:2019oag, BELLE:2019xld, ATLAS:2018gqc, LHCb:2020gog, Belle-II:2023esi}. The $B \to K$ form factors are based on the Lattice results from the FLAG Lattice Average \cite{FlavourLatticeAveragingGroupFLAG:2024oxs} and the LCSR results from \cite{Gubernari:2018wyi}. For the $B \to K^*$ and $B_s \to \phi$ form factors, we take the results for the coefficients of the BSZ parametrization from the Ref. \cite{Gubernari:2023puw}, which is based on the Lattice results from \cite{Horgan:2013hoa, Horgan:2015vla} and the LCSR results from \cite{Gubernari:2018wyi, Gubernari:2020eft}. The inputs on the CKM parameters are taken from Ref. \cite{ckmfitter}. The methodology of the fit is the same as discussed in Ref. \cite{Biswas:2020uaq}\footnote{All the fits carried out in this analysis and the subsequent analyses are performed using a \emph{Mathematica\textsuperscript \textregistered} package~\cite{OptEx}.}. As benchmark scenarios, we have considered the DM mass $M_{\chi}$ = (250, 500, 750, 1000) GeV for the hard cutoff scales $\Lambda$ = (1,2) TeV. The results are presented for the product couplings scaled by $\Lambda^4$. Since the neutrino is only left-handed in the SM, for the combined fits with $b \to s \ell \ell$ and $b \to s \nu \bar{\nu}$ observables, we haven't considered the right-handed leptonic operator $ \mathcal{O}_{e\chi}$. The results of the global fit considering $ b \to s \ell \ell $ and $b \to s \nu \nu $ observables are shown in Table \ref{tab:fitresults}. We find that the product of quark and lepton couplings is $\mathcal{O}(10^{-13}-10^{-14})$, and the absolute values of the product couplings decrease in general as the mass of the dark fermion $M_{\chi}$ increases suggesting that the NP parameter space becomes more constrained with increasing DM mass, a pattern also seen in the previous sections. The product couplings involving the right-handed quark operators $\mathcal{O}_{d\chi}$ are comparatively better constrained than their left-handed counterparts $\mathcal{O}_{q\chi}$. 

\begin{table}[t!]
	\centering
	\rowcolors{1}{blue!7}{green!7}
	\renewcommand{\arraystretch}{1.4}
	\setlength{\tabcolsep}{13pt}
	\begin{tabular}{|c|c|c|c|}
		\hline
		\rowcolor{violet!10}
		\multicolumn{4}{|c|}{$b \to s \ell \ell$}\\
		\hline
		\hline
		\rowcolor{blue!7}
		Hard Cutoff & DM mass & Couplings & Fit Result \\
		$\Lambda$ [TeV] & $M_{\chi}$ [GeV] & $\mathcal{C}_{q(d)\chi} \, \mathcal{C}_{e \chi} / \Lambda^4 $ &  $[\rm GeV^{-4}]$ \\
		\hline 
		\cellcolor{lime!10}&  & $\mathcal{C}_{q\chi, 23}^{VL(R)} \, \mathcal{C}_{e \chi ,  22}^{VL(R)}$ & $(-26.38 \pm 10.83) \times 10^{-14}$ \\
		\cellcolor{lime!10}&\multirow{-2}{*}{\cellcolor{green!10}250} & $\mathcal{C}_{d\chi, 23}^{VL(R)} \, \mathcal{C}_{e \chi ,  22}^{VL(R)}$ & $(-12.32 \pm 13.28) \times 10^{-14}$ \\
		\cline{2-4}
		\cline{2-4}
		\cellcolor{lime!10}& \cellcolor{blue!7} & $\mathcal{C}_{q\chi, 23}^{VL(R)} \, \mathcal{C}_{e \chi ,  22}^{VL(R)}$ & $(-13.19 \pm 5.96) \times 10^{-14}$ \\
		\cellcolor{lime!10}&\multirow{-2}{*}{500} & $\mathcal{C}_{d\chi, 23}^{VL(R)} \, \mathcal{C}_{e \chi ,  22}^{VL(R)}$ & $(-6.16 \pm 6.82)\times 10^{-14}$\\
		\cline{2-4}
		\cline{2-4}
		\cellcolor{lime!10}&  & 
		$\mathcal{C}_{q\chi, 23}^{VL(R)} \, \mathcal{C}_{e \chi ,  22}^{VL(R)}$ & $(-14.12 \pm 6.26) \times 10^{-14}$ \\
		\multirow{-6}{*}{\cellcolor{lime!10}1} &\multirow{-2}{*}{\cellcolor{green!10}750} &  
		$\mathcal{C}_{d\chi, 23}^{VL(R)} \, \mathcal{C}_{e \chi ,  22}^{VL(R)}$ & $(-6.60 \pm 7.27) \times 10^{-14}$ \\
		\hline \hline
		\cellcolor{lime!10}& \cellcolor{blue!7} & 
		$\mathcal{C}_{q\chi, 23}^{VL(R)} \, \mathcal{C}_{e \chi ,  22}^{VL(R)}$ & $(-17.59 \pm 7.43) \times 10^{-14}$ \\
		\cellcolor{lime!10}& \multirow{-2}{*}{250} 
		& $\mathcal{C}_{d\chi, 23}^{VL(R)} \, \mathcal{C}_{e \chi ,  22}^{VL(R)}$ & $(-8.21 \pm 8.95) \times 10^{-14}$ \\
		\cline{2-4}
		\cline{2-4}
		\cellcolor{lime!10} &  & 
		$\mathcal{C}_{q\chi, 23}^{VL(R)} \, \mathcal{C}_{e \chi ,  22}^{VL(R)}$ & $(-6.60 \pm 3.49) \times 10^{-14}$ \\
		\cellcolor{lime!10} &\multirow{-2}{*}{\cellcolor{green!10}500} &  
		$\mathcal{C}_{d\chi, 23}^{VL(R)} \, \mathcal{C}_{e \chi ,  22}^{VL(R)}$ & $(-3.08 \pm 3.74)\times 10^{-14}$\\
		\cline{2-4}
		\cline{2-4}
		\cellcolor{lime!10}    & \cellcolor{blue!7} & 
		$\mathcal{C}_{q\chi, 23}^{VL(R)} \, \mathcal{C}_{e \chi ,  22}^{VL(R)}$ & $(-4.14 \pm 2.26) \times 10^{-14}$ \\
		\cellcolor{lime!10} & \multirow{-2}{*}{750} &  
		$\mathcal{C}_{d\chi, 23}^{VL(R)} \, \mathcal{C}_{e \chi ,  22}^{VL(R)}$ & $(-1.93 \pm 2.63) \times 10^{-14}$ \\
		
		\cline{2-4}
		\cline{2-4}
		\cellcolor{lime!10}  & &   $\mathcal{C}_{q\chi, 23}^{VL(R)} \, \mathcal{C}_{e \chi ,  22}^{VL(R)}$ & $(-3.30 \pm 1.86) \times 10^{-14}$ \\
		\multirow{-8}{*}{\cellcolor{lime!10} 2} & \multirow{-2}{*}{\cellcolor{green!10}1000} &  
		$\mathcal{C}_{d\chi, 23}^{VL(R)} \, \mathcal{C}_{e \chi ,  22}^{VL(R)}$ & $(-1.54 \pm 2.23) \times 10^{-14}$\\
		\hline
	\end{tabular} 
	\caption{Fit results from  global fit of $ b \to s \ell \ell $ observables. The results are presented for the product couplings scaled by $\Lambda^4$ corresponding to DM mass $M_{\chi} = (250, 500, 750, 1000)$ GeV and the hard cutoff scales $\Lambda  = (1,\,2)$  TeV.}
	\label{tab:fitresults2}
\end{table}
 
We have also performed a fit with only the $b \to s \ell \ell$ observables for constraining the product couplings involving the right-handed leptonic operator $\mathcal{O}_{e \chi}$, the results of which are shown in Table~\ref{tab:fitresults2}. We have checked that the values of the product couplings involving the left-handed leptonic operators $\mathcal{O}_{l \chi}$ are similar to those in Table~\ref{tab:fitresults}, so we don't show them in this table. Comparing both the tables, we find that the product couplings involving the operator $\mathcal{O}_{e \chi}$ [Table \ref{tab:fitresults2}] are comparatively less constrained and have larger error bars as compared to the couplings involving the left-handed leptonic operator $\mathcal{O}_{l \chi}$ [Table \ref{tab:fitresults}], though the order of magnitudes obtained are more or less similar. The product couplings involving both right handed quark and lepton couplings are consistent with zero. 

Next, we also consider the measurements of the differential branching fractions for the $B \to \pi \mu^+ \mu^-$ decays \cite{LHCb:2015hsa}. The $B \to \pi$ form factors are based on the Lattice results from the Refs. \cite{FermilabLattice:2015cdh,Flynn:2015mha,Colquhoun:2022atw} and the LCSR results from \cite{Gubernari:2018wyi,Leljak:2021vte}. The results of the global fit involving the $b \to d \ell \ell$ observables are shown in Table~\ref{tab:fitresults1}. For these product couplings, we mostly get zero-consistent solutions because of the compatibility between the experimental measurements and the SM prediction. The product couplings involving the operators $\mathcal{O}_{e \chi}$ have larger uncertainties as compared to the product couplings involving the operators $\mathcal{O}_{\ell \chi}$, as also seen in Tables~\ref{tab:fitresults} and \ref{tab:fitresults2}.

\begin{table}[htb!]
	\centering
	\rowcolors{1}{blue!7}{green!10}
	\renewcommand{\arraystretch}{1.4}
	\setlength{\tabcolsep}{8pt}
	\begin{tabular}{|c |c|c|c|}
		\hline
		\rowcolor{violet!10}
		\multicolumn{4}{|c|}{$b \to d \ell \ell $ }\\
		\hline
		\hline
		\rowcolor{blue!7}
		Hard Cutoff  & DM mass & Couplings & Fit Result \\
		$\Lambda$ [TeV] & $M_{\chi}$ [GeV] & $\mathcal{C}_{q(d) \chi} \mathcal{C}_{\ell(e)\chi }/ \Lambda^4 $ & $[\rm GeV^{-4}]$  \\
		\hline  
		\cellcolor{lime!10} & & $\mathcal{C}_{q\chi, 13}^{VL(R)} \, \mathcal{C}_{\ell \chi ,  22}^{VL(R)}$, $\mathcal{C}_{d\chi, 13}^{VL(R)} \, \mathcal{C}_{\ell \chi ,  22}^{VL(R)}$ & $(2.95 \pm 4.64) \times 10^{-14}$ \\
		
		\cellcolor{lime!10} & \multirow{-2}{*}{\cellcolor{green!10}250} & $\mathcal{C}_{q\chi, 13}^{VL(R)} \, \mathcal{C}_{e \chi ,  22}^{VL(R)}$, $\mathcal{C}_{d\chi, 13}^{VL(R)} \, \mathcal{C}_{e \chi ,  22}^{VL(R)}$ & $(-5.48 \pm 21.50) \times 10^{-14}$ \\
		\cline{2-4}
		\cline{2-4}
		\cellcolor{lime!10}& \cellcolor{blue!7} & $\mathcal{C}_{q\chi, 13}^{VL(R)} \, \mathcal{C}_{\ell \chi ,  22}^{VL(R)}$, $\mathcal{C}_{d\chi, 13}^{VL(R)} \, \mathcal{C}_{\ell \chi ,  22}^{VL(R)}$ & $(1.47 \pm 2.32) \times 10^{-14}$ \\
		\cellcolor{lime!10} & \multirow{-2}{*}{500} & $\mathcal{C}_{q\chi, 13}^{VL(R)} \, \mathcal{C}_{e \chi ,  22}^{VL(R)}$, $\mathcal{C}_{d\chi, 13}^{VL(R)} \, \mathcal{C}_{e \chi ,  22}^{VL(R)}$ & $(-2.74 \pm 10.75) \times 10^{-14}$\\
		\cline{2-4}
		\cline{2-4}
		\cellcolor{lime!10} &  & 
		$\mathcal{C}_{q\chi, 13}^{VL(R)} \, \mathcal{C}_{\ell \chi ,  22}^{VL(R)}$, $\mathcal{C}_{d\chi, 13}^{VL(R)} \, \mathcal{C}_{\ell \chi ,  22}^{VL(R)}$ & $(1.58 \pm 2.49) \times 10^{-14}$ \\
		
		\multirow{-6}{*}{\cellcolor{lime!10}1}  & \multirow{-2}{*}{\cellcolor{green!10}750} &  
		$\mathcal{C}_{q\chi, 13}^{VL(R)} \, \mathcal{C}_{e \chi ,  22}^{VL(R)}$, $\mathcal{C}_{d\chi, 13}^{VL(R)} \, \mathcal{C}_{e \chi ,  22}^{VL(R)}$ & $(-2.93 \pm 11.5) \times 10^{-14}$ \\
		\hline \hline
		\cellcolor{lime!10}  & \cellcolor{blue!7} & 
		$\mathcal{C}_{q\chi, 13}^{VL(R)} \, \mathcal{C}_{\ell \chi ,  22}^{VL(R)}$, $\mathcal{C}_{d\chi, 13}^{VL(R)} \, \mathcal{C}_{\ell \chi ,  22}^{VL(R)}$ & $(1.97 \pm 3.10) \times 10^{-14}$ \\
		
		\cellcolor{lime!10} & \multirow{-2}{*}{250}  &  
		$\mathcal{C}_{q\chi, 13}^{VL(R)} \, \mathcal{C}_{e \chi ,  22}^{VL(R)}$, $\mathcal{C}_{d\chi, 13}^{VL(R)} \, \mathcal{C}_{e \chi ,  22}^{VL(R)}$ & $(-3.65 \pm 14.3) \times 10^{-14}$ \\
		\cline{2-4}
		\cline{2-4}
		\cellcolor{lime!10} &  & 
		$\mathcal{C}_{q\chi, 13}^{VL(R)} \, \mathcal{C}_{\ell \chi ,  22}^{VL(R)}$, $\mathcal{C}_{d\chi, 13}^{VL(R)} \, \mathcal{C}_{\ell \chi ,  22}^{VL(R)}$ & $(0.74 \pm 1.16) \times 10^{-14}$ \\
		
		\cellcolor{lime!10} &\multirow{-2}{*}{\cellcolor{green!10}500} &  
		$\mathcal{C}_{q\chi, 13}^{VL(R)} \, \mathcal{C}_{e \chi ,  22}^{VL(R)}$, $\mathcal{C}_{d\chi, 13}^{VL(R)} \, \mathcal{C}_{e \chi ,  22}^{VL(R)}$ & $(-1.37 \pm 5.37) \times 10^{-14}$\\
		\cline{2-4}
		\cline{2-4}
		\cellcolor{lime!10}  & \cellcolor{blue!7} & 
		$\mathcal{C}_{q\chi, 13}^{VL(R)} \, \mathcal{C}_{\ell \chi ,  22}^{VL(R)}$, $\mathcal{C}_{d\chi, 13}^{VL(R)} \, \mathcal{C}_{\ell \chi ,  22}^{VL(R)}$ & $(0.46 \pm 0.73) \times 10^{-14}$ \\
		
		\cellcolor{lime!10} & \multirow{-2}{*}{750} &  
		$\mathcal{C}_{q\chi, 13}^{VL(R)} \, \mathcal{C}_{e \chi ,  22}^{VL(R)}$, $\mathcal{C}_{d\chi, 13}^{VL(R)} \, \mathcal{C}_{e \chi ,  22}^{VL(R)}$ & $(-0.86 \pm 3.37) \times 10^{-14}$ \\
		\cline{2-4}
		\cline{2-4}
		\cellcolor{lime!10}  & &   $\mathcal{C}_{q\chi, 13}^{VL(R)} \, \mathcal{C}_{\ell \chi ,  22}^{VL(R)}$, $\mathcal{C}_{d\chi, 13}^{VL(R)} \, \mathcal{C}_{\ell \chi ,  22}^{VL(R)}$ & $(0.37 \pm 0.58) \times 10^{-14}$ \\
		
		\multirow{-8}{*}{\cellcolor{lime!10} 2} & \multirow{-2}{*}{\cellcolor{green!10}1000} &  
		$\mathcal{C}_{q\chi, 13}^{VL(R)} \, \mathcal{C}_{e \chi ,  22}^{VL(R)}$, $\mathcal{C}_{d\chi, 13}^{VL(R)} \, \mathcal{C}_{e \chi ,  22}^{VL(R)}$ & $(-0.68 \pm 2.69) \times 10^{-14}$\\
		\hline
	\end{tabular} 
	\caption{Fit results from global fit of $ b \to d \ell \ell$ observables. The results are presented for the product couplings scaled by $\Lambda^4$ corresponding to DM mass $M_{\chi}$ = (250, 500, 750, 1000) GeV and the hard cutoff scales $\Lambda$ = (1,2) TeV.}
	\label{tab:fitresults1}
\end{table}
Note that the branching fractions $\mathcal{B}(B_s^0\to\mu^+\mu^-)$ and $\mathcal{B}(B^0\to\mu^+\mu^-)$ also play an important role in constraining the allowed dSMEFT parameter space. To illustrate the impact of the dark matter mass on these constraints, in Fig.~\ref{fig:rare_plots} of Appendix~\ref{secapp:onlybsll}, we show the variation of the relevant couplings as a function of $M_\chi$. The figure clearly demonstrates the sensitivity of the branching fractions to the dark matter mass. Furthermore, the allowed ranges of the couplings obtained for smaller values of $M_\chi$ are in good agreement with those reported in the corresponding tables, thereby providing a consistency check of our analysis.

\subsection{Dark Matter Observables} \label{sec:darkmatter}
In our operator basis, summarized in Table~\ref{tab:operator_lists}, each dSMEFT operator involves a dark-sector fermion $\chi$ coupled to SM fields. We identify $\chi$ as a fermionic WIMP dark matter candidate stabilized by a $\mathbb{Z}_2$ symmetry, under which $\chi$ is odd and all SM fields are even. The EFT description is assumed to be valid for $\Lambda > M_\chi$, ensuring that the thermal freeze-out dynamics occur within the regime of validity of the effective theory. 

We assume that the relic abundance of $\chi$ is produced through the standard thermal freeze-out mechanism. In the early Universe, $\chi$ remains in thermal equilibrium with the SM bath through annihilations into SM particles. As the temperature drops below $M_\chi$, the interaction rate decreases and eventually falls below the Hubble expansion rate, leading to freeze-out and a relic abundance of dark matter. Requiring successful thermal freeze-out therefore imposes a lower bound on the dSMEFT couplings, which can be estimated by demanding that the interaction rate exceed the Hubble rate prior to freeze-out. Thermal equilibrium is maintained as long as the interaction rate exceeds the expansion rate of the Universe \cite{Kolb:1990vq},
\begin{equation}
 \langle \sigma v \rangle \,  n_{\chi}^{\rm eq} > H,
\end{equation}
where $H = 1.66 \sqrt{g_*}\,T^2/M_P$ is the Hubble expansion rate, $g_*$ denotes the effective number of relativistic degrees of freedom, and $T$ is the temperature of the Universe, and $n_{\chi}^{\rm eq}$ is the equilibrium number density. 
The evolution of the DM number density is governed by the Boltzmann equation \cite{Lee:1977ua},
\begin{equation}
\frac{dn_\chi}{dt} + 3H n_\chi
= - \langle \sigma v \rangle
\left( n_\chi^2 - \left(n_{\chi}^{\mathrm{eq}}\right)^2 \right),
\end{equation}
where $n_{\chi}$ is the DM number density. The dynamics of freeze-out are controlled by the thermally averaged annihilation cross section $\langle \sigma v \rangle$, which encodes the strength of DM annihilation into SM particles. This quantity can be expanded in powers of the relative velocity of the annihilating DM particles as \cite{Kolb:1990vq}
\begin{equation}
\langle \sigma v \rangle = a + b\,\langle v^2 \rangle + \mathcal{O}(v^4)~,
\end{equation}
where $a$ and $b$ correspond to the $s$-wave and $p$-wave contributions, respectively. For fermionic DM with pure vector interactions, the annihilation into fermion pairs contains an unsuppressed $s$-wave component, implying that the leading contribution to freeze-out is velocity independent. Same contributions are obtained for pure axial-vector interactions between SM and DM particles. For the other mixed currents, the thermal averaged cross-section is velocity suppressed \cite{Berlin:2014tja}. Freeze-out occurs when the annihilation rate $\Gamma_\chi = n_\chi \langle \sigma v \rangle$ drops below the Hubble expansion rate, after which the comoving number density becomes effectively constant.

Through the dimension-6 interactions considered in this work, the DM annihilates into SM fermions at tree level, thereby contributing to the relic density. Since the annihilation proceeds at tree level within this setup, the relic density calculation does not introduce additional ultraviolet sensitivities in the EFT framework. For the chosen operators, the DM particle $\chi$ couples exclusively to $b$, $d$, and $s$ quarks, as well as to muons. Consequently, the relic abundance is determined by annihilation into these final states. The relevant tree-level annihilation channels include $\chi \bar{\chi} \to \{ b\bar{s},\, b\bar{d},\, \mu^{+}\mu^{-} \} + \mathrm{h.c.}$. In addition to the singlet fermion operators, the doublet operators also contribute to these channels and further induce additional annihilation modes arising from the SU(2) counterparts. In particular, this leads to channels such as $\chi \bar{\chi} \to \{ t\bar{c},\, t\bar{u} \} + \mathrm{h.c.}$, where the top quark couples to the DM only through left-handed interactions. For the DM to annihilate into top quarks, its mass must lie within the kinematically allowed region: $2M_{\chi} \geq m_{t} + m_{u_i}$, where $u_{i} \in \{u, c\}$. The corresponding Feynman diagrams are shown in Fig.~\ref{fig:Feynman_relic}. The resulting relic abundance is required to satisfy the Planck 2018 constraint \cite{Planck:2018vyg}:
\begin{equation}
    \Omega_{\rm DM} h^2 = 0.1200 \pm 0.0012 \,.
\end{equation}
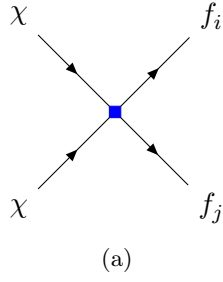
\begin{figure}
	\centering
	\subfloat[]{\begin{tikzpicture}
		\begin{feynman}
			\vertex (a1){\(\chi\)};
			\vertex [square dot, blue, above right=1.8cm of a1](a2){};
			\vertex [above left=1.8cm of a2](a3){\(\chi\) };
			\vertex [above right=1.8cm of a2](a4){ \( f_i\)};
			\vertex [below right=1.8cm of a2](a5){\(f_j\)};
			
			\diagram* {
				(a1) -- [fermion,arrow size=1pt] (a2) ,
				(a3)-- [fermion,arrow size=1.pt] (a2),
				(a2) --[fermion,arrow size=1.pt](a5),
				(a2) --[fermion,arrow size=1.pt](a4),
			};
		\end{feynman}
	\end{tikzpicture}}
    \caption{Feynman diagram of the DM annihilating to SM fermions, contributing to the relic density. The final state fermion pairs $f_{i} \bar{f}_j$ stand for the pairs: $\{ b \bar s, \, b \bar d , \, \mu^{+} \mu^{-}, \, t \bar{c} , \, t \bar u \} + h.c.$ }
    \label{fig:Feynman_relic}
\end{figure}
\paragraph{\underline{Thermal equilibrium of the DM}:}
As discussed, the DM $\chi$ in our scenario is a WIMP candidate, which attains its present relic density via thermal freeze-out. For this mechanism to work, the DM must be in thermal equilibrium with the SM plasma in the early Universe and follow its equilibrium number density. The condition for maintaining thermal equilibrium is given by
\begin{equation} 
n_{\chi}^{\rm eq} \langle\sigma v \rangle_{\rm tot} \gtrsim H(T)\,,
\end{equation}
where \(H(T)\) denotes the Hubble expansion rate at temperature \(T\). At later times, as the temperature drops, the interaction rate becomes comparable to the expansion rate, $n_{\chi}^{\rm eq}\langle\sigma v\rangle \sim H$, leading to the departure from equilibrium and eventual freeze-out of the DM. This condition provides a lower bound on the interaction strength of a given operator to ensure that the DM was in thermal equilibrium in the early Universe. Since our analysis is performed in a single-operator framework, we impose this condition separately for each operator.
\begin{figure}[t]
    \centering
    \subfloat[]{\includegraphics[scale=0.18]{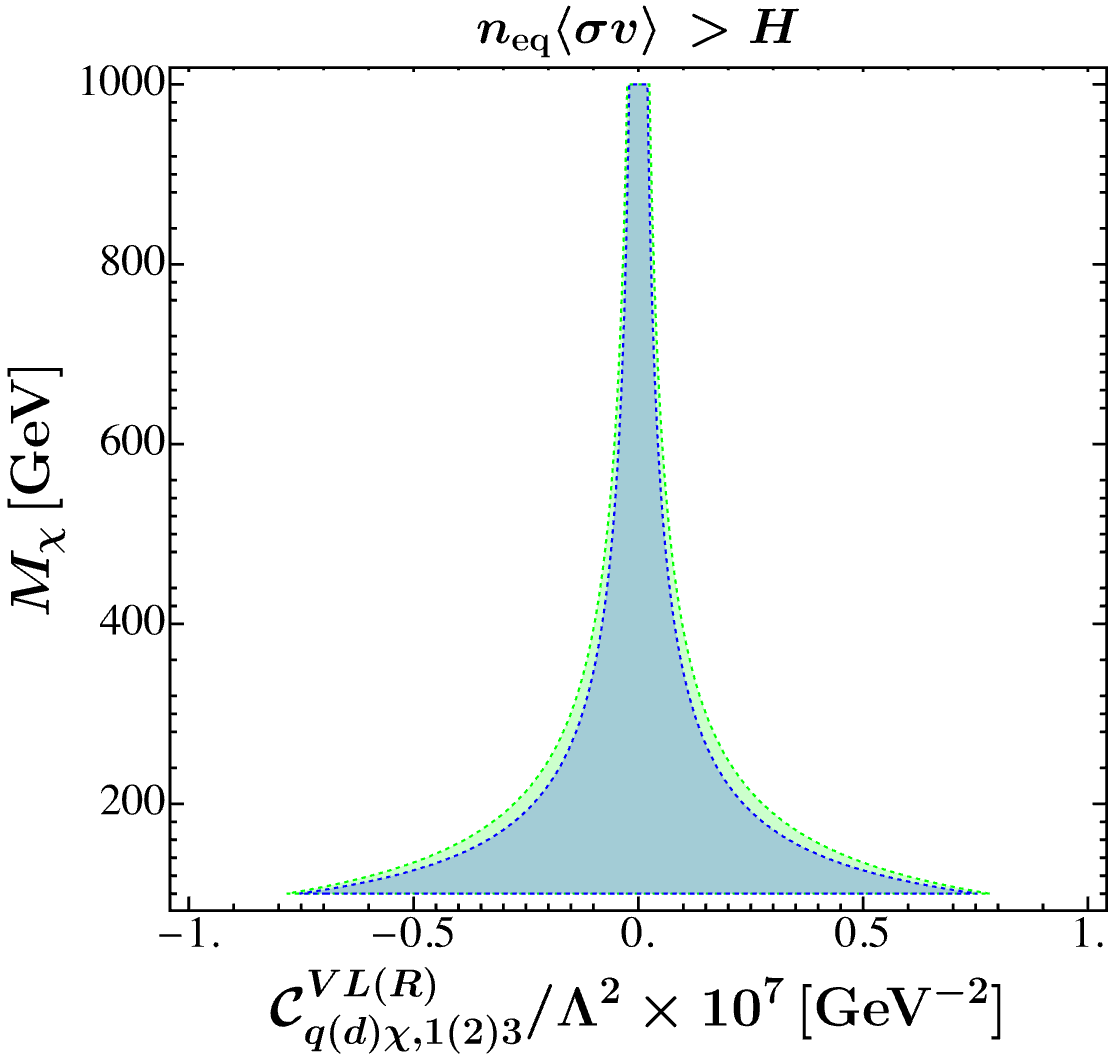}}~~
    \subfloat[]{\includegraphics[scale=0.18]{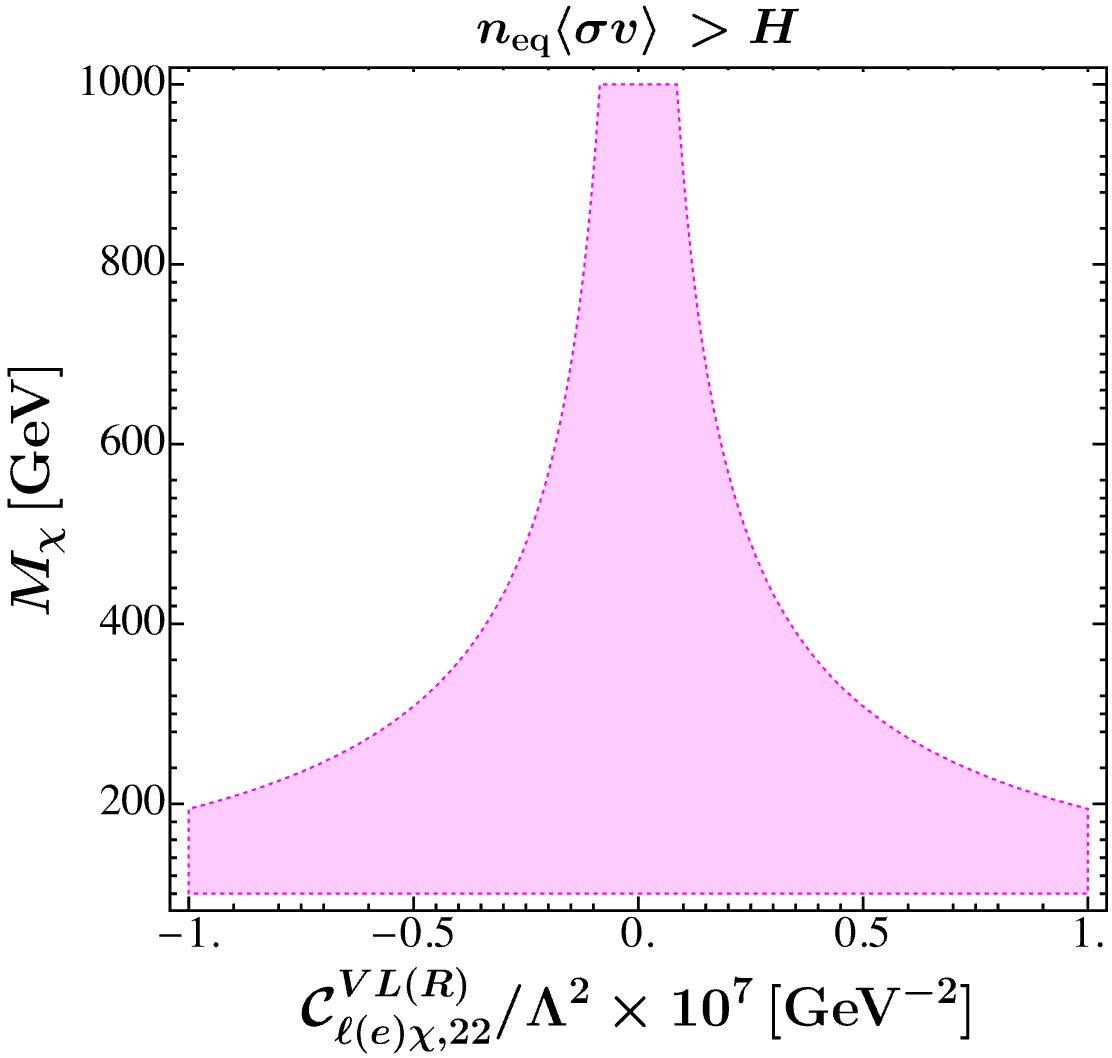}}
    \caption{Excluded region from the condition of the DM to be in thermal equilibrium with the SM bath in the early universe. The left plot is for DM coupling with left-handed quarks (blue) and right-handed quarks (green), whereas the plot in the right panel is for the leptons. }
    \label{fig:thermal_equi}
\end{figure}

Figure~\ref{fig:thermal_equi} shows the disallowed parameter space in the mass-coupling plane arising from the requirement that the DM remains in thermal equilibrium with the SM in the early Universe. The constraint is evaluated near the freeze-out temperature, where the DM departs from equilibrium with the thermal bath. The left panel corresponds to DM interactions with quarks, while the right panel shows the case for leptons. In the left panel, the blue and green regions represent the disallowed parameter space for left-handed and right-handed quark couplings, respectively, and we observe that both cases lead to almost similar excluded regions. In contrast, the disallowed region for muons in the right panel is comparatively wider than that for quarks. For a representative DM mass of $M_{\chi} \sim 500\,\mathrm{GeV}$, the leptonic couplings are disallowed for $|\mathcal{C}_{\ell (e)}^{VL(R)} / \Lambda^2| \leq 0.2 \times 10^{-7}\,\mathrm{GeV}^{-2}$, whereas for quarks at the same DM mass we obtain a stronger constraint, $|\mathcal{C}_{q (d)}^{VL(R)} / \Lambda^2| \leq 0.05 \times 10^{-7}\,\mathrm{GeV}^{-2}$. We also observe that the disallowed region becomes narrower with increasing DM mass. Therefore, for the DM to behave as a WIMP and achieve the observed relic density via the freeze-out mechanism, the couplings must lie above these lower bounds.

\paragraph{\underline{Direct Detection of Dark Matter} :}
DM can be probed through its elastic scattering with nuclei in direct detection experiments. At the microscopic level, this corresponds to the scattering process $\chi N \to \chi N$, where $N$ denotes a nucleon. In these experiments, DM particles from the Galactic halo interact with detector nuclei, leading to nuclear recoils that can be measured in low-background underground facilities. The scattering process is typically classified into spin-independent (SI) and spin-dependent (SD) interactions, depending on the underlying interaction structure. Spin-independent interactions lead to coherent scattering off the entire nucleus and are enhanced by the nuclear mass, making them particularly sensitive in current experiments, while spin-dependent interactions depend on the spin content of the nucleus and are comparatively less constrained. For the sub-TeV mass range of the DM, recent experimental results from direct detection experiments such as XENONnT \cite{XENON:2025vwd, XENON:2023cxc}, LZ \cite{LZ:2024zvo, LZ:2022lsv}, and PandaX-4T \cite{PandaX:2024qfu, PandaX:2023ejt} have provided increasingly stringent bounds on the spin-independent DM-nucleon scattering cross section, thereby placing strong constraints on the parameter space of particle DM models. Among these, LZ \cite{LZ:2024zvo} provides the most stringent constraints. The projection bound from the future XLZD experiment is available in \cite{XLZD:2024nsu}.
\begin{figure}
	\centering
    \subfloat[]{\begin{tikzpicture}
		\begin{feynman}
			\vertex (a1){\(\chi\)};
			\vertex [square dot, blue, above right=2cm of a1](a2){};
			\vertex [above left=1.8cm of a2](a3){\(\chi\) };
			\vertex [above right=0.8cm of a2](a4);
			\vertex [below right=0.8cm of a2](a5);
            \vertex [above right=0.9cm of a4](a6){\( u(d)\)};
			\vertex [below right=0.9cm of a5](a7){\(u(d)\)};
			
			\diagram* {
				(a1) -- [fermion,arrow size=0.7pt] (a2) ,
				(a3)-- [fermion,arrow size=0.7pt] (a2),
				(a2) --[fermion,arrow size=0.7pt,edge label'={\(b(t)\)}](a5) --[fermion, arrow size = 0.7 pt](a7),
                (a5) --[boson,edge label'={\(W\)}](a4),
				(a2) --[fermion,arrow size=0.7pt,edge label={\( d (u) \)}](a4) --[fermion, arrow size = 0.7 pt](a6),
				
			};
		\end{feynman}
	\end{tikzpicture}\label{fig:Feyn_direct_vertex}}~~~~~
    \subfloat[]{\begin{tikzpicture}
		\begin{feynman}
			\vertex (a1){\(\chi\)};
			\vertex [square dot, blue, above right=2cm of a1](a2){};
			\vertex [above left=1.8cm of a2](a3){\(\chi\) };
			\vertex [above right=1.8cm of a2](a4){\(d(u)\)};
            \vertex [below right=0.6cm of a2](a5);
            \vertex [below right=0.6cm of a5](a6);
            \vertex [below right=0.6cm of a6](a7){\(d(u)\)};
			
			\diagram* {
				(a1) -- [fermion,arrow size=0.7pt](a2),
                (a3) --[fermion,arrow size=0.7pt](a2) --[fermion,arrow size=0.7pt](a4),
                (a7) --[fermion, arrow size=0.7pt](a6) --[fermion, arrow size=0.7pt, edge label={\(t(b)\)}](a5) --[fermion, arrow size=0.7pt, edge label={\(b(t)\)}](a2),
                (a6) --[boson, edge label'={\(W\)}, half right, looseness=2.54](a5),
			};
		\end{feynman}
	\end{tikzpicture}\label{fig:Feyn_direct_leg}}
    \caption{Feynman diagrams contributing to the DM-nucleon scattering cross-sections $\chi u \to \chi u$ and $\chi d \to \chi d$ via vertex correction  (Fig.~\ref{fig:Feyn_direct_vertex}) and external leg correction (Fig.~\ref{fig:Feyn_direct_leg}) via one-loop processes.}
    \label{fig:Feyn_direct}
\end{figure}
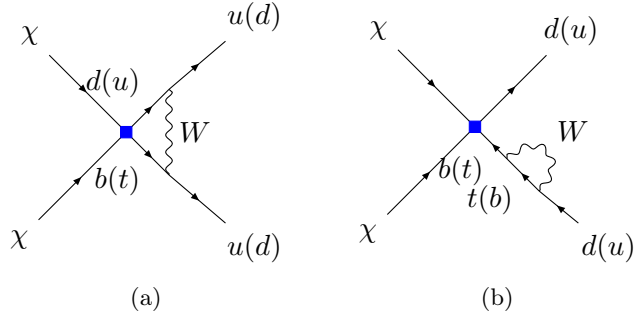

In our analysis, the same dSMEFT WCs that govern the DM relic density also contribute to direct detection observables, providing a complementary probe of the parameter space. Since the DM interactions considered in this work are flavor-violating, contributions to direct detection arise only at the one-loop level. The corresponding Feynman diagrams are shown in Fig.~\ref{fig:Feyn_direct}. We obtain the scattering  $\chi \, u \to \chi\, u$ and $\chi \, d  \to \chi \,  d$ via a vertex correction penguin diagram and a wave-function renormalization. Due to the presence of both vector and axial-vector interactions, our framework generates contributions to both spin-independent and spin-dependent scattering cross sections. However, given that current experimental bounds on spin-dependent interactions are significantly weaker than those on spin-independent interactions, we focus only on the spin-independent cross section and the corresponding constraints.

\paragraph{\underline{Indirect Detection of Dark Matter} :}
DM can also be probed through indirect detection experiments, which search for the products of DM annihilation or decay in astrophysical environments. In regions with high DM density, such as the Galactic center or dwarf spheroidal galaxies, DM particles can annihilate into SM particles, producing observable fluxes of gamma rays, cosmic rays, and neutrinos. The expected signals depend on the annihilation cross section, the DM mass, and the distribution of DM in the halo. Among these, gamma-ray observations provide particularly strong constraints due to their relatively clean propagation. Current observations from experiments such as Fermi Large
Area Telescope \cite{Fermi-LAT:2015att,Fermi-LAT:2016afa}, High Energy Stereoscopic System  (H.E.S.S.) \cite{HESS:2016mib}, and  Cherenkov Telescope Array (CTA) \cite{Silverwood:2014yza} have placed stringent bounds on the annihilation cross section for a wide range of DM masses and final states, thereby offering complementary constraints to those obtained from relic density and direct detection studies.

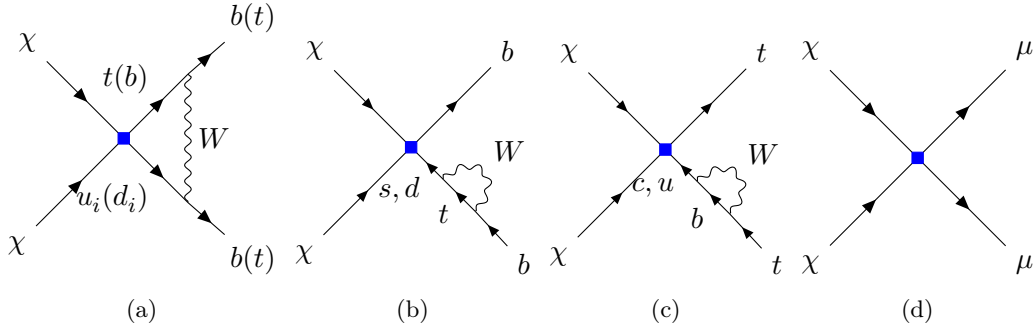
\begin{figure}
	\centering
    \subfloat[]{\begin{tikzpicture}
		\begin{feynman}
			\vertex (a1){\(\chi\)};
			\vertex [square dot, blue, above right=2cm of a1](a2){};
			\vertex [above left=1.8cm of a2](a3){\(\chi\) };
			\vertex [above right=1.2cm of a2](a4);
			\vertex [below right=1.2cm of a2](a5);
            \vertex [above right=0.6cm of a4](a6){\(b(t)\)};
			\vertex [below right=0.6cm of a5](a7){\(b(t)\)};
			
			\diagram* {
				(a1) -- [fermion,arrow size=1.2pt] (a2) ,
				(a3)-- [fermion,arrow size=1.2pt] (a2),
				(a2) --[fermion,arrow size=1.2pt,edge label'={\(u_i(d_i)\)}](a5) --[fermion, arrow size = 1.2 pt](a7),
                (a5) --[boson,edge label'={\(W\)}](a4),
				(a2) --[fermion,arrow size=1.2pt,edge label={\(t(b)\)}](a4) --[fermion, arrow size = 1.2 pt](a6),
				
			};
		\end{feynman}
	\end{tikzpicture}}
    \subfloat[]{\begin{tikzpicture}
		\begin{feynman}
			\vertex (a1){\(\chi\)};
			\vertex [square dot, blue, above right=2cm of a1](a2){};
			\vertex [above left=1.8cm of a2](a3){\(\chi\) };
			\vertex [above right=1.8cm of a2](a4){\(b\)};
            \vertex [below right=0.6cm of a2](a5);
            \vertex [below right=0.6cm of a5](a6);
            \vertex [below right=0.6cm of a6](a7){\(b\)};
			
			\diagram* {
				(a1) -- [fermion,arrow size=1pt](a2),
                (a3) --[fermion,arrow size=1pt](a2) --[fermion,arrow size=1pt](a4),
                (a7) --[fermion, arrow size=1pt](a6) --[fermion, arrow size=1pt, edge label={\(t\)}](a5) --[fermion, arrow size=1pt, edge label={\(s,d\)}](a2),
                (a6) --[boson, edge label'={\(W\)}, half right, looseness=2.54](a5),
			};
		\end{feynman}
	\end{tikzpicture}\label{fig:Feyn_indirect_1}}
    \subfloat[]{\begin{tikzpicture}
		\begin{feynman}
			\vertex (a1){\(\chi\)};
			\vertex [square dot, blue, above right=2cm of a1](a2){};
			\vertex [above left=1.8cm of a2](a3){\(\chi\) };
			\vertex [above right=1.8cm of a2](a4){\(t\)};
            \vertex [below right=0.6cm of a2](a5);
            \vertex [below right=0.6cm of a5](a6);
            \vertex [below right=0.6cm of a6](a7){\(t\)};
			
			\diagram* {
				(a1) -- [fermion,arrow size=1pt](a2),
                (a3) --[fermion,arrow size=1pt](a2) --[fermion,arrow size=1pt](a4),
                (a7) --[fermion, arrow size=1pt](a6) --[fermion, arrow size=1pt, edge label={\(b\)}](a5) --[fermion, arrow size=1pt, edge label={\(c,u\)}](a2),
                (a6) --[boson, edge label'={\(W\)}, half right, looseness=2.54](a5),
			};
		\end{feynman}
	\end{tikzpicture}\label{fig:Feyn_indirect_2}}
    \subfloat[]{\begin{tikzpicture}
		\begin{feynman}
			\vertex (a1){\(\chi\)};
			\vertex [square dot, blue, above right=2.0cm of a1](a2){};
			\vertex [above left=2.0cm of a2](a3){\(\chi\) };
			\vertex [above right=2.0cm of a2](a4){ \( \mu \)};
			\vertex [below right=2.0cm of a2](a5){\(\mu\)};
			
			\diagram* {
				(a1) -- [fermion,arrow size=1.2pt] (a2) ,
				(a3)-- [fermion,arrow size=1.2pt] (a2),
				(a2) --[fermion,arrow size=1.2pt](a5),
				(a2) --[fermion,arrow size=1.2pt](a4),
				
			};
		\end{feynman}
	\end{tikzpicture}\label{fig:Feyn_indirect_3}}~~
	\caption{Feynman diagrams for DM indirect detection. From left to right, the diagrams depict DM annihilation into $b \bar{b}$, $t \bar{t}$, and $\mu^{+} \mu^{-}$, respectively.}
    \label{fig:Feyn_indirect}
\end{figure}

In our scenario, the DM interacts with the $b$, $d$, and $s$ quarks and their doublet components. It also interacts with muons. The bounds on the annihilation cross sections come from the channels $\chi \bar{\chi} \to (b \bar{b},\, t \bar{t},\, \mu^{+} \mu^{-})$. The corresponding Feynman diagrams are shown in Fig.~\ref{fig:Feyn_indirect}. The same couplings contribute to both $\chi \bar{\chi} \to b \bar{b}$ and $\chi \bar{\chi} \to t \bar{t}$. The $t \bar{t}$ channel contributes only when kinematically allowed (additional contributions to $b \bar b$ comes from the $\mathcal{C}_{d\chi, i3}$ operators). The annihilation into quarks occurs at the one-loop level. It includes contributions from wave-function renormalization. The annihilation into muons occurs at the tree level. Therefore, the quark channels are suppressed compared to the muon channel. The muon channel can dominate the constraints. Hence, the lepton couplings can be further constrained from indirect detection.   

\section{Combined Constraints from Dark Matter and Flavor Physics}\label{sec:DM_flavor_combined}
As discussed in the previous sections, the dSMEFT operators listed in Table~\ref{tab:operator_lists} contribute to a wide range of low-energy flavor observables while simultaneously impacting DM phenomenology through the presence of the $\mathbb{Z}_2$-odd fermion $\chi$. This dual sensitivity makes it possible to constrain the same underlying operator coefficients using both flavor and DM observables, an avenue that has remained largely unexplored in the literature. In this work, we therefore perform a combined analysis of low-energy flavor data and DM constraints to derive complementary bounds on the dSMEFT parameter space and assess the strengths and limitations of the different operator scenarios. 

The relevant flavor processes have been discussed in the previous section, where bounds on combinations of quark and lepton WCs were obtained from a global analysis of semileptonic FCNC observables. The DM sector provides additional and independent constraints through the relic abundance, as well as direct- and indirect-detection observables discussed in Sec.~\ref{sec:darkmatter}. Unlike flavor observables, which typically constrain products of quark and lepton couplings, DM observables probe the WCs individually. Furthermore, neutral meson mixing imposes complementary restrictions on the quark-sector operators, both individually and through their combinations. By combining these distinct probes, we obtain a more comprehensive determination of the allowed parameter space than is possible from flavor or DM observables alone. For convenience, we present our results in terms of the rescaled combinations $\mathcal{C}_i/\Lambda^2$, which have mass dimension $[\mathrm{GeV}^{-2}]$, rather than displaying the cutoff scale explicitly. In our entire analysis, the DM mass is varied in the range $M_\chi \in (0.1 - 1)$ TeV.

It is important to note that neutral meson mixing probes only the quark-DM operators, whereas the $b \to s\ell\ell$ observables depend on both quark-DM and lepton-DM operators. As a result, the allowed parameter space and the correlations among the WCs depend sensitively on the operator content of the theory. A comprehensive study therefore requires identifying the regions of parameter space that simultaneously satisfy flavor and DM constraints. To this end, we consider a sequence of increasingly general operator scenarios, ranging from purely hadronic interactions to scenarios involving both quark and lepton operators: 
\begin{itemize} 
\item \textbf{Single quark-DM operator scenario:} 
Only a single quark-DM operator is present at a time. The dominant constraints arise from neutral meson mixing and the DM relic abundance. Direct- and indirect-detection limits are comparatively weak since they are generated only at the loop level.

 \item \textbf{Two quark-DM operators scenario:} 
 Two quark-DM operators are simultaneously switched on. As in the previous case, neutral meson mixing and relic-density requirements provide the most stringent bounds. However, certain operator combinations can induce cancellations in the meson-mixing amplitudes, substantially enlarging the allowed parameter space. 
 
 \item \textbf{Mixed quark-DM and lepton-DM operators scenario:}
  One quark-DM operator and one lepton-DM operator are simultaneously present. In addition to the DM constraints, rare semileptonic flavor observables provide powerful probes of the parameter space. The lepton operator also opens tree-level DM annihilation into charged leptons, thereby enhancing the sensitivity of indirect-detection searches. At the same time, the meson-mixing constraints on the quark operator remain active, leading to a non-trivial interplay between flavor and DM observables.
  
   \item \textbf{Multiple flavor-DM operators scenario:} 
   Two quark-DM operators and one lepton-DM operator are simultaneously present. This is the most general setup considered in our analysis and incorporates all relevant flavor and DM constraints. The presence of two quark operators allows for possible cancellations in neutral meson mixing, while the lepton operator introduces additional constraints from rare semileptonic decays and tree-level DM annihilation. Consequently, this scenario exhibits the richest interplay between flavor and DM phenomenology and provides the most comprehensive assessment of the viable parameter space.

 \end{itemize}
\subsection{Single quark-DM operator scenario}
\begin{figure}[t]
    \centering
    \subfloat[]{\includegraphics[scale=0.22]{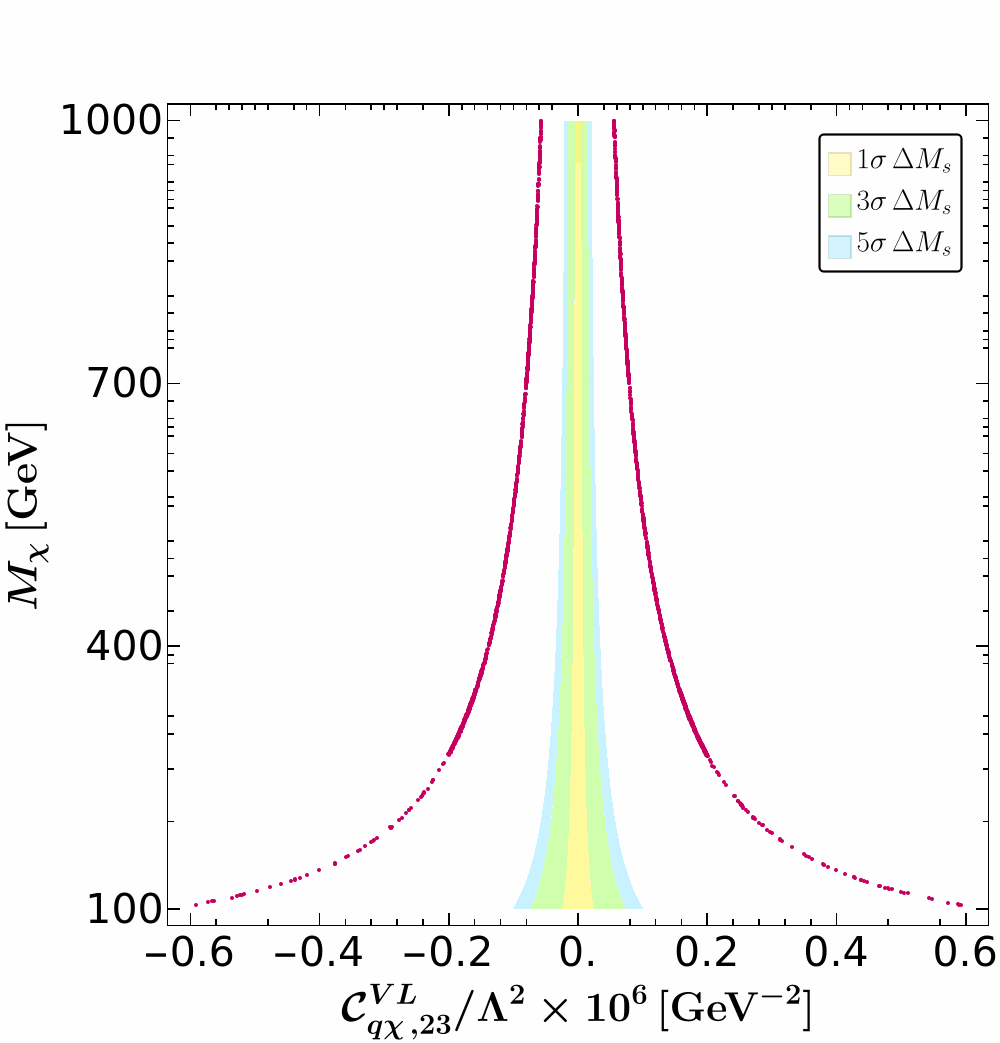}
    \label{fig:DM_flavor_oneOp1}}~~
    \subfloat[]{\includegraphics[scale=0.22]{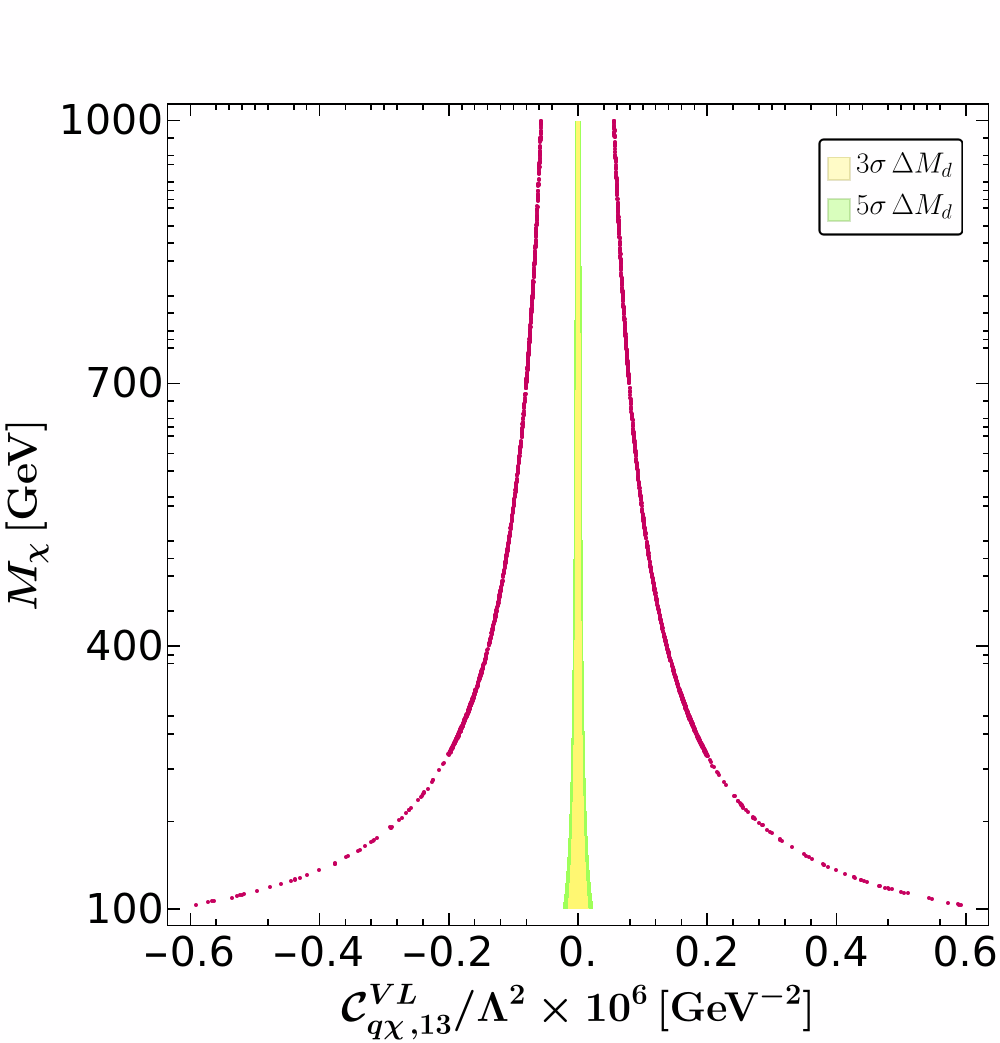}\label{fig:DM_flavor_oneOp2}}
    \caption{Allowed solution from the dark sector constraints (relic), shown in red, and flavor observables (meson mixing), in the mass-coupling plane of the DM. No region in the parameter space simultaneously satisfies the flavor and the dark sector constraints.}
    \label{fig:DM_flavor_oneOp}
\end{figure}

The simplest scenario in which both low-energy flavor observables and DM constraints contribute is the single quark-DM operator scenario. In this case, the operator is constrained by neutral meson mixing and the DM relic density. The parameter space allowed by these constraints is shown in Fig.~\ref{fig:DM_flavor_oneOp}, where only one quark-DM operator is switched on at a time. The model is described by only two free parameters: the corresponding WC and the DM mass, $M_{\chi}$. Consequently, the relic density requirement determines a one-dimensional solution in the mass-coupling plane, appearing as a line rather than an extended region. Since the annihilation cross section scales as $\sigma \propto \mathcal{C}_{i}^{2}$, the relic density solution is symmetric under $\mathcal{C}_{i}\rightarrow-\mathcal{C}_{i}$. In this scenario, the relic density provides the only relevant DM constraint. Both direct and indirect detection bounds are relatively weak because they arise only at the one-loop level and therefore impose no additional restrictions on the parameter space.

Figure~\ref{fig:DM_flavor_oneOp1} shows the results for the operator $\mathcal{O}_{q\chi,23}^{VL}$, whose flavor constraint arises from $B_s^0-\bar{B}_s^0$ mixing. The red line denotes the relic density solution, while the yellow, green, and blue regions correspond to the parameter space allowed by the neutral meson mixing observable at the $1\sigma$, $3\sigma$, and $5\sigma$ confidence levels, respectively. No overlap is observed between the relic density solution and the regions allowed by meson mixing, indicating that no coupling value can simultaneously satisfy both flavor and DM constraints. A similar behavior is found for the operator $\mathcal{O}_{q\chi,13}^{VL}$. As discussed previously, no viable solution exists when the meson mixing observables are restricted to their $1\sigma$ ranges. Therefore, only the $3\sigma$ and $5\sigma$ allowed regions are shown. Nevertheless, even after relaxing the flavor constraints, no overlap with the relic-density solution is observed. The same qualitative conclusions hold for the corresponding right-handed operators and the down-type quark operators. To avoid unnecessary repetition, these results are not shown explicitly.

We therefore conclude that the single-quark-operator scenario is excluded once dark sector and flavor constraints are imposed simultaneously. Since no viable parameter space survives in this framework, we now consider scenarios involving two quark operators simultaneously, in which additional degrees of freedom may allow compatible solutions.
Since in our case the DM-nucleon cross section lies below the neutrino floor, it cannot be probed in current direct detection experiments. Therefore, we need to consider alternative ways to probe the existence of DM and its interactions with visible particles, such as indirect detection, where one studies the annihilation of DM into SM particles using gamma-ray observations. In our scenario, DM can annihilate into $b \bar{b}$, $t \bar{t}$ final states. 
\subsection{Two quark-DM operators scenario} \label{sec:2-quark}
Since the single quark operator scenario cannot simultaneously satisfy the relic density and neutral meson mixing constraints, we now turn to the scenario in which two quark-DM operators are switched on simultaneously. As discussed in Sec.~\ref{sec:mixing}, the simultaneous presence of two quark operators can relax the neutral meson mixing constraints through cancellations among their contributions. For a given quark transition, there are four independent quark operators, leading to a total of six possible operator combinations. In the following, we present the results for three representative combinations and briefly discuss the remaining possibilities.
\begin{figure}[t]
    \centering
    \subfloat[]{\includegraphics[width=0.5\linewidth]{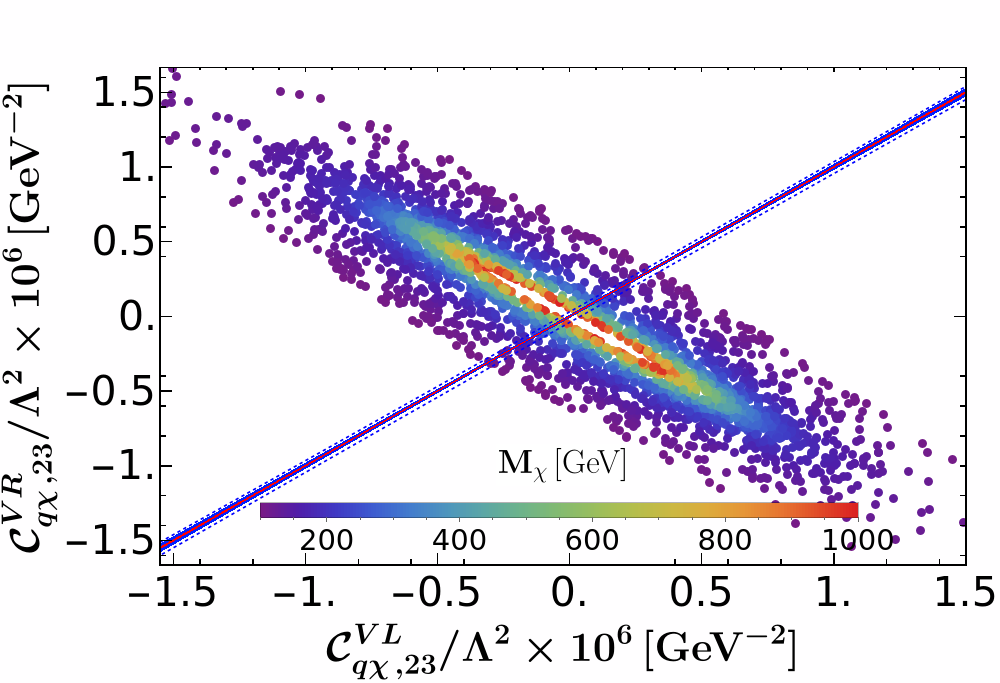}
    \label{fig:plot_DM_flavor_2qrk_1}}~~
    \subfloat[]{\includegraphics[width=0.5\linewidth]{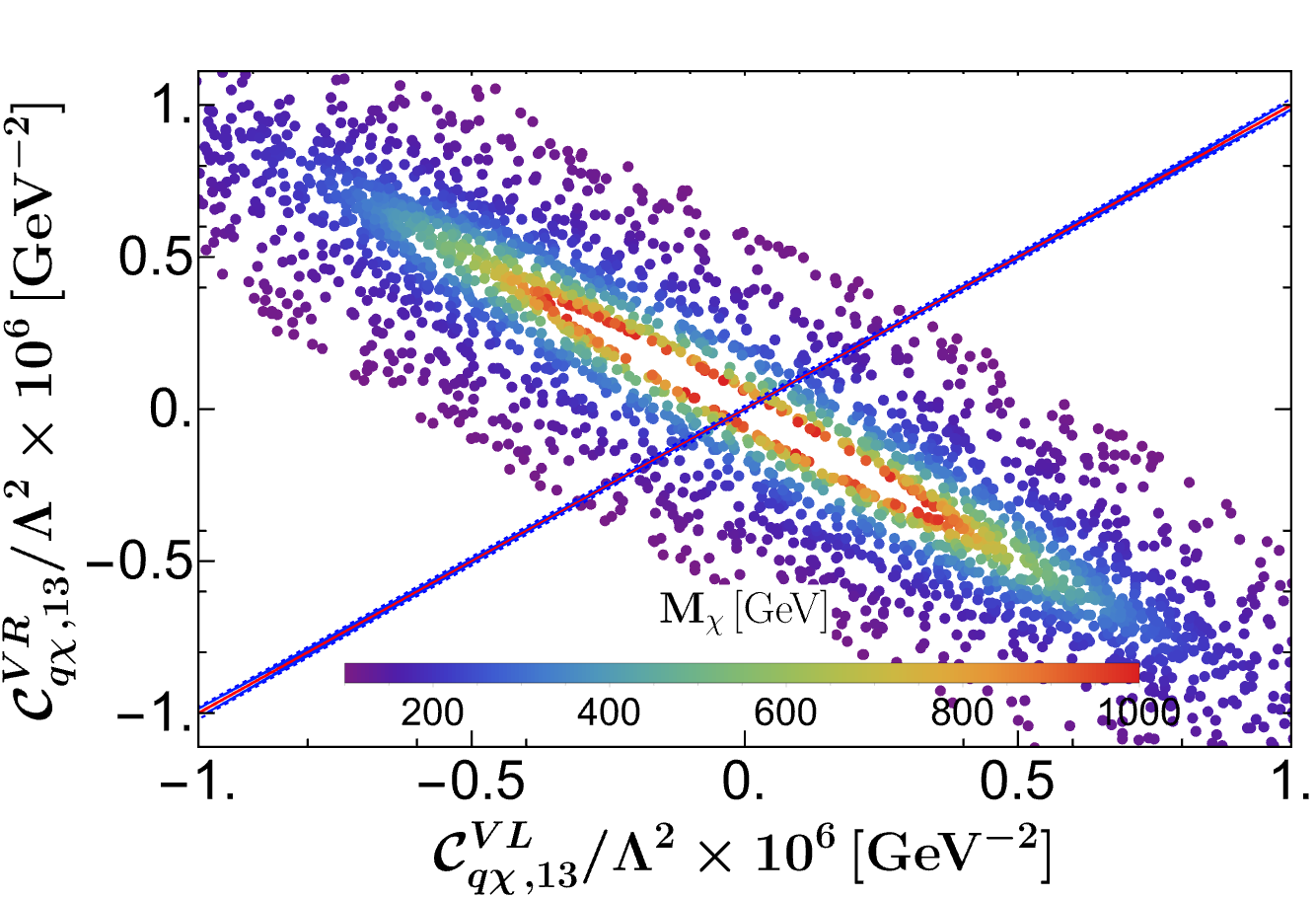}\label{fig:plot_DM_flavor_2qrk_2}}\\
    
    \subfloat[]{\includegraphics[width=0.5\linewidth]{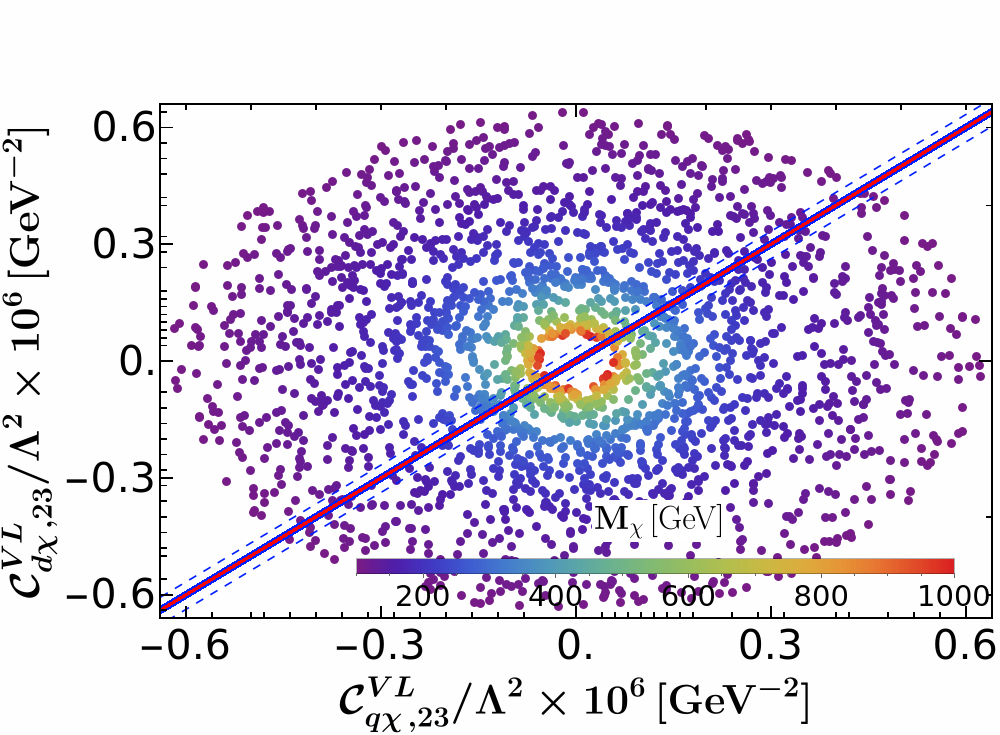}
    \label{fig:plot_DM_flavor_2qrk_3}}~~
    \subfloat[]{\includegraphics[width=0.5\linewidth]{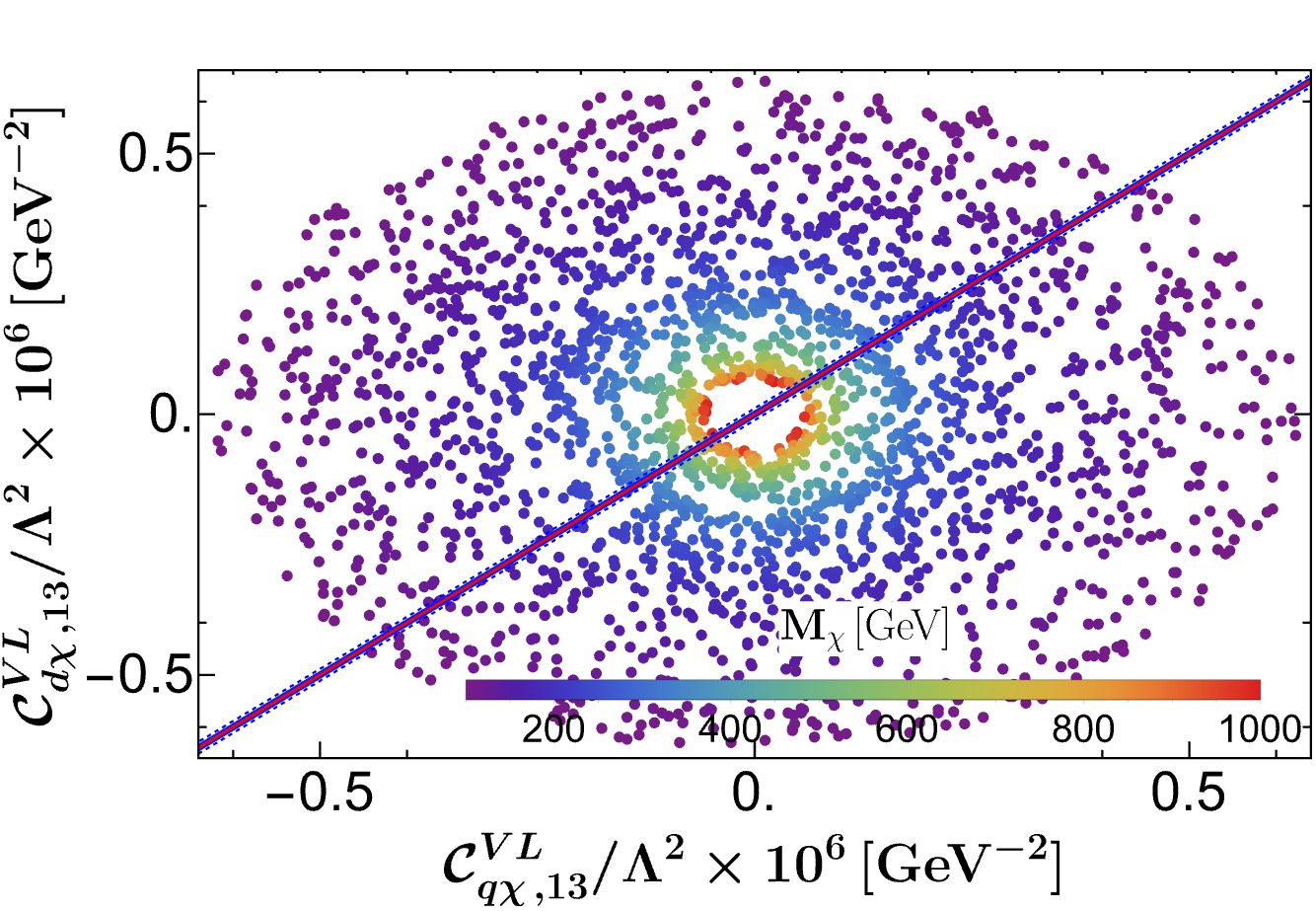}\label{fig:plot_DM_flavor_2qrk_4}}\\

    \subfloat[]{\includegraphics[width=0.5\linewidth]{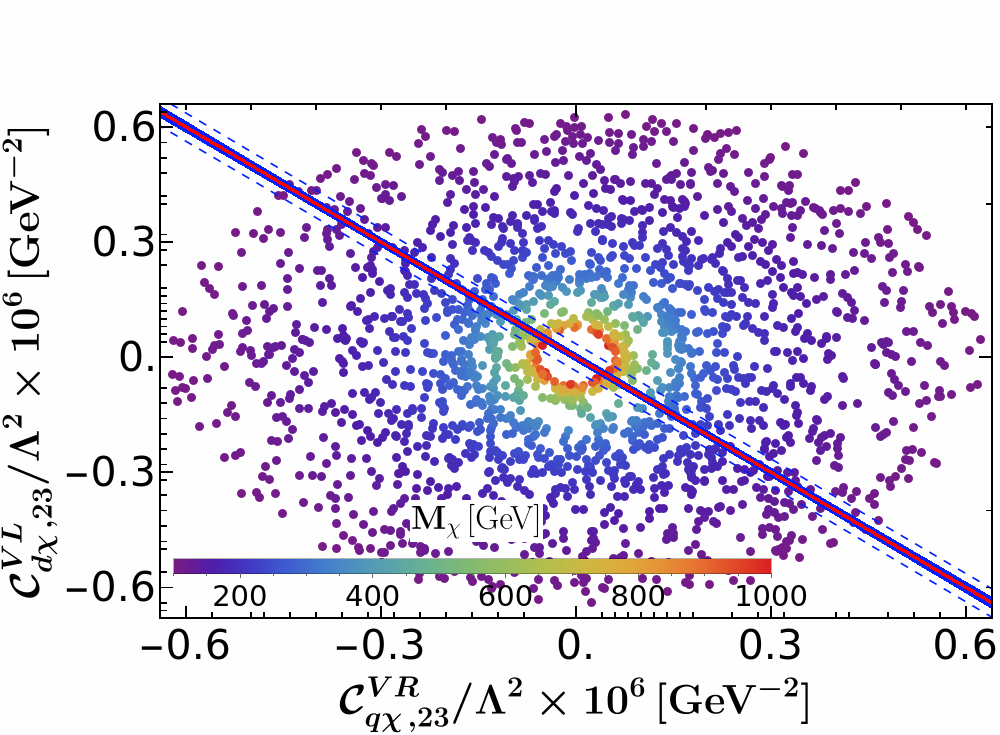}
    \label{fig:plot_DM_flavor_2qrk_5}}~~
    \subfloat[]{\includegraphics[width=0.5\linewidth]{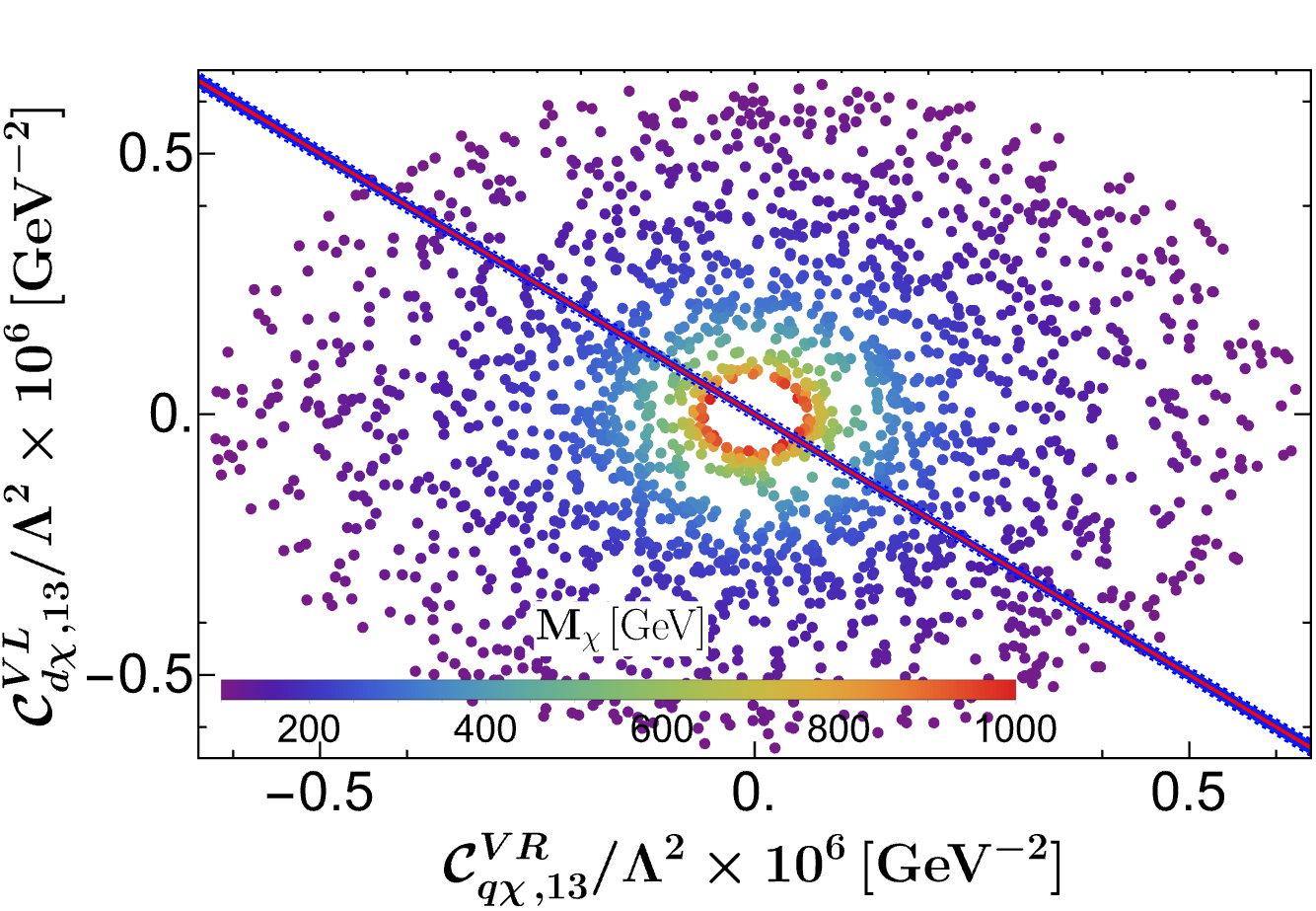}\label{fig:plot_DM_flavor_2qrk_6}}\\
    \caption{Allowed parameter space from the DM relic density, direct detection, indirect detection, and neutral meson mixing processes in the two-quark operator scenarios. The solid (dashed) lines are the allowed solutions from $B_s^0-\bar{B}_s^0$ mixing process at 1$\sigma$ (3$\sigma$) levels. For $B^0 - \bar{B}^0$ mixing we have taken $3 \sigma \, (5 \sigma)$ uncertainties of the observable. }
    \label{fig:plot_DM_flavor_2qrk}
\end{figure}

Figures~\ref{fig:plot_DM_flavor_2qrk_1} and \ref{fig:plot_DM_flavor_2qrk_2} display the allowed parameter space in the $\mathcal{C}_{q\chi,i3}^{VL}-\mathcal{C}_{q\chi,i3}^{VR}$ plane, with $i=2,1$, respectively. The relic-density constraint allows a relatively broad region of parameter space and exhibits a clear correlation between the two WCs. In particular, the DM constraints favor regions in which positive values of one coefficient are accompanied by negative values of the other. This behavior originates from the requirement of partial cancellations among the different contributions to the DM annihilation and direct-detection amplitudes, which are necessary to obtain the observed relic abundance while satisfying current experimental bounds.

The meson-mixing constraints are shown for $m_\chi=(250,1000)$ GeV. For the $B_s^0-\bar{B}_s^0$ system, the solid and dashed contours correspond to the $1\sigma$ and $3\sigma$ allowed regions, respectively, while for the $B^0-\bar{B}^0$ system only the $3\sigma$ (solid) and $5\sigma$ (dashed) regions are displayed. As noted earlier, no viable parameter space exists within the $1\sigma$ interval for the latter case. The meson-mixing bounds permit only a narrow band in the parameter space, reflecting the strong correlation required between the two WCs in order to suppress the new-physics contributions to the mixing amplitudes. Similar behavior is observed in the $\mathcal{C}_{d\chi,13}^{VL}-\mathcal{C}_{d\chi,13}^{VR}$ scenario. Although the relic-density constraint is compatible with comparatively large values of the WCs, indirect-detection constraints do not provide any significant additional restriction since the relevant annihilation channels are generated only at the one-loop level.

Figures~\ref{fig:plot_DM_flavor_2qrk_3} and \ref{fig:plot_DM_flavor_2qrk_4} show the allowed regions in the $\mathcal{C}_{q\chi,i3}^{VL}-\mathcal{C}_{d\chi,i3}^{VL}$ plane, with
$i = 2, 1$, respectively. In contrast to the previous scenario, the relic-density requirement does not generate a pronounced correlation between the two WCs. Instead, it mainly imposes an upper limit of approximately $|\mathcal{C}_{q(d)\chi,i3}^{VL}| \lesssim 0.6\times10^{-6}\,\mathrm{GeV}^{-2}$. Consequently, smaller WCs fail to produce a sufficiently large annihilation cross section and are excluded by the relic-density upper bound, whereas larger values are generally favored, particularly for lighter DM masses. The solid and dashed contours denote the neutral-meson-mixing constraints following the convention described above. Unlike the $\mathcal{C}_{q\chi,i3}^{VL}-\mathcal{C}_{q\chi,i3}^{VR}$ case, a substantial overlap between the flavor- and DM-allowed regions is observed, demonstrating that both classes of constraints can be simultaneously satisfied. The corresponding $\mathcal{C}_{q\chi,i3}^{VR}-\mathcal{C}_{d\chi,i3}^{VR}$ scenario exhibits qualitatively identical behavior.

Figures~\ref{fig:plot_DM_flavor_2qrk_5} and \ref{fig:plot_DM_flavor_2qrk_6} present the allowed parameter space in the mixed-chirality planes. As in the previous scenario, the relic-density constraint does not induce any significant correlation between the two WCs. The neutral-meson-mixing constraints, however, lead to a distinctly different pattern. In particular, positive values of one Wilson coefficient favor negative values of the other. This behavior originates from the term proportional to $\mathcal{C}_{d\chi,i3}^{VL}\mathcal{C}_{q\chi,i3}^{VR}$ in Eq.~\eqref{eq:DeltaM_NP}, which enters the mixing amplitude with a 
positive sign relative to the remaining contributions. Since the experimentally allowed region requires an enhancement of the new-physics contribution, the product $\mathcal{C}_{d\chi,i3}^{VL}\mathcal{C}_{q\chi,i3}^{VR}$ must be negative. 
Consequently, the meson-mixing constraints select parameter regions in which the two WCs have opposite signs. This behavior differs from that observed in Figs.~\ref{fig:plot_DM_flavor_2qrk_3} and \ref{fig:plot_DM_flavor_2qrk_4}, where all contributions enter the mixing amplitude with the same sign. In those cases, the data favor WCs with the same sign, leading to a positive correlation between the allowed couplings. Despite the different correlation pattern, sizable regions satisfying both flavor and DM constraints remain viable. Finally, replacing the left-handed DM current with a right-handed one yields qualitatively identical results, indicating that the overall conclusions are largely insensitive to the chirality of the DM current.

Overall, the two quark-DM operators scenario admits viable regions of parameter space that simultaneously satisfy the constraints from neutral meson mixing and DM phenomenology. The interplay between the two quark operators can significantly relax the meson-mixing bounds through cancellations among different new-physics contributions, thereby opening parameter regions that would otherwise be excluded in the single-operator case. Nevertheless, in the absence of a lepton-DM operator, rare semileptonic FCNC processes such as $b\to s\ell^+\ell^-$ remain unaffected, and the associated flavor anomalies or constraints cannot be addressed within this framework. This motivates the consideration of the next scenario involving both quark-DM and lepton-DM operators, where low-energy semileptonic flavor observables and DM observables can be studied simultaneously. 

From the correlation plots shown in Fig.~\ref{fig:plot_DM_flavor_2qrk}, one can infer the typical size of the WCs consistent with both flavor and DM constraints. For relatively light DM masses, the combined analysis requires 
$$|\mathcal{C}_{q(d)\chi,i3}^{VL}|,\, |\mathcal{C}_{q(d)\chi,i3}^{VR}| \lesssim 5\times10^{-7}\ {\rm GeV}^{-2}, \qquad i=1,2. $$
As the DM mass increases, the allowed parameter space shrinks considerably. In particular, for $m_\chi \gtrsim 400~{\rm GeV}$, the flavor and DM constraints become increasingly restrictive, and the maximum allowed values of the WCs are typically reduced to the ${\cal O}(10^{-8})~{\rm GeV}^{-2}$ range. These results demonstrate the strong complementarity between flavor and DM observables and highlight the importance of a combined analysis in probing the viable dSMEFT parameter space.

\subsection{Mixed quark-DM and lepton-DM operators scenario}
To accommodate the constraints from rare semileptonic decays, we now investigate the phenomenologically relevant scenario in which the DM interacts through one quark operator and one lepton operator. In this framework, the free parameters are the dSMEFT WCs $(\mathcal{C}_i)$, the effective cutoff scale $\Lambda$, and the DM mass $M_\chi$. Since the semileptonic observables depend only on the product of the quark and lepton couplings, our results are presented in terms of $\mathcal{C}_{q(d)\chi}\mathcal{C}_{\ell(e)\chi}$. However, the DM observables constrain the individual WCs, thereby allowing us to infer correlations between the quark and lepton couplings. In contrast, meson mixing receives contributions solely from the quark operator, providing complementary constraints on the parameter space. The complementarity of these observables enables a correlated determination of the individual WCs and identifies the parameter space simultaneously consistent with flavor and DM constraints. In the following, we discuss the interplay of relic density, direct detection, indirect detection, rare semileptonic decays, and meson mixing constraints.

\begin{figure}[t]
	\centering
	\subfloat[]{\includegraphics[width=0.5\linewidth]{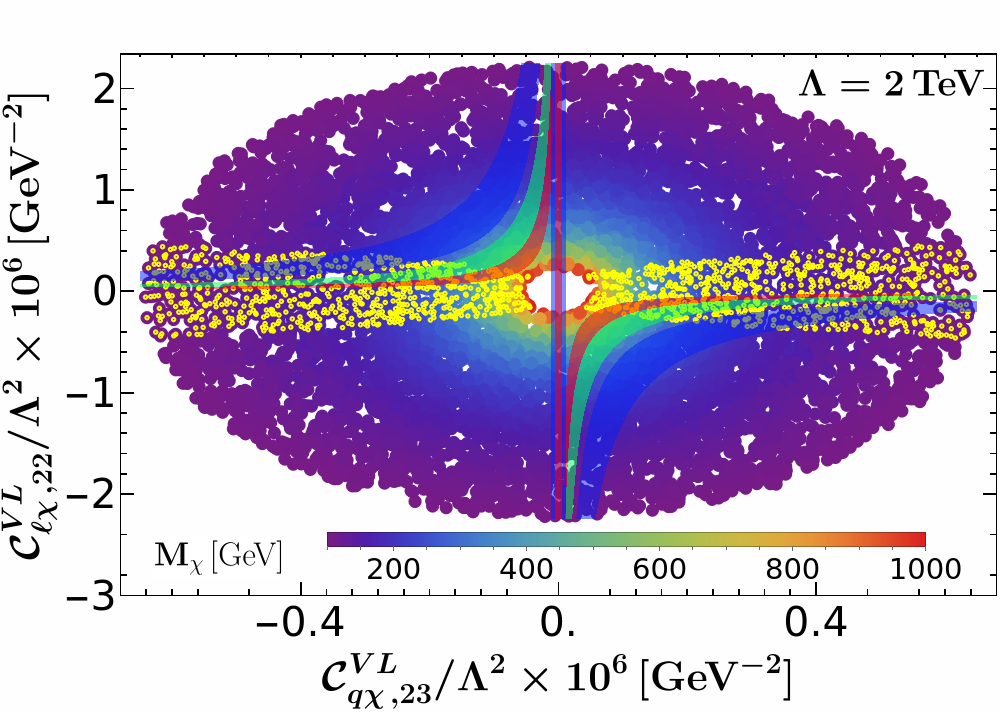}
		\label{fig:DM_flavor_1}}~~
	\subfloat[]{\includegraphics[width=0.5\linewidth]{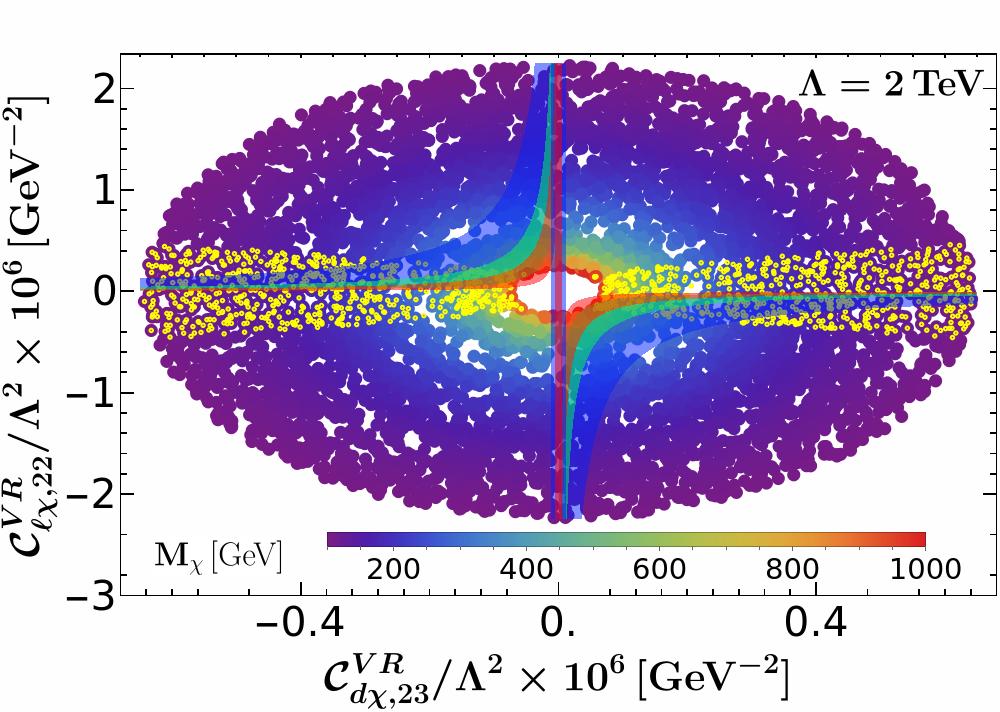}\label{fig:DM_flavor_2}}\\
	\subfloat[]{\includegraphics[width=0.5\linewidth]{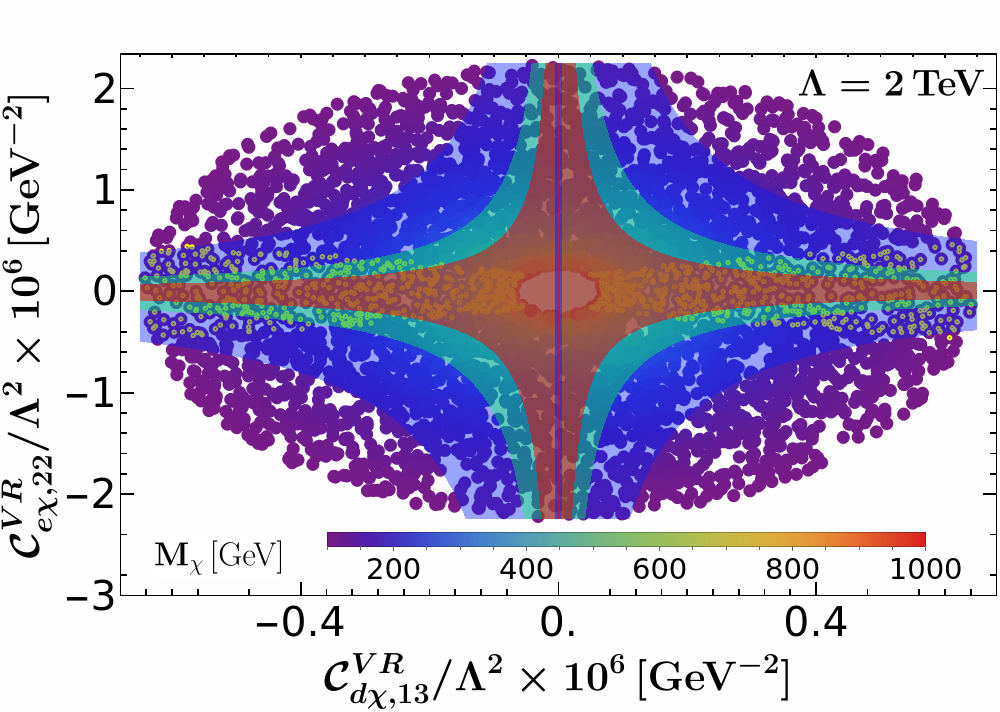}\label{fig:DM_flavor_3}}
	\caption{Allowed parameter space from the combined analysis of DM and flavor observables in the one-quark one-lepton operator scenario. The scatter points in the plots show allowed solutions from relic density. The yellow region is allowed by indirect detection bounds, the hyperbolic regions are allowed by semileptonic rare decays, and the vertical bands are the allowed solutions from neutral meson mixing processes. }
	\label{fig:plot_DM_flavor}
\end{figure}

\paragraph{\underline{DM Relic Density} :}
In Fig.~\ref{fig:plot_DM_flavor}, we present the parameter space consistent with the observed DM relic abundance in the one-quark and one-lepton operator scenario. Figure~\ref{fig:DM_flavor_1} shows the correlation between the WCs $\mathcal{C}^{VL}_{q\chi,23}$ and $\mathcal{C}^{VL}_{\ell\chi,22}$, corresponding to interactions between left-handed quark and lepton currents, respectively. Since the quark and lepton couplings are left-handed, DM can annihilate into $b\bar{s}$, $t\bar{c}$, and their Hermitian conjugates, as well as into a $\mu^+\mu^-$ pair.

Only weak correlations are observed between the quark and lepton WCs, indicating that the relic density does not strongly prefer any particular combination of these couplings. The quark Wilson coefficient is constrained slightly more strongly than the corresponding lepton coefficient because of the color enhancement associated with quark final states. In contrast, a strong correlation is observed between the WCs and the DM mass, with the allowed parameter space becoming increasingly restricted as the DM mass increases. Varying the DM mass over the range $(0.1-1)$~TeV, we obtain:
\begin{align}\label{eq:bounds_DM_cvLqX23_cvLlX22}
    1.43 \times 10^{-9} \leq \,  &|\mathcal{C}_{q\chi,23}^{VL}/\Lambda^2| \, \leq 6.41 \times 10^{-7} \rm \, GeV^{-2}\,, \nonumber  \\
    1.03 \times 10^{-8} \leq  \, & |\mathcal{C}_{\ell\chi,22}^{VL}/\Lambda^2| \, \leq  2.22 \times 10^{-6} \rm \, GeV^{-2}\,.
\end{align}
As expected, for a fixed DM mass, the allowed ranges of the WCs are more restrictive than the combined bounds quoted above. Since the relic abundance is inversely proportional to the thermally averaged annihilation cross section, WCs smaller than the lower limits in Eq.~\eqref{eq:bounds_DM_cvLqX23_cvLlX22} result in an overabundance of DM, whereas larger values lead to an underabundance.

In the present framework, DM annihilation proceeds at tree level, and the relic density depends only on the DM mass and the dSMEFT WCs. Consequently, similar bounds are obtained for the coupling plane $\mathcal{C}^{VL}_{q\chi,13} - \mathcal{C}^{VL}_{\ell\chi,22}$, and we do not present those results separately. Likewise, changing the chirality of the DM current does not produce any noticeable change in the allowed parameter space.

The remaining panels of Fig.~\ref{fig:plot_DM_flavor} exhibit similar features. Since only weak correlations are observed between the quark and lepton operators in all cases, the resulting bounds on the WCs are nearly identical for the different operator combinations. Therefore, we do not discuss each scenario individually. Furthermore, operators involving quark doublets and singlets lead to almost identical parameter spaces, with only a slight difference in the allowed bounds. This difference originates from the larger annihilation contribution into down-type quarks for the doublet operators compared to the channels involving the top quark.

For all the scenarios shown in Fig.~\ref{fig:plot_DM_flavor}, the spin-independent direct detection (SIDD) cross section is generated only at the one-loop level, whereas the relic density receives contributions at the tree level. Consequently, the direct detection cross section remains highly suppressed, with values of the order of $(10^{-51}-10^{-52})\,\mathrm{cm}^2$ throughout the relic-allowed parameter space. These predictions lie well below both the current experimental limits and the neutrino floor. Therefore, the entire parameter space consistent with the observed relic density automatically satisfies the present direct detection constraints. Even for WCs of order unity, $\mathcal{C}_{q(d)\chi}\sim\mathcal{O}(1)$, the predicted direct detection cross section remains far below the current exclusion limits.

\paragraph{\underline{DM Indirect Detection Bound} :}
The contributions of the operators under consideration to the indirect detection processes are illustrated by the Feynman diagrams in Fig.~\ref{fig:Feyn_indirect}. As shown, the quark operators contribute to DM annihilation only at the one-loop level, whereas the lepton (muon) operator contributes already at tree level, the same order at which it contributes to the relic density. Consequently, we expect the indirect detection limits to impose significantly stronger constraints on the lepton operator than on the quark operator. In particular, DM annihilation into quark pairs is loop suppressed, while annihilation into $\mu^+\mu^-$ proceeds at tree level and therefore dominates the indirect detection signal.

Figure~\ref{fig:plot_DM_flavor} shows the parameter space allowed after imposing the combined constraints from the DM relic density, SIDD and indirect detection. The yellow points in each panel denote the region consistent with the indirect detection limits. The strongest constraints arise from the annihilation channel into muon pairs. Consequently, the indirect detection limits further reduce the allowed ranges of the lepton WCs. Hence, the bounds on the quark operators will remain as that of the relic-allowed and we obtain the following upper bound on the lepton couplings as:
\begin{align}\label{eq:indirect_couplings}
    \left|\mathcal{C}_{\ell(e)\chi,22}^{VL(R)} \big/ \Lambda^2\right|
    &\leq 0.6 \times 10^{-6}~\mathrm{GeV}^{-2}\,.
\end{align}

These limits correspond to the maximum allowed values of the WCs. As observed in the relic density analysis, the allowed couplings exhibit a strong correlation with the DM mass. Consequently, the indirect detection bounds become increasingly stringent for larger DM masses, while they relax towards lower DM masses.

\paragraph{\underline{All Constraints Combined} :}
As discussed in the previous section, the relic density determines both upper and lower limits on the individual WCs, while the indirect detection bounds further restrict the allowed values of the lepton couplings. We now investigate the impact of incorporating the constraints from rare semileptonic decays and neutral meson mixing on the parameter space obtained from combined DM constraints. The corresponding bounds on the Wilson coefficient combinations from semileptonic decays are summarized in Tables~\ref{tab:fitresults}, \ref{tab:fitresults2} and \ref{tab:fitresults1}. In Fig.~\ref{fig:plot_DM_flavor}, the hyperbolic regions denote the parameter space allowed by the semileptonic decays for a hard cutoff scale of $\Lambda=2$~TeV, whereas the vertical bands correspond to the regions allowed by the neutral meson mixing observables.

Figure~\ref{fig:DM_flavor_1} presents the allowed parameter space in the coupling plane $\mathcal{C}_{q\chi,23}^{VL}-\mathcal{C}_{\ell\chi,22}^{VL}$, where the flavor constraints are considered at the $1\sigma$ level. The overlap between the hyperbolic region and the yellow points corresponds to the parameter space simultaneously satisfying the semileptonic and indirect detection constraints. The stringent constraint from $B_s^0-\bar{B}_s^0$ mixing severely restricts the allowed values of $\mathcal{C}_{q\chi,23}^{VL}$, leaving only a very narrow range of quark couplings. Consequently, no parameter space remains that can simultaneously satisfy the relic density, indirect detection, semileptonic decay, and meson mixing constraints at the $1\sigma$ level.

Figure~\ref{fig:DM_flavor_2} shows the corresponding analysis in the coupling plane $\mathcal{C}_{d\chi,23}^{VR}-\mathcal{C}_{\ell\chi,22}^{VR}$. The semileptonic constraints differ slightly from those in the previous scenario, while the remaining constraints remain essentially unchanged. Although a small overlap between the semileptonic and meson mixing constraints appears at the $1\sigma$ level, this region is excluded once the indirect detection bounds are imposed. Therefore, this operator combination also fails to accommodate all the flavor and cosmological constraints simultaneously.

In the previous two cases, we discussed the quark-lepton WCs corresponding to the $b \to s \ell \ell$ transitions. We now consider the $b \to d \ell \ell$ transition, shown in Fig.~\ref{fig:DM_flavor_3}. As discussed in Sec.~\ref{sec:mixing}, the measured value of $\Delta M_d$ differs from the SM prediction by $1.35\sigma$. Therefore, the allowed regions from both the $b \to d\,\ell\ell$ observables and $\Delta M_d$ are presented at the $2\sigma$ confidence level. Unlike the $b \to s\,\ell\ell$ transition, the Wilson coefficient combinations obtained from the $b \to d\,\ell\ell$ observables are consistent with zero at the $2\sigma$ level, implying that vanishing NP couplings are also allowed. Similarly, the $\Delta M_d$ observable permits only a narrow region around the origin. Consequently, the semileptonic and meson mixing observables can be simultaneously satisfied for very small quark WCs. However, as discussed previously, such small quark couplings lead to an overabundance of DM relic density and are therefore excluded. Hence, no parameter space simultaneously satisfies all the flavor and DM constraints in this scenario.

The above discussion corresponds to taking the flavor observables at $1\sigma$ for the $b\to s \ell \ell$ transitions and at $2\sigma$ for the $b\to d \ell \ell$ transitions. Relaxing the flavor constraints to $3\sigma$ enlarges the parameter space allowed by the semileptonic decays and meson mixing. However, the surviving solutions correspond to very small quark WCs, which lead to an overabundance of DM relic density and are therefore excluded. Consequently, even after adopting more conservative flavor uncertainties, no common parameter space simultaneously satisfies all flavor and cosmological constraints. The same conclusion is reached for a cutoff scale of $\Lambda=1$~TeV, and therefore we do not present those results separately.

The absence of a viable parameter space in the one-quark and one-lepton operator scenario motivates extending our analysis to the case of two-quark and one-lepton operators. As discussed in Sec.~\ref{sec:2-quark}, the two-quark operator scenario can satisfy the DM and meson mixing constraints but is irrelevant for the semileptonic anomalies. We therefore investigate in the next section whether the inclusion of two quark operators together with a lepton operator can simultaneously accommodate the flavor and cosmological observables.

\subsection{Multiple flavor-DM operators scenario}
As the one-quark one-lepton operator scenario fails to simultaneously satisfy all the dark-sector and flavor constraints, we now extend our analysis to the two-quark one-lepton operator scenario. The introduction of an additional quark operator provides extra degrees of freedom, which may open up regions of the parameter space compatible with all the experimental constraints. As in the previous scenario, we first discuss the constraints arising from the dark sector, followed by the impact of the flavor observables on the allowed parameter space. Before presenting the detailed analysis, it is useful to summarize the role of the different observables. The DM relic density constrains the individual as well as the combined values of the operator couplings. The indirect detection limits primarily constrain the lepton operator, while their impact on the quark operators remains comparatively weak since the corresponding contributions arise only at the one-loop level. The rare semileptonic decays constrain specific combinations of the quark and lepton operators, whereas neutral meson mixing depends solely on the quark operators, placing bounds on both their individual contributions and their combinations. Since this scenario involves three independent operators, a significantly larger set of coupling combinations becomes available, allowing us to explore whether the additional degrees of freedom can accommodate all the dark-sector and flavor constraints simultaneously. For the following figures corresponding to the three WCs, we take the hard-cutoff scale at $\Lambda = 2$ TeV.

\begin{figure}[t]
    \centering
    \subfloat[]{\includegraphics[width=0.5\linewidth]{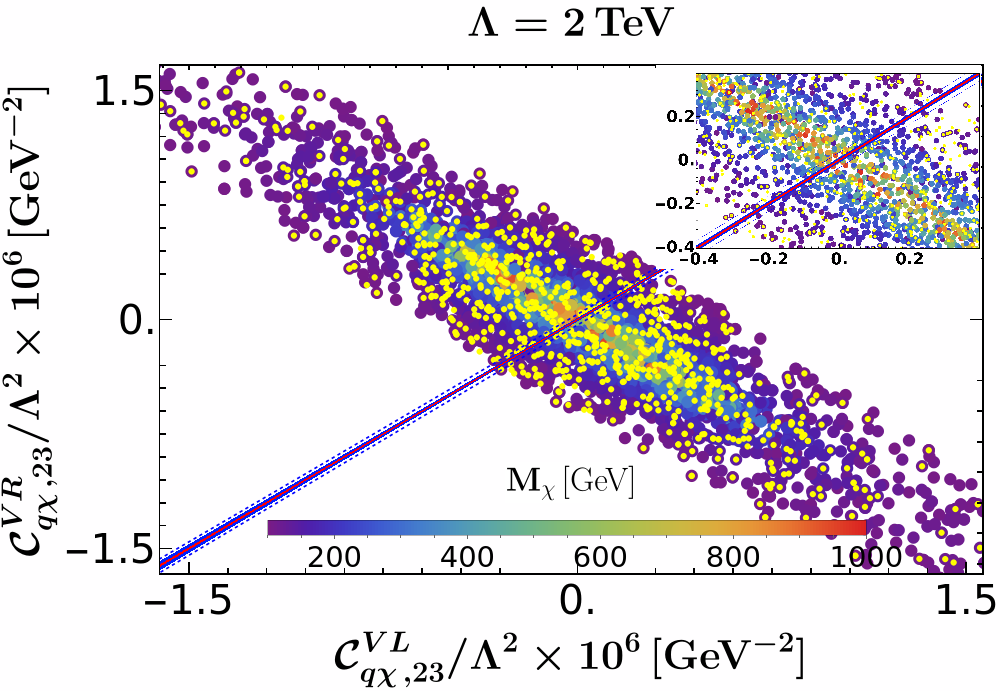}
    \label{fig:plot_cvLqX23_cvRqX23_cvLlX22_1}}~~
    \subfloat[]{\includegraphics[width=0.475\linewidth]{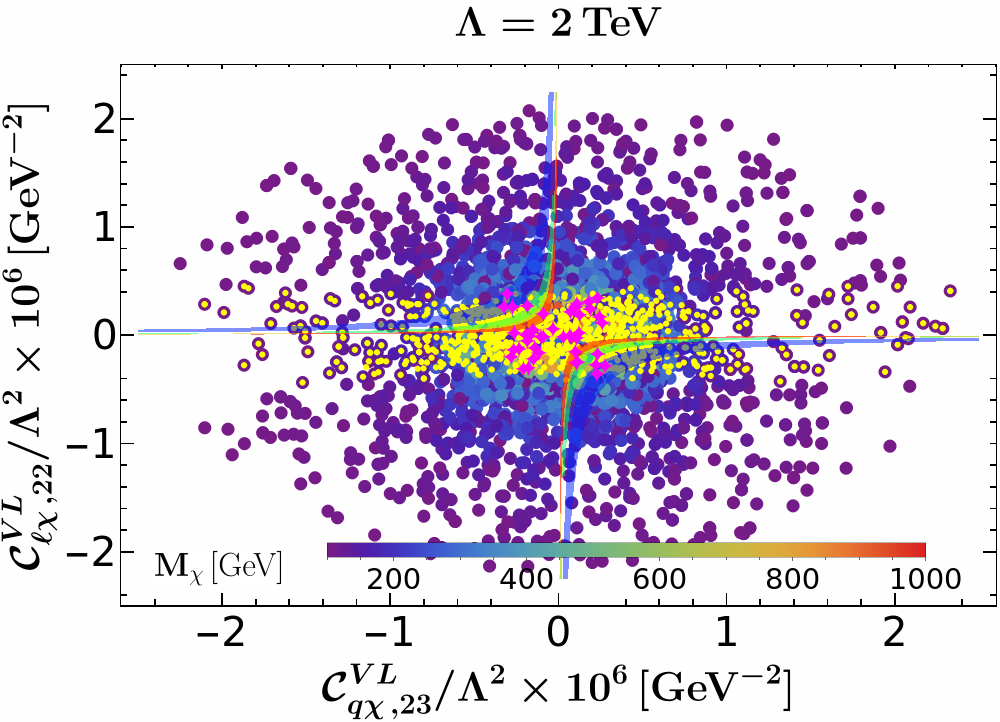}\label{fig:plot_cvLqX23_cvRqX23_cvLlX22_2}}
    \caption{Allowed parameter space from the DM relic density, direct detection, indirect detection, and low-energy flavor observables in the two-quark one-lepton operator scenario. The scatter points in the plots show allowed solutions from relic density. The yellow region is allowed by indirect detection bounds, the hyperbolic regions are allowed by semileptonic rare decays, and the solid (dashed) bands are the allowed solutions from the neutral meson mixing process at 1$\sigma$ (3$\sigma$) levels. The magenta markers on the right plot indicate the parameter points consistent with $\Delta_s$ observable within 3$\sigma$ uncertainty.}
    \label{fig:plot_cvLqX23_cvRqX23_cvLlX22}
\end{figure}
\begin{figure}[htbp!]
    \centering
    \includegraphics[width=0.5\linewidth]{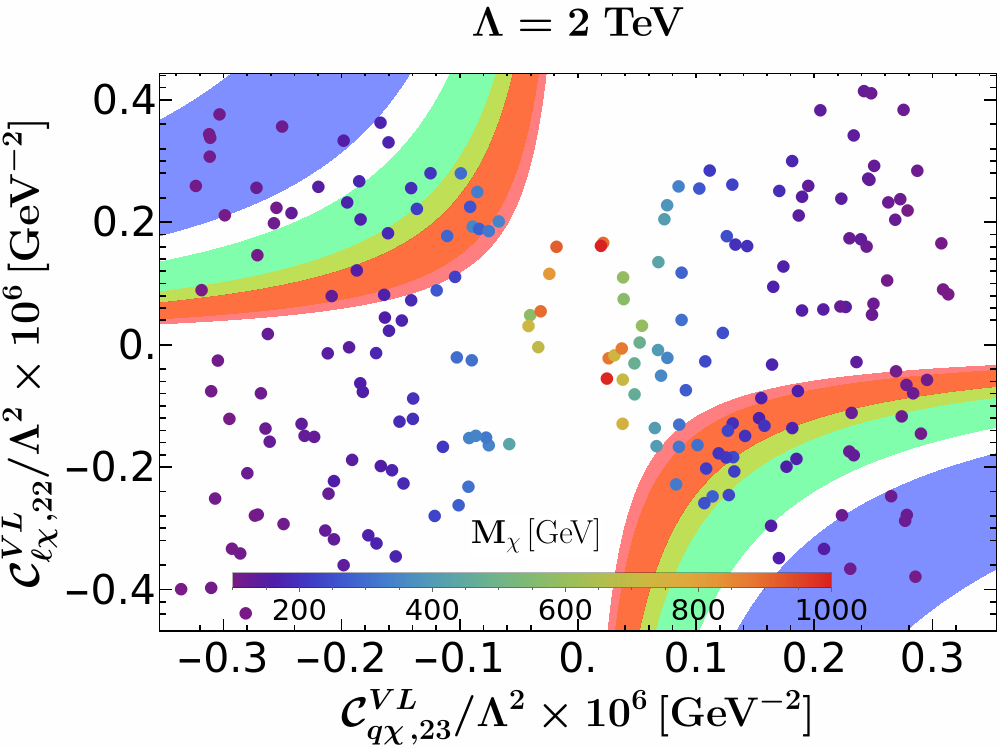}
    \caption{Final allowed parameter space in the quark–lepton coupling plane for the scenario with the WCs $\mathcal{C}_{q\chi,23}^{VL}$, $\mathcal{C}_{q\chi,23}^{VR}$, and $\mathcal{C}_{\ell\chi,22}^{VL}$ considered simultaneously. The scatter points satisfy the constraints from the DM relic density, indirect detection, and neutral meson mixing ($3\sigma$) relevant to this scenario. The color of the points indicates the DM mass. The colored bands show the corresponding semileptonic constraints, with the different colors representing different DM masses, as indicated by the color bar.}
    \label{fig:plot_cvLqX23_cvRqX23_cvLlX22_zoomed}
\end{figure}
The allowed parameter space obtained by simultaneously varying the three WCs $\mathcal{C}_{q\chi,23}^{VL}$, $\mathcal{C}_{q\chi,23}^{VR}$, and $\mathcal{C}_{\ell\chi,22}^{VL}$ is shown in Fig.~\ref{fig:plot_cvLqX23_cvRqX23_cvLlX22}. Compared to the one-quark one-lepton operator scenario, the allowed region in the $\mathcal{C}_{q\chi,23}^{VL}$--$\mathcal{C}_{q\chi,23}^{VR}$ plane, displayed in Fig.~\ref{fig:plot_cvLqX23_cvRqX23_cvLlX22_1}, is significantly enlarged. This is a consequence of the simultaneous presence of two quark-DM operators with the same quark-current chirality but opposite DM-current chiralities, which provides additional freedom in satisfying the various phenomenological constraints. 

The colored points represent parameter regions consistent with the observed relic abundance, with different colors corresponding to different values of the DM mass. The relic-density constraint exhibits a pronounced negative correlation between $\mathcal{C}_{q\chi,23}^{VL}$ and $\mathcal{C}_{q\chi,23}^{VR}$, reflecting the requirement of relative cancellations among the different contributions to the DM annihilation and scattering amplitudes. The yellow points denote the subset of solutions that additionally satisfy the indirect-detection bounds. These points are distributed throughout the entire relic-density allowed region, indicating that indirect-detection observables do not impose any significant additional restriction on the quark WCs in this scenario. 

The solid and dashed bands correspond to the regions allowed by neutral meson mixing at the $1\sigma$ and $3\sigma$ confidence levels, respectively. In contrast to the DM constraints, the meson-mixing observables favor solutions in which $\mathcal{C}_{q\chi,23}^{VL}$ and $\mathcal{C}_{q\chi,23}^{VR}$ have the same sign. This preference arises from the structure of the new-physics contributions to the mixing amplitude, for which constructive additions among the different operator contributions are required to reproduce the experimentally preferred region. Consequently, the flavor and DM constraints tend to pull the WCs in opposite directions: the relic-density constraint prefers opposite-sign solutions, whereas meson mixing prefers same-sign solutions. The simultaneous satisfaction of both requirements therefore restricts the viable parameter space to a relatively small overlap region in the $\mathcal{C}_{q\chi,23}^{VL}$--$\mathcal{C}_{q\chi,23}^{VR}$ plane. 

This figure thus provides a clear illustration of the complementarity between DM and flavor observables. While the presence of an additional quark-DM operator substantially enlarges the parameter space compared with the one-quark one-lepton scenario, the combined analysis reveals that only a limited subset of this region remains compatible with both the DM and meson-mixing constraints.

Figure~\ref{fig:plot_cvLqX23_cvRqX23_cvLlX22_2} shows the allowed parameter space in the $\mathcal{C}_{q\chi,23}^{VL}$-$\mathcal{C}_{\ell\chi,22}^{VL}$ plane. It can be seen that the indirect detection bound further restricts the allowed range of the lepton coupling compared to the relic density allowed region, while leaving the allowed range of the quark coupling essentially unchanged. This behavior is expected since the quark contribution to the indirect detection process arises only at the one-loop level. As this plane involves the lepton operator, the constraints from the rare semileptonic decays can also be imposed. The hyperbolic bands correspond to different DM masses, as indicated by the color bar, and are obtained from the fit results listed in Table~\ref{tab:fitresults}. Although the neutral meson mixing constraint cannot be displayed directly in this plane, the parameter points consistent with the $\Delta_s$ observable within the $3\sigma$ uncertainty and dark-sector constraints are explicitly indicated by the magenta dots. It can be seen that a small region of the parameter space satisfies the relic density, indirect detection, neutral meson mixing, and rare semileptonic decay constraints simultaneously. Therefore, unlike the one-quark one-lepton operator scenario, the introduction of an additional quark operator opens up a region of parameter space that is consistent with all the dark-sector and flavor constraints.

Figure~\ref{fig:plot_cvLqX23_cvRqX23_cvLlX22_zoomed} is similar to Fig.~\ref{fig:plot_cvLqX23_cvRqX23_cvLlX22_2}, but shows a zoomed-in region to better illustrate the correlations among the couplings and the resulting bounds on the WCs. To better sample the allowed solution region, we generate a larger number of points in this region and present the corresponding zoomed-in parameter space. The scatter points satisfy the DM relic density, indirect detection, and $B_s^0-\bar{B}_s^0$ mixing constraints, $\Delta_s$. Here, we consider the allowed $3\sigma$ range of the meson-mixing constraint; however, the resulting parameter space remains qualitatively similar when the more stringent $1\sigma$ range is imposed, as can be seen from Fig.~\ref{fig:plot_cvLqX23_cvRqX23_cvLlX22_1}. The colored bands correspond to the semileptonic $B$-decay constraints discussed previously, while the points lying on a colored band and having the same color as that band represent the solutions that satisfy all the constraints considered in our analysis. The dominant constraint on the quark coupling arises from neutral meson mixing, while the indirect detection bound imposes a strong constraint on the lepton coupling. The semileptonic constraints further establish a strong correlation between the quark and lepton couplings. The resulting upper bound on the Wilson coefficient is
$|\mathcal{C}_{q\chi,23}^{VL(R)}/\Lambda^2|\lesssim3.5\times10^{-7}~\mathrm{GeV}^{-2}$, and $|\mathcal{C}_{\ell \chi, 22}^{VL} / \Lambda^2| \lesssim 4.5 \times 10^{-7}\, \rm ~GeV^{-2}$. These bounds are obtained for a DM mass of $\sim 100$~GeV. For larger DM masses, the bounds on both couplings become stronger, as shown by the color variation in the plot.

\begin{figure}[t]
    \centering
    \subfloat[]{\includegraphics[width=0.5\linewidth]{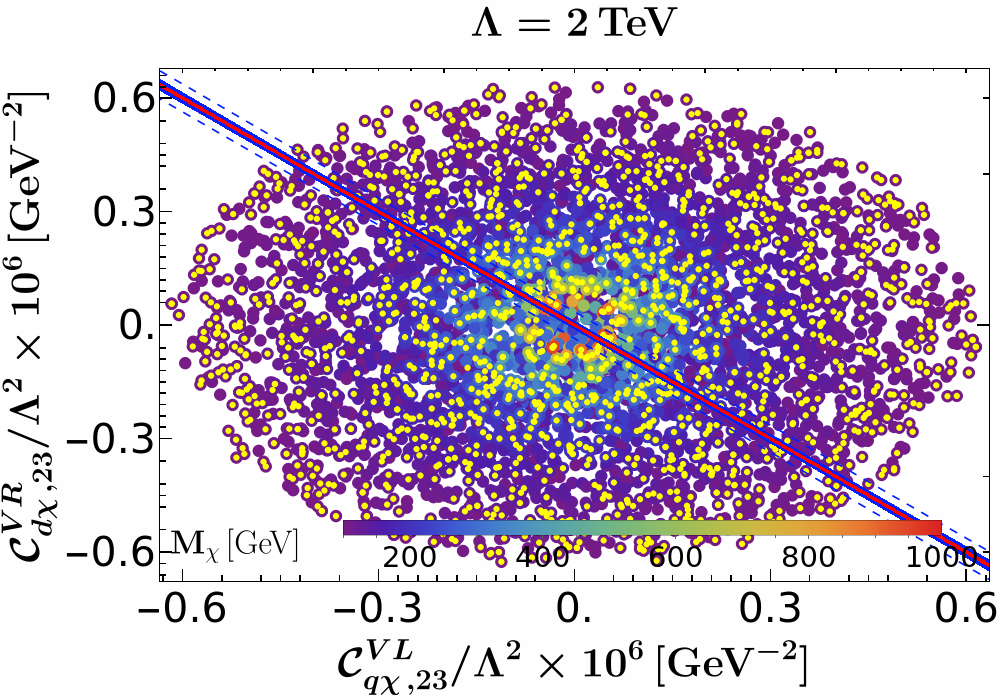}
    \label{fig:plot_cvLqX23_cvRdX23_cvRlX22_1}}~~
    \subfloat[]{\includegraphics[width=0.48\linewidth]{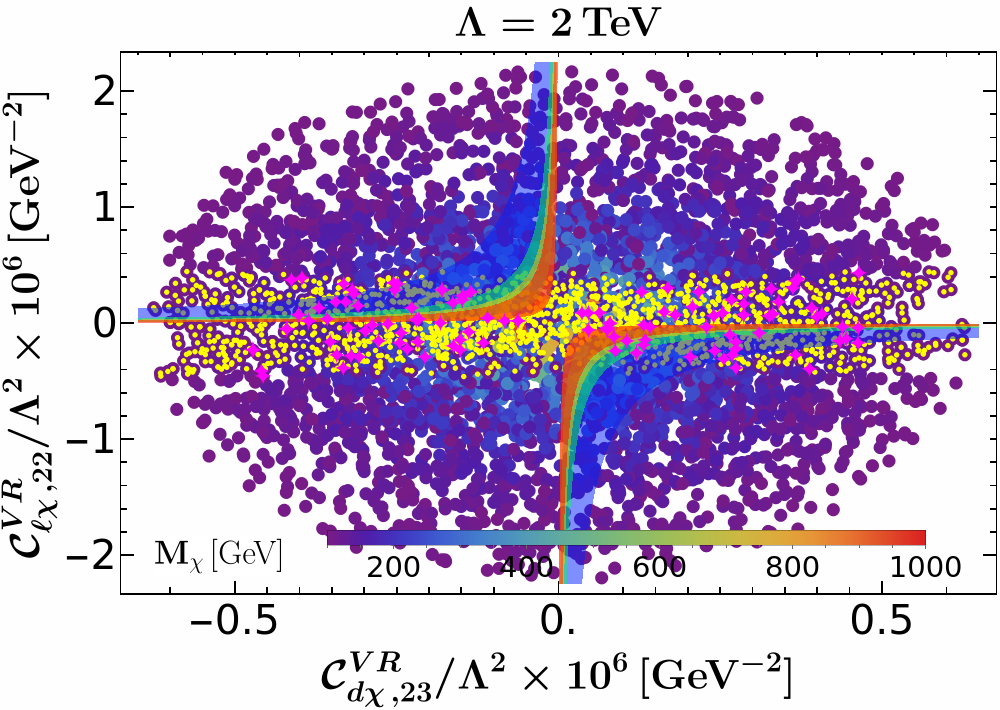}\label{fig:plot_cvLqX23_cvRdX23_cvRlX22_2}}
    \caption{Allowed parameter space from the DM relic density, direct detection, indirect detection, and low-energy flavor observables. The description of the various constraints is similar to Fig. \ref{fig:plot_cvLqX23_cvRqX23_cvLlX22}.}
    \label{fig:plot_cvLqX23_cvRdX23_cvRlX22}
\end{figure}
\begin{figure}[htbp!]
    \centering
    \includegraphics[width=0.5\linewidth]{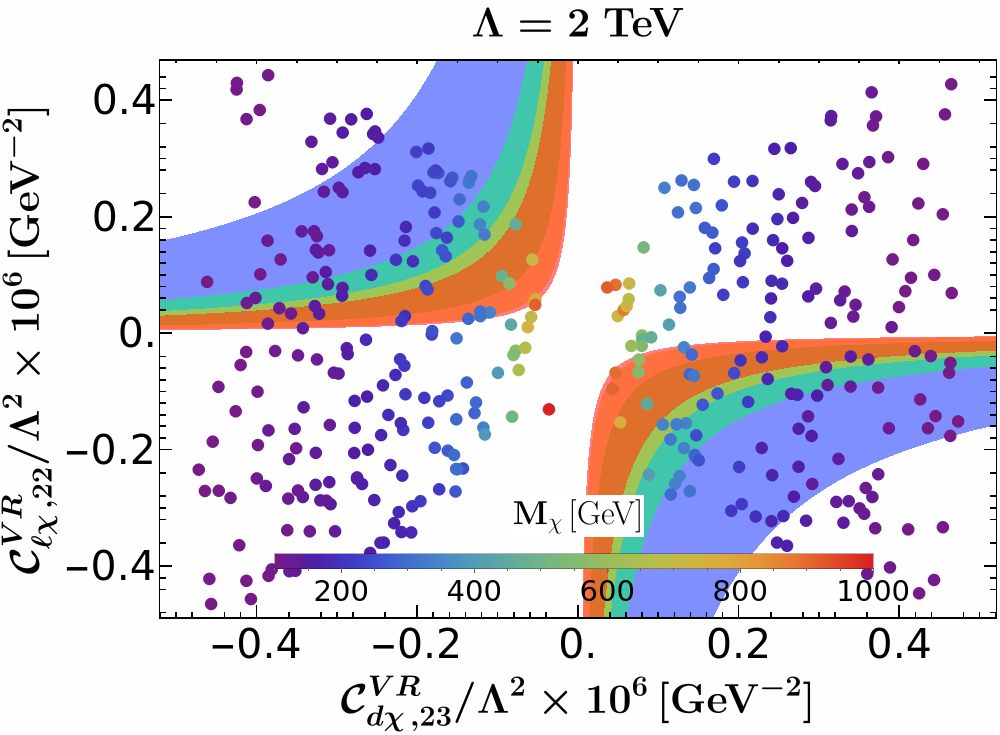}
    \caption{Final allowed parameter space in the quark–lepton coupling plane for the scenario with the WCs $\mathcal{C}_{q\chi,23}^{VL}$, $\mathcal{C}_{d\chi,23}^{VR}$, and $\mathcal{C}_{\ell\chi,22}^{VR}$ considered simultaneously. The description of the various constraints is similar to Fig.~\ref{fig:plot_cvLqX23_cvRqX23_cvLlX22_zoomed}.}
    \label{fig:plot_cvLqX23_cvRdX23_cvRlX22_zoomed}
\end{figure}
The allowed parameter space obtained by simultaneously varying the three WCs $\mathcal{C}_{q\chi,23}^{VL}$, $\mathcal{C}_{d\chi,23}^{VR}$, and $\mathcal{C}_{\ell\chi,22}^{VR}$ is presented in Fig.~\ref{fig:plot_cvLqX23_cvRdX23_cvRlX22}. Figure~\ref{fig:plot_cvLqX23_cvRdX23_cvRlX22_1} displays the corresponding constraints in the $\mathcal{C}_{q\chi,23}^{VL}$--$\mathcal{C}_{d\chi,23}^{VR}$ plane. In contrast to the scenario shown in Fig.~\ref{fig:plot_cvLqX23_cvRqX23_cvLlX22_1}, the relic-density constraint does not induce any pronounced correlation between the two quark WCs. Instead, the dominant feature is a clear dependence on the DM mass, with different mass ranges populating distinct regions of parameter space.

An important observation is that the relic-density constraint allows even very small values of the quark WCs. This is because the annihilation rate of the DM can remain sufficiently large through interactions mediated by the lepton operator, provided the corresponding lepton Wilson coefficient takes an appropriate value. Consequently, the observed relic abundance can be reproduced even when the contributions from the quark operators are strongly suppressed. As in the previous scenario, the subset of points satisfying the indirect-detection constraints, shown in yellow, populate the entire relic-density allowed region. This demonstrates that indirect-detection observables do not impose any significant additional restrictions on the quark WCs. The solid and dashed bands denote the regions allowed by the $B_s^0-\bar{B}_s^0$ mixing observable at the $1\sigma$ and $3\sigma$ confidence levels, respectively. Consistent with the discussion of the two-quark operator scenario, the meson-mixing constraints favor WCs of opposite signs.  Consequently, comparatively large values of both WCs are allowed simultaneously, leading to a broad region compatible with meson-mixing data.

Most importantly, a sizeable overlap between the flavor- and DM-allowed regions is obtained, demonstrating that both sets of constraints can be satisfied simultaneously over a broad range of WCs. Since this plane involves only quark-DM operators, the constraints arising from rare semileptonic decays cannot be represented directly in this parameter space. Their impact is instead illustrated in the following figure, where one of the axes corresponds to a lepton-DM Wilson coefficient.

Figure~\ref{fig:plot_cvLqX23_cvRdX23_cvRlX22_2} shows the allowed parameter space in the $\mathcal{C}_{d\chi,23}^{VR}$-$\mathcal{C}_{\ell\chi,22}^{VR}$ plane. The yellow points represent the parameter points that satisfy the indirect detection constraint. It can be seen that, while the quark coupling remains essentially unconstrained by indirect detection, the allowed range of the lepton coupling is restricted to $|\mathcal{C}_{\ell\chi,22}^{VR}/\Lambda^2| \lesssim 2 \times 10^{-7}~\mathrm{GeV}^{-2}$. The hyperbolic regions correspond to the parameter space allowed by the rare semileptonic decays arising from the $b\to s\ell\ell$ and $b\to s\nu\bar{\nu}$ transitions. As in the previous case, the magenta points denote the parameter points that simultaneously satisfy the dark-sector constraints and the neutral meson mixing bound. It can be seen that a comparatively large region of the parameter space remains consistent with all the constraints considered, making this scenario considerably less constrained than the previous two-quark one-lepton case.

Figure~\ref{fig:plot_cvLqX23_cvRdX23_cvRlX22_zoomed} shows the same coupling plane as Fig.~\ref{fig:plot_cvLqX23_cvRdX23_cvRlX22}. Here, we generate additional points in the common solution region across the different sectors of observables to better illustrate the correlations among the couplings and the resulting bounds. The description of the scatter points and the constraints they satisfy is identical to that in Fig.~\ref{fig:plot_cvLqX23_cvRqX23_cvLlX22_zoomed}. In this case, the resulting bounds on the couplings are $(|\mathcal{C}_{d\chi,23}^{VR}/\Lambda^2|,|\mathcal{C}_{q\chi,23}^{VL}/\Lambda^2|) \lesssim 4.5\times10^{-7}~\mathrm{GeV}^{-2}$ and $|\mathcal{C}_{\ell\chi,22}^{VR}/\Lambda^2| \lesssim 4.5\times10^{-7}~\mathrm{GeV}^{-2}$ for $M_{\chi}$ $\sim 100$ GeV. Higher mass corresponds to more stringent bounds.

\begin{figure}[t]
    \centering
    \subfloat[]{\includegraphics[width=0.5\linewidth]{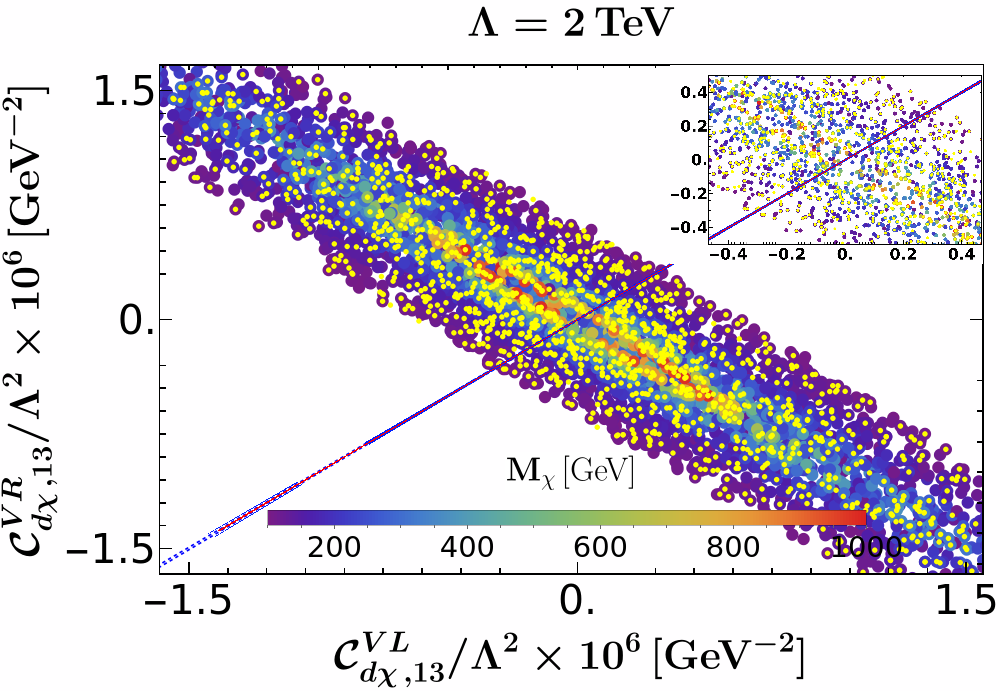}
    \label{fig:plot_cvLdX13_cvRdX13_cvReX22_1}}~~
    \subfloat[]{\includegraphics[width=0.48\linewidth]{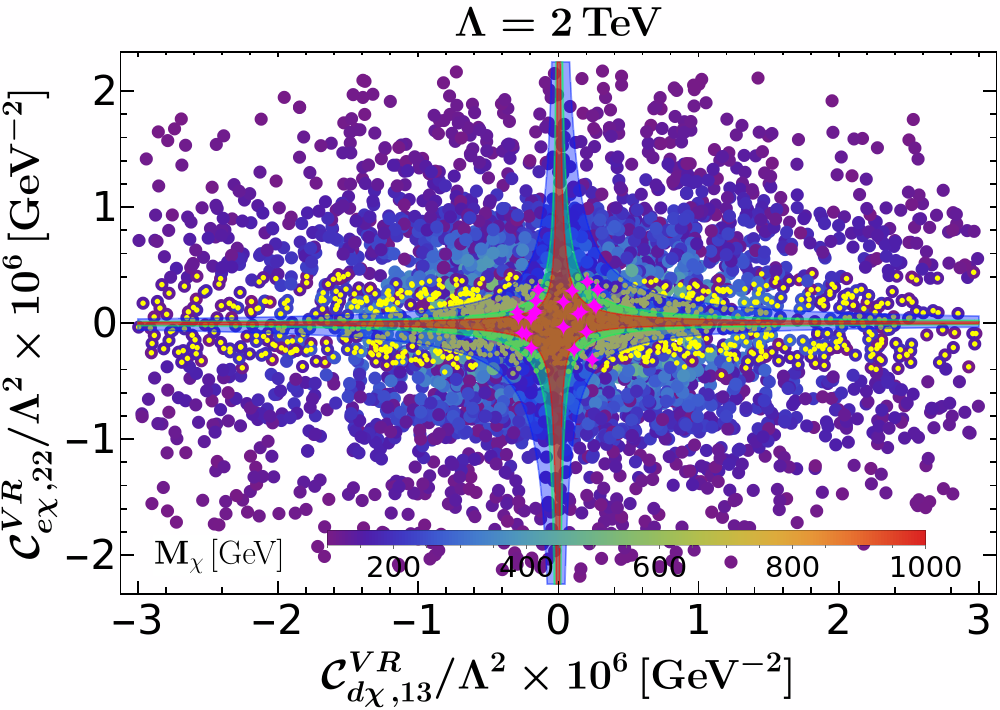}\label{fig:plot_cvLdX13_cvRdX13_cvReX22_2}}
    \caption{Allowed parameter space from the DM relic density, direct detection, indirect detection, and low-energy flavor observables. The description of the various constraints is similar to Fig. \ref{fig:plot_cvLqX23_cvRqX23_cvLlX22}.}
    \label{fig:plot_cvLdX13_cvRdX13_cvReX22}
\end{figure}
\begin{figure}[htbp!]
    \centering
    \includegraphics[width=0.5\linewidth]{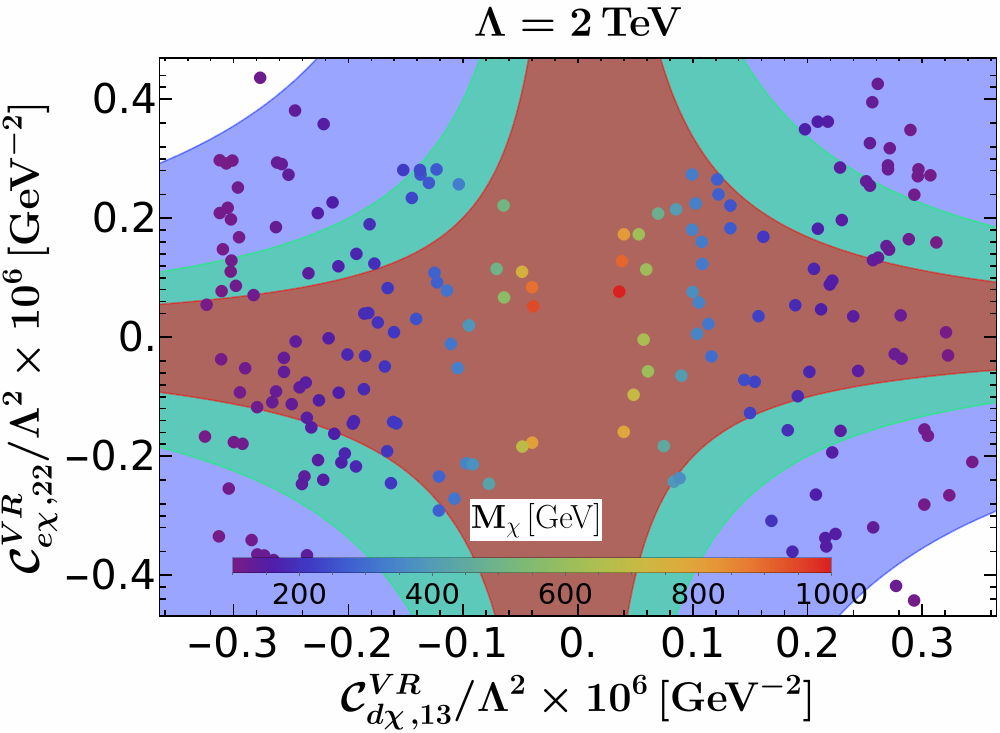}
    \caption{Final allowed parameter space in the quark-lepton coupling plane for the scenario with the WCs $\mathcal{C}_{d\chi,13}^{VL}$, $\mathcal{C}_{d\chi,13}^{VR}$, and $\mathcal{C}_{e\chi,22}^{VR}$ considered simultaneously. The description of the various constraints is similar to that in Fig.~\ref{fig:plot_cvLqX23_cvRqX23_cvLlX22_zoomed}. Here, we have considered the allowed $3\sigma$ range of the relevant neutral meson mixing observable, $\Delta M_d$, for $B^0-\bar{B}^0$ mixing. }
    \label{fig:plot_cvLdX13_cvRdX13_cvReX22_zoomed}
\end{figure}
The previous two scenarios focused on the $b\to s \ell \ell$ transition. We now consider the $b\to d \ell \ell$ transition, for which the allowed parameter space is shown in Fig.~\ref{fig:plot_cvLdX13_cvRdX13_cvReX22}. In this scenario, the three WCs, $\mathcal{C}_{d\chi,13}^{VL}$, $\mathcal{C}_{d\chi,13}^{VR}$, and $\mathcal{C}_{e\chi,22}^{VR}$, are varied simultaneously. Here, the two quark operators have the same quark chirality while the DM current has the opposite chirality, and the lepton operator is right-handed. The relevant flavor observables are the $b\to d\ell\ell$ transitions and the $B^0-\bar{B}^0$ mixing observable. Figure~\ref{fig:plot_cvLdX13_cvRdX13_cvReX22_1} shows the allowed parameter space in the $\mathcal{C}_{d\chi,13}^{VL}$-$\mathcal{C}_{d\chi,13}^{VR}$ plane. The colored points satisfy the relic density constraint and exhibit a strong negative correlation between the two quark couplings. The yellow points denote the parameter points that additionally satisfy the indirect detection constraint. Similar to the previous scenarios, the indirect detection bound does not impose any additional restriction on the quark couplings. The solid and dashed bands correspond to the parameter space allowed by the neutral meson mixing observable $\Delta M_d$ at the $2\sigma$ and $3\sigma$ confidence levels, respectively. As discussed earlier, the measured value of $\Delta M_d$ is not consistent with the SM prediction at the $1\sigma$ level, and therefore we use the $2\sigma$ uncertainty as our reference. Owing to the small experimental uncertainty on $\Delta M_d$, the corresponding NP contribution is also very stringently constrained. Consequently, for the parameter range shown in the figure, the $2\sigma$ allowed region appears almost as a line. Since the relic density and meson mixing constraints favor opposite correlations between the two quark couplings, only a small overlap region remains that satisfies both the dark-sector and meson mixing constraints.

Figure~\ref{fig:plot_cvLdX13_cvRdX13_cvReX22_2} shows the allowed parameter space in the $\mathcal{C}_{d\chi,13}^{VR}$-$\mathcal{C}_{e\chi,22}^{VR}$ plane. As in the previous scenarios, the indirect detection bound primarily constrains the lepton coupling while leaving the quark coupling essentially unchanged. The hyperbolic bands represent the parameter space allowed by the rare semileptonic $b\to d\ell\ell$ decays (Table \ref{tab:fitresults1}). Unlike the previous two scenarios, the global fit to the corresponding WCs is consistent with the SM prediction, implying that the NP contribution is also consistent with zero. As a result, the allowed regions from the semileptonic observables are qualitatively different from those obtained in the $b\to s$ transition. The magenta points denote the parameter points that satisfy the neutral meson mixing constraint together with all the dark-sector constraints. Therefore, the magenta points lying on the semileptonic allowed bands correspond to the parameter space that simultaneously satisfies all the constraints considered in this work.

Figure~\ref{fig:plot_cvLdX13_cvRdX13_cvReX22_zoomed} provides a detailed view of the common allowed region shown in Fig.~\ref{fig:plot_cvLdX13_cvRdX13_cvReX22}, with an increased density of points to better resolve the correlations among the couplings and the corresponding constraints on the WCs. Here, the relevant neutral meson mixing observable is considered within its allowed $3\sigma$ range; however, imposing the more stringent $2\sigma$ range leads to a similar result, as can be seen from Fig.~\ref{fig:plot_cvLdX13_cvRdX13_cvReX22_1}. It is worth noting that the constraints from the semileptonic $b\to d\mu\mu$ transitions are weaker compared to the previous cases, with the corresponding solutions remaining zero-consistent at $1\sigma$. Consequently, almost the entire region allowed by the other constraints remains viable after imposing the semileptonic meson-decay constraints. The resulting bounds on the coupling parameter space are $|\mathcal{C}_{d\chi,13}^{VL(R)}/\Lambda^2|\lesssim3.5\times10^{-7}~\mathrm{GeV}^{-2}$ and $|\mathcal{C}_{e\chi,22}^{VR}/\Lambda^2|\lesssim4.5\times10^{-7}~\mathrm{GeV}^{-2}$. These bounds correspond to a DM mass of $\sim 100$~GeV. As the DM mass increases, the bounds on both couplings become more stringent, as demonstrated in the plot. In all the above plots, the vanishing-WC solution is not allowed, as the meson mixing constraints require both quark couplings to be simultaneously large or small, while the small-coupling solution cannot satisfy the observed relic density.  

\subsection{Possible Model Realizations of the Effective Operators} \label{sec:UV}
The four-fermion operators considered in this work, summarized in Table~\ref{tab:operator_lists}, can arise from a variety of well-motivated simplified models and UV scenarios upon integrating out heavy degrees of freedom. While our analysis is performed in a model-independent dSMEFT framework with a fermionic DM candidate, identifying possible underlying realizations provides a useful connection between the EFT constraints and the parameters of specific high-energy models. In particular, the bounds obtained on the effective operators can be translated into constraints on the masses and couplings of the corresponding mediators. The relevant interactions can arise either at tree level or through loop effects, while different constructions can accommodate the DM sector, the flavor structure, or both. In the following, we discuss several representative realizations of the effective interactions considered in our analysis, starting with simplified models and subsequently considering more UV-motivated gauge extensions.

A representative class of simplified models capable of generating the effective interactions considered here is provided by leptoquark (LQ) models~\cite{Dorsner:2016wpm}. In particular, vector leptoquarks such as $U_{1}\,(3,1,2/3)$ and $U_{3}\,(3,3,2/3)$ generate the relevant semileptonic interactions at tree level and have therefore been extensively studied in the literature~\cite{Becirevic:2016yqi, Kosnik:2012dj, DiLuzio:2017vat, Bhaskar:2021pml, Belanger:2022kvj, Bauer:2015knc, Hiller:2014yaa, Becirevic:2015asa}. Scenarios in which fermionic DM interacts with SM fermions through vector leptoquark exchange can also directly realize the type of four-fermion interactions considered in our dSMEFT framework~\cite{Mandal:2018czf, Mohamadnejad:2019wqb}. Upon integrating out the heavy leptoquark, these interactions can be matched onto the corresponding dSMEFT operators, relating their WCs to the underlying leptoquark masses and couplings. The tree-level nature of this matching allows the constraints obtained in our analysis to be directly translated into bounds on the corresponding model parameters.


Other simplified models can also give rise to the effective interactions of interest. Extensions of the SM containing vector-like quarks (VLQs) together with a scalar or vector mediator can generate four-fermion interactions involving a fermionic DM candidate, which can be matched onto the corresponding dSMEFT operators~\cite{Blanke:2017fum, Belfatto:2025ids, Agrawal:2014aoa}. The operators considered in our work can also be directly realized within $t$-channel fermionic DM models, where integrating out the heavy scalar or vector mediator generates effective interactions between DM and SM quarks~\cite{Goyal:2016zeh, Arina:2025zpi, Arina:2020tuw, Cho:2019fmt, Bhattacharya:2025mlg, Kolay:2026wca}. Depending on the chiral structure of the underlying couplings, these interactions can map onto the vector operators involving $\chi$ considered in our dSMEFT analysis. Similarly, simplified models with a new neutral vector mediator, commonly referred to as $Z^\prime$ models, provide another well-studied realization of the relevant effective interactions~\cite{Davighi:2021oel, Altmannshofer:2016brv, Bishara:2018vix, Jacques:2016dqz}. Integrating out the $Z^\prime$ generates semileptonic operators relevant for $b\to d_i\ell^+\ell^-$ transitions, while a coupling of the $Z^\prime$ to the DM sector can simultaneously generate the corresponding dSMEFT operators involving $\chi$.

The above constructions treat the relevant mediators as simplified degrees of freedom. The same effective interactions can also arise within extended gauge theories, where the mediator is embedded in a larger gauge structure. A well-motivated class of such scenarios is provided by extensions of the SM with an additional $\rm U(1)_x$ gauge symmetry and a fermionic DM candidate. Depending on the choice of charge assignments required for anomaly cancellation, different scenarios such as $xL_i-yL_j$ can arise. These models can give rise to vector interactions with SM fermions, and the exchange of the associated neutral gauge boson can generate the effective operators considered in our analysis. Such contributions can arise at the tree level, while loop effects can also contribute to the corresponding flavor transitions. These scenarios have been explored in various studies in the literature~\cite{Borah:2020swo, Bause:2021prv, Ko:2019tts, Altmannshofer:2016jzy, Bhatia:2017tgo, Baek:2017sew, Datta:2017pfz, Celis:2015eqs, Altmannshofer:2016oaq, Crivellin:2015hha, Alves:2015mua, Allanach:2015gkd}. Specific realizations include, for example, 2HDM frameworks with a gauged $L_{\mu}-L_{\tau}$ symmetry~\cite{Crivellin:2015mga, Duy:2026kxv}.

Another well-motivated class of UV scenarios are the 331 models, based on the gauge group $SU(3)_C \times SU(3)_L \times U(1)_X$, which provide an extension of the SM in which gauge anomaly cancellation requires the number of fermion generations to be three~\cite{Pisano:1992bxx}. These models naturally predict flavor non-universal gauge interactions and tree-level FCNCs mediated by a heavy neutral gauge boson~\cite{Buras:2012dp, Descotes-Genon:2017ptp, Buras:2014yna, Escalona:2025jla}. The tree-level exchange of this boson generates effective semileptonic vector operators of the form considered in our analysis. A different realization is provided by left-right symmetric models, based on the gauge group $SU(2)_L \times SU(2)_R \times U(1)_{B-L}$, which generate contributions to $b \to d_i \ell^+ \ell^-$ transitions at the one-loop level through penguin and box diagrams involving right-handed gauge bosons and scalar states. These loop effects induce effective semileptonic vector operators through radiative corrections~\cite{Dcruz:2023mvf, Zhang:2007da, Morozumi:2024mit, Beall:1981ze}.

The above examples demonstrate that the effective interactions considered in our analysis can arise from a variety of simplified models and UV-motivated scenarios, with both the dSMEFT operators involving the DM and the semileptonic operators relevant for flavor transitions being realizable in suitable constructions. While these scenarios provide possible connections between the EFT and underlying high-energy theories, in this work, we remain within the model-independent dSMEFT framework without assuming any particular UV realization.

From our analysis, we find that the minimal operator setup capable of simultaneously satisfying all currently available flavor and dark matter constraints consists of two quark-DM operators together with a single lepton-DM operator, although scenarios with additional operators remain viable. This result highlights the importance of going beyond the conventional single-operator approach and performing a correlated multi-operator analysis. Indeed, the viable parameter space often emerges only through non-trivial correlations among different WCs, which can significantly relax individual constraints while remaining consistent with all observables. Our study therefore demonstrates that a combined analysis of flavor and DM observables is essential for reliably identifying the allowed regions of the dSMEFT parameter space.

Such correlated operator structures arise naturally in a variety of ultraviolet-complete new-physics scenarios, where several effective operators are generated simultaneously rather than individually. We have discussed a number of representative UV realizations above. In particular, extensions of the Standard Model involving an additional neutral gauge boson $Z^\prime$, originating from an enlarged gauge symmetry such as an extra $U(1)_X$, together with a fermionic DM candidate, can generate the required quark-DM and lepton-DM operator structures and naturally induce the correlations among WCs observed in our analysis \cite{Buras:2012dp, Descotes-Genon:2017ptp, Buras:2014yna, Escalona:2025jla, Dcruz:2023mvf, Zhang:2007da, Morozumi:2024mit, Beall:1981ze}. Similar effective interactions can also be realized in simplified models, such as vector leptoquark scenarios with fermionic DM, through tree-level or loop-induced processes~\cite{Dorsner:2016wpm, Becirevic:2016yqi, Kosnik:2012dj, DiLuzio:2017vat, Bhaskar:2021pml, Belanger:2022kvj, Bauer:2015knc, Hiller:2014yaa, Becirevic:2015asa, Mandal:2018czf, Mohamadnejad:2019wqb}. Consequently, the multi-operator scenarios preferred by our phenomenological analysis are not merely an effective-field-theory possibility but are well motivated from the perspective of realistic UV completions. The tightly constrained regions identified in this work, therefore, provide valuable benchmark targets for future flavor, dark matter, and collider searches aimed at uncovering the underlying origin of flavor-DM interactions.

\subsection{Prediction of the $t \to c \ell \ell (\nu \nu)$ and $b \to d \nu \nu $ branching ratios}
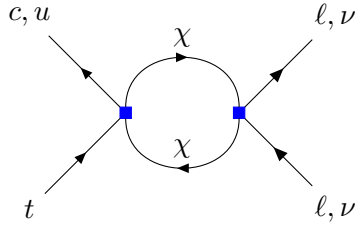
\begin{figure}[H]
		\centering 
		\begin{tikzpicture}
			\begin{feynman}
			\vertex (a1){\( t\)};
			\vertex [square dot,blue,above right=1.8cm of a1](a2){};
			\vertex [above left=1.8cm of a2](a3){\(c, u\)};
			\vertex [square dot,blue,right=1.5cm of a2](a4){};
			\vertex [above right=1.8cm of a4](a5){\(\ell, \nu\)};
			\vertex [below right=1.8cm of a4](a6){\(\ell, \nu\)};
			
			\diagram* {
				(a1) -- [fermion,arrow size=1pt] (a2) -- [fermion,arrow size=1pt] (a3),
				(a2) --[fermion,arrow size=1pt,  half left, looseness=1.5,edge label={\(\chi\)}](a4) --[fermion,arrow size=1pt,  half left, looseness=1.5,edge label'={\(\chi\)}](a2),
				(a6) --[fermion,arrow size=1.2pt](a4) --[fermion,arrow size=1.2pt](a5),
			};
			\end{feynman}
			\end{tikzpicture}
    \caption{Feynman diagram contributing to the top-FCNC process such as $ t \to u_{i}\, \ell \ell (\nu \bar{\nu})$ via double insertion of the dim-6 dSMEFT operators.} \label{fig:t2cll_loop}
\end{figure}

The dimension-6 FCNC operators that we have considered in our analysis are constrained by various low-energy observables. The final allowed parameter space from low-energy flavor and dark-sector observables, and the correlations among the masses and couplings, are discussed in detail in section \ref{sec:DM_flavor_combined}. In this work, we have considered dSMEFT operators that are SMEFT basis extended by a dark sector Dirac fermion $\chi$. So the operators have to be invariant under the SM gauge group, i.e., $\rm SU(3)_c \times SU(2)_L \times U(1)_Y$. The operators of the first two rows of the Table~\ref{tab:operator_lists} show the interaction of the quark and the lepton doublets interacting with the DM bilinear, which gives us the interaction operator (with quarks) like $(\bar{b}_{L} \gamma_{\mu}d_{iL})(\bar{\chi}\gamma^{\mu}P_{L,R}\chi)$. The operator $(\bar{t}_{L} \gamma_{\mu}u_{iL})(\bar{\chi}\gamma^{\mu}P_{L,R}\chi)$, where $(u_i \in u,c)$ which contributes to $t \to c(u) \ell \ell$ and $t \to c(u) \nu \bar{\nu}$ transitions also has the same Wilson coefficient. Similarly, the leptonic operator $\mathcal{Q}^{V,A}_{\ell \chi }$ will contribute to the interaction $(\bar{\nu} \gamma_{\mu} P_L \nu)(\bar{\chi} \gamma^{\mu} P_{L, R} \chi)$. So, from the constraints of the low-energy observables, we can actually predict the branching ratio of the processes related to the top quark decays. In our scenario, we can get these processes via double insertion of the dim-6 operators. The corresponding Feynman diagram can be found in Fig.~\ref{fig:t2cll_loop}. 
We can write the effective interaction of the $t \to u_i \ell \ell (\nu \nu)$ decays as:
\begin{equation}
   \mathcal{L}^{\rm eff}_{t \to u_i \ell \ell (\nu \nu)}  \supset \sum_{j,k=L,R}\mathcal{C}^{\ell}_{V_{jk}} (\bar{u}_i \gamma_{\mu} P_{j}t)(\bar{\ell} \gamma^{\mu}P_{k}\ell)  + \sum_{j=L,R} \mathcal{C}^{\nu}_{V_{jL}} (\bar{u}_i \gamma_{\mu} P_{j}t)(\bar{\nu} \gamma^{\mu}P_{L}\nu)  \,. 
\end{equation}
By ignoring the masses of the light quarks and leptons,  the branching ratio in this effective basis can be written as:
\begin{equation}
    \mathcal{B}[t \to u_i \ell \ell (\nu \nu)] = \frac{m_t^5}{1536 \pi^3 \Gamma_t}  \sum_{j,k=L,R}|\mathcal{C}^{\ell}_{V_{jk}}|^2 (3 |\mathcal{C}^{\nu}_{V_{jL}}|^2 )~, \,
\end{equation}
where $\Gamma_t$ is the total decay width of the top quark, and is given in Table \ref{tab:input}. Experimental upper limits on these decay processes are not available, while the SM predictions are several orders smaller,
ranging from $\mathcal{O}(10^{-17})$ to $\mathcal{O}(10^{-14})$ \cite{Frank:2006ku}. Using the fit results from Table \ref{tab:fitresults}, we get predictions of the branching ratios for these processes of the order of $\mathcal{O}(10^{-17}-10^{-14})$, which is of the same order as the SM predictions.

The dSMEFT operators considered
in this work also contribute to the $b \to d \nu \bar{\nu}$ processes like $B^{+(0)} \to \pi^{+(0)} \nu \bar{\nu}$ and $B^{+(0)} \to \rho^{+(0)} \nu \bar{\nu}$ decays. The $B \to \rho$ form factors are based on the LCSR results from Refs. \cite{Gubernari:2018wyi,Bharucha:2015bzk,Gao:2019lta}. For one of the benchmark points given in Table \ref{tab:fitresults1}, we get the predictions as 
\begin{align}
\rm BR(B^0 \to \pi^0 \nu \bar{\nu}) &= (8.79 \pm 0.38) \times 10^{-6}, \nonumber \\ 
\rm BR(B^+ \to \pi^+ \nu \bar{\nu}) &= (9.46 \pm 0.41) \times 10^{-6}, \\ \nonumber
\rm BR(B^0 \to \rho^0 \nu \bar{\nu}) &= (2.17 \pm 0.13) \times 10^{-5}, \\ \nonumber  
\rm BR(B^+ \to \rho^+ \nu \bar{\nu}) &= (2.34 \pm 0.14) \times 10^{-5},	
\end{align}
 which are lower than the experimental upper limits shown in Table \ref{tab:invisible_data}.

\section{Summary} \label{sec:summary}
In this work, we investigated the phenomenology of dimension-6 dSMEFT operators involving a singlet $\mathbb{Z}_2$-odd fermionic dark matter candidate. Focusing on operators coupled to third-generation quarks and second-generation leptons, we explored the interplay between low-energy flavor observables and dark matter phenomenology within a unified effective-field-theory framework. These operators generate flavor-changing neutral-current processes at one loop while simultaneously governing DM annihilation and detection, providing a unique opportunity to probe the same underlying interactions through complementary observables.

Using neutral meson mixing, rare dileptonic, semileptonic, and invisible $B$-meson decays, we derived stringent constraints on the dSMEFT parameter space. We found that neutral meson mixing provides the strongest flavor constraints, particularly on individual quark operators, whereas rare and semileptonic decays probe products of quark and lepton couplings and therefore provide complementary sensitivity. On the DM side, we studied the relic abundance together with direct and indirect detection observables. The relic-density requirement strongly constrains the WCs and determines the viable mass-dependent parameter space, while direct-detection limits are generally weak due to their loop-induced nature. Indirect detection, especially through the $\mu^+\mu^-$ channel, imposes important constraints on the lepton-sector couplings.

By combining flavor and DM observables, we performed a systematic analysis of four representative operator scenarios. We found that neither the single-quark operator scenario nor the one-quark one-lepton operator scenario can simultaneously satisfy all flavor and DM constraints. The two-quark operator scenario admits viable parameter regions due to correlations among the WCs that relax the meson-mixing constraints. Most importantly, the multi-operator flavor-DM scenario, involving two quark-DM and one lepton-DM operators, successfully accommodates all current flavor and DM constraints, although the allowed parameter space is highly restricted. The surviving regions exhibit non-trivial correlations among the WCs that would be missed in a single-operator analysis.

For $M_\chi\simeq100~\mathrm{GeV}$, the combined analysis requires $ |\mathcal{C}_{\ell(e)\chi,22}^{VL(R)}/\Lambda^2| \lesssim 4.5 \times 10^{-7}~\mathrm{GeV}^{-2}, $ together with $ |\mathcal{C}_{q\chi,23}^{VL}/\Lambda^2 \, (\mathcal{C}_{d\chi,13}^{VR}/\Lambda^2 )| \lesssim 4.5 (3.5) \times10^{-7}~\mathrm{GeV}^{-2}, $ while combinations involving $\mathcal{C}_{q\chi,23}^{VR}$ and $\mathcal{C}_{d\chi,23}^{VR}$ can attain somewhat larger values owing to non-trivial correlations among the WCs. More generally, the combined flavor and DM constraints imply $ |\mathcal{C}_{q(d)\chi,i3}^{VL(R)}| \lesssim 4.5\times10^{-7}~\mathrm{GeV}^{-2} $ for light DM masses, with the allowed values decreasing to the $\leq 10^{-7}~\mathrm{GeV}^{-2}$ range for heavier DM. Very small WCs $\lesssim 10^{-8} \, \rm GeV^{-2}$ is also disfavored by the DM relic density. 

Using the allowed parameter space, we predicted the branching fractions of several FCNC processes, including $t\to c(u)\ell^+\ell^-$ and $b\to d\nu\bar{\nu}$, obtaining values as large as ${\cal O}(10^{-14})$ and ${\cal O}(10^{-5})$, respectively. We also discussed possible ultraviolet completions capable of generating the effective operators considered here. Overall, our study highlights the crucial role of combined flavor and dark matter analyses, demonstrates the importance of multi-operator effects, and identifies well-motivated benchmark regions that can be probed in future flavor, dark matter, and collider experiments.
\subsubsection*{Acknowledgment}
LK would like to thank Subhajit Kala and Dipankar Pradhan for various insightful discussions on several topics. LK acknowledges financial support from the DAE-BRNS YSRP grant No.~57/20/02/2024 and IIT Gandhinagar under project grant No. OTH/R\&D/13467. IR acknowledges financial support from the NSERC of Canada. 
\appendix	
\section{Loop functions} \label{app:loop}
The functions $\mathbb{C}_{i}s$ are given below:
\begin{align}\label{eq:loop_contribution} 
\mathbb{C}_{1} = &  \frac{1}{128 \pi ^2 m_B^2} \Bigg[ m_B^2 \left(2 \left(m_B^2-8 M_{\chi }^2\right) \log \left(\frac{\Lambda ^2}{M_{\chi }^2}\right)+3 m_B^2-16 M_{\chi }^2\right)  \nonumber \\ 
& +2 \sqrt{m_B^4-8 m_B^2 M_{\chi }^2} \left(m_B^2-4 M_{\chi }^2\right) \log \left(\frac{\sqrt{m_B^4-8 m_B^2 M_{\chi }^2}-m_B^2+4 M_{\chi }^2}{4 M_{\chi }^2}\right) \Bigg] \,, \nonumber \\
\mathbb{C}_{2} = & \frac{ M_{\chi }^2}{8 \pi ^2 m_B^2}  \Bigg[ m_B^2 \left(\log \left(\frac{\Lambda ^2}{M_{\chi }^2}\right)+2\right)+\sqrt{m_B^4-8 m_B^2 M_{\chi }^2} \log \left(\frac{\sqrt{m_B^4-8 m_B^2 M_{\chi }^2}-m_B^2+4 M_{\chi }^2}{4 M_{\chi }^2}\right)\Bigg] \,, \nonumber \\
\mathbb{C}_{3}= & \mathbb{C}_{2} \,, \nonumber \\ 
\mathbb{C}_{4}  = & \mathbb{C}_{1} \,.
\end{align}
In the above expression, $M_{\chi}$ denotes the mass of the DM particle $\chi$, and $\Lambda$ is the hard cutoff scale used in this analysis.

\section{Bounds from Rare FCNC Decays}\label{secapp:onlybsll}
\begin{figure}[htbp!]
	\centering
	\subfloat[]{\includegraphics[width=0.4\linewidth]{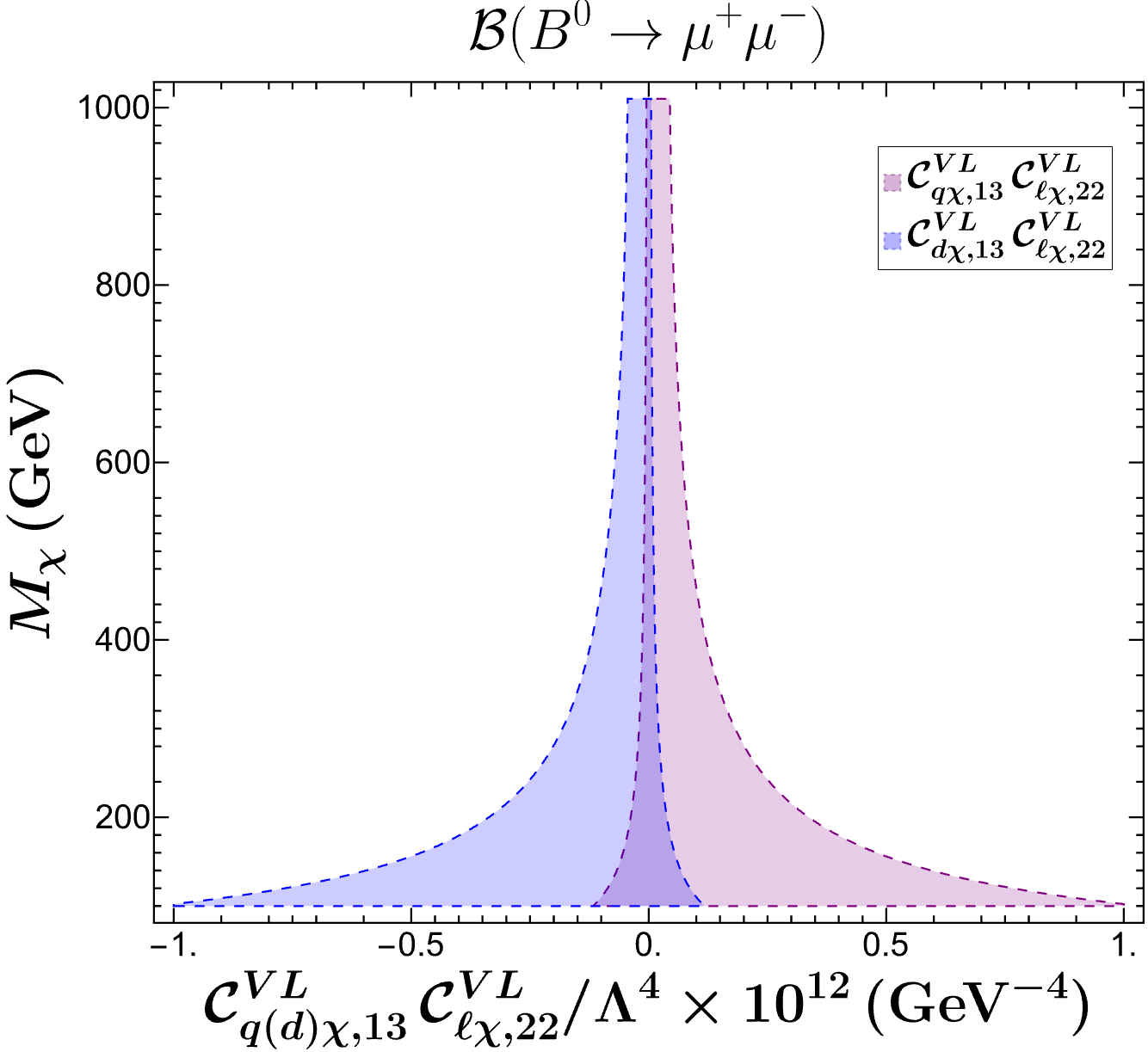}}~
	\subfloat[]{\includegraphics[width=0.4\linewidth]{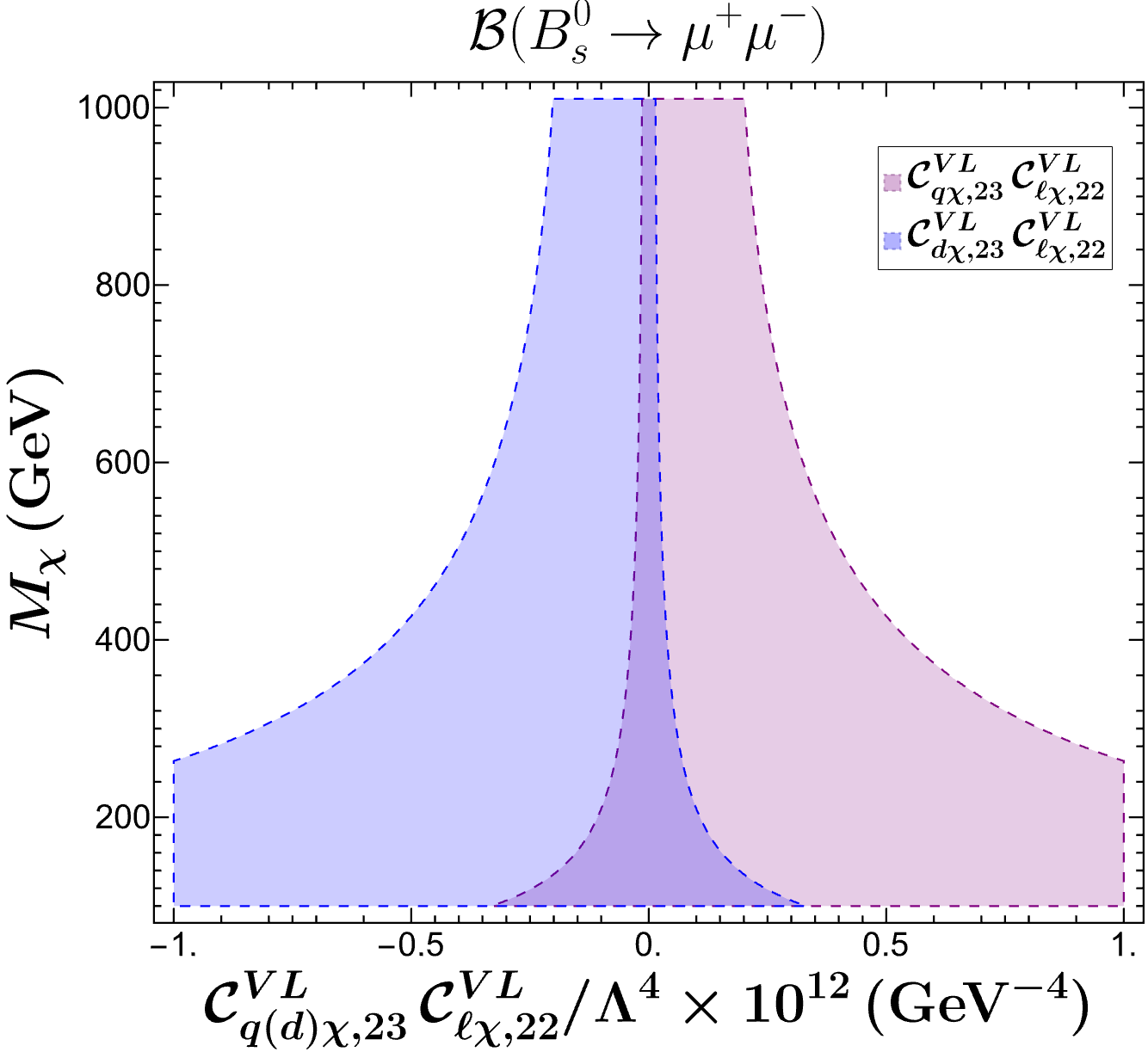}}~
	\caption{Bounds from the rare decays $B_{(s)}^{0} \to \mu^{+} \mu^{-}$ in the mass-coupling plane for a few coupling combinations. }
	\label{fig:rare_plots}
\end{figure}

Figure~\ref{fig:rare_plots} shows the allowed regions in the mass-coupling plane obtained from the rare decays $B_{(s)}^0 \to \mu^{+}\mu^{-}$. The $y$-axis denotes the DM mass, while the $x$-axis represents the product of the quark and lepton couplings. The purple and blue regions correspond to left- and right-handed quark currents, respectively. Specifically, the purple region represents the coupling combinations $\mathcal{C}_{q\chi, i3}^{VL}\mathcal{C}_{\ell(e)\chi, 22}^{VL(R)}$ and $\mathcal{C}_{q\chi, i3}^{VR}\mathcal{C}_{\ell(e)\chi, 22}^{VR(L)}$, whereas the blue region corresponds to $\mathcal{C}_{d\chi, i3}^{VL}\mathcal{C}_{\ell(e)\chi, 22}^{VR(L)}$ and $\mathcal{C}_{d\chi, i3}^{VR}\mathcal{C}_{\ell(e)\chi, 22}^{VL(R)}$, where $i=(1,2)$ corresponds to the $B^{0}$ and $B_s^{0}$ mesons, respectively. For DM masses around $100~\mathrm{GeV}$, the allowed values of the coupling products are of the order of $\mathcal{O}(10^{-12})\,\mathrm{GeV}^{-4}$. As the DM mass increases, the constraints become progressively stronger, improving by nearly two orders of magnitude for masses around $1~\mathrm{TeV}$. An interesting feature is the asymmetry between positive and negative values of the coupling products. In the purple region, negative values are more strongly constrained than positive values, whereas the opposite is observed in the blue region, reflecting different interference patterns between the SM and new-physics contributions. Similar to neutral meson mixing, the decay $B^{0}\to\mu^{+}\mu^{-}$ provides stronger constraints than $B_s^{0}\to\mu^{+}\mu^{-}$. In our framework, the new-physics contribution is not subject to CKM suppression. Consequently, the smaller measured branching ratio of $B^{0}\to\mu^{+}\mu^{-}$ translates into more stringent bounds on the allowed mass-coupling parameter space.

\bibliographystyle{JHEP} 
\bibliography{dark_SMEFT}

\end{document}